Ministry of Education and Science of the Russian Federation

Lobachevsky University
The National Research University of Nizhny Novgorod

Anatoly M. Bragov, Andrey K. Lomunov

**Use of the Kolsky Bar Method for Studying
High-Rate Deformation Processes in
Materials of Various Physical Natures**
(originally published in Russian in 2017)

*Monograph*

Publishing House of Nizhny Novgorod State University
2020

**Summary**

An analysis is presented of the applicability of the Kolsky technique for studying the high-rate deformation processes of structural materials under compression. Errors in the method caused by inertia and friction are estimated. Modifications of the Kolsky method for studying the properties of materials subjected to various stress-strain states are considered. Original modifications of the method are presented that significantly expand the range of mechanical properties that can be investigated. Issues connected with using the Kolsky method to study the properties of brittle materials are analyzed.

The original monograph version in 2017 was based on research funded by project RSCF 15-19-10032. The production of a new version of the monograph, including a substantial update of the content and conversion to English version of the figures, formulas and tables, has been completed with the financial support of the Ministry of Science and Higher Education of the Russian Federation (task 0729-2020-0054).

The authors are sincerely grateful to Stephen Walley for his help (using Yandex) in translating the monograph into English and updating the review of worldwide Kolsky bar studies.

*Printed with the approval of the*
*Scientific and Methodological Council of the Research School*
*"Computers and Experimental Mechanics"*

Table of Contents

## INTRODUCTION

Many modern load-bearing and protective structures used in the military, aerospace, energy and civilian sectors can be exposed to the intense short-duration effects of explosions, shocks or impact both during normal use and as a result of terrorist acts and man-made or natural disasters. In recent years, the problems of ensuring the safety of military and industrial facilities, nuclear power plants, large public buildings, as well as the container transport of explosives, toxic and radioactive substances have become particularly urgent. Materials of various physical nature can act as damping components of protective structures and containers: metals, wood, polymers, composites, ceramics, porous refractories, etc.

Numerical codes such as, for example. ANSYS, LS-DYNA, and ABAQUS are widely used to calculate the stress-strain state and strength of structures in such situations. One of the most important components of these codes that governs the accuracy of the calculations are the constitutive relations and fracture criteria or materials. These describe the complex behaviour of materials such as strain hardening and strain rate effects. They also need to be able to take account of the history of changes in the loading, deformation, anisotropy, etc. To populate material behaviour models and fracture criteria with the necessary parameters and constants, an extensive database is required of the dynamic properties of materials of various physical types (metals and their alloys, polymers and their composites, ceramics, wood, soils, rocks, concrete etc.). However, these properties have not been fully studied, and the results are often incomplete, contradictory or absent. This is especially true for new promising nano-, polymer, composite and ceramic materials.

The study of the patterns of behaviour of materials of various physical types for a wide range of changes in temperature, strain rate and load amplitude is one of the urgent problems in the experimental mechanics of deformable solids. A special place in such research is occupied by the study of the effect of strain rate and strain rate change on the physical and mechanical properties of materials for strain rates in the range $10^2$ - $10^5$ s$^{-1}$. A systematic study of these effects was begun at the beginning of the last century by the Hopkinsons, father and son. Since then, a huge number of different dynamic experiments have been performed worldwide and many different materials have been tested. Regular conferences and symposiums on high-rate deformation (SCCM, IUTAM, EUROMECH, ISIE, DYMAT, etc.) show that interest in the study of the behaviour of materials under dynamic loading is not diminishing. This is due both to the appearance of new materials and also the improvement of numerical methods that allow more complex mechanical behaviour to be modelled. This in turn leads to the need to develop new experimental methods to populate such models with the necessary parameters.

# 1. USING THE KOLSKY METHOD TO STUDY THE PROCESSES OF HIGH-RATE DEFORMATION OF MATERIALS

The systematic study of the dynamic properties of materials began in the late 19th to early 20th centuries with the research of father and son John and Bertram Hopkinson [1-6]. In Russia, the study of the properties of materials at high strain rates was begun in the 1930s by Nikolai N. Davidenkov [7-12] and was continued by F.F. Wittman, N.A. Zlatin, V.A. Stepanov, Yu.Ya. Voloshenko-Klimovitskii, A.A. Ilyushin, V.S. Lensky, and R.A. Vasin, etc. Abroad, the works of R.M. Davies, A. Nadai, J. Duffy, G.I. Taylor, H. Kolsky, J.D. Campbell, W. Lindholm, J. Bell, and A. Kobayashi are devoted to the solution of this issue. In recent times, systematic studies of the processes involved in the high-rate deformation of various materials may be found in the publications of S.A. Novikov, G.V. Stepanov, A.P. Bolshakov, A.M. Bragov, C. Albertini, J.E. Field, G.T. Gray, W.G. Proud, J.R. Klepaczko, J. Harding, F.E. Hauser, G. Gary, Z. Rosenberg, S. Nemat-Nasser, and others.

Unfortunately, in the 1970s in the USSR only machines for conducting quasistatic tests under various loading conditions were mass-produced. In those years, there were no international or Russian standards for performing dynamic experiments, and therefore no high-rate testing machines were manufactured. So the main attention of researchers was oriented towards creating a methodology for such tests. Areas of interest included loading devices, means of measuring forces, displacements and strains, as well as techniques for obtaining dynamic deformation diagrams. A large number of different loading devices were designed and built, along with gauges for measuring forces and strains. However, the development of new loading methods is still ongoing, as is as the search for non-invasive techniques for the reliable measurement of stresses and strains that do not alter the mechanical properties of materials under pulsed loading.

The most common methods for obtaining dynamic stress-strain curves are: (i) drop-weight tests (tensile or compressive) for which $\dot{\varepsilon} = const.$ [13-15]; (ii) the cam plastometer [16, 17]; (iii) the methods of H. Kolsky [18] and G.I. Taylor [19]. There are also less commonly used methods, such as studying the expansion of ring specimens under the influence of internal pressure (for example, by exploding a wire in water [20, 21], the method of A.A. Ilyushin and V.S. Lensky, based on the theory of propagation of one-dimensional elastic-plastic waves [22-24] and the distribution of residual strains along the length of the specimen, etc. [25].

Commercial pile drivers can be adapted for performing dynamic drop-weight tests, but their main disadvantages are: (i) low strain rates in the specimen if they are used without propulsion); (ii) repeated blows of the drop-weight on the anvil or specimen; (iii) a complex pattern of wave propagation in the anvil, which is not considered when analysing the deformation of the specimen. Other disadvantages of traditional drop-weight tests include a lack of understanding of the actual processes by which the specimen deforms over time, which results in errors in the calculation of stress-strain curves, especially at large deformations.

A significant disadvantage of drop-weight tests is the inability to set, maintain and control the desired loading law during a test. This disadvantage was partially overcome by the development of the cam plastometer [16, 17], a type of rotary moving-weight machine. The active element of the device is a cam with a logarithmic profile, which results in an almost constant strain rate.

The presence of wave phenomena, as well as the inhomogeneity of stress caused by friction on the end surfaces, local buckling and stress concentration at the points of attachment of the specimen to loading or measurement devices, are a significant drawback in the testing of long rod specimens. To eliminate these shortcomings, Hoggatt & Recht proposed the study of thin annular or long tubular specimens dynamically internally loaded [26]. Under the action of an axisymmetric radial pressure, a uniaxial stress state is realized in a thin ring, whereas in a long tubular specimen plane strain is obtained, both at high strain rates. These methods have the following important advantages [20]:
   - wave processes the specimen can be neglected;
   - a uniform stress distribution is created in the ring;
   - the recording of pressure (load) and strain can be performed using devices that are not physically attached to the specimen and therefore do not affect the stress and strain fields in the specimen.

The loading of ring specimens can be performed using mechanical multi-sector devices (Figure 1) [27] by the pressure of a liquid or a gas applied to the specimen either directly or through an intermediate elastic element [28-30]. These systems provide a maximum strain rate of about $10$ $s^{-1}$. To increase the strain rate range up to $10^4$ $s^{-1}$, loading with an explosive blast can be used [31-35]. In this case, the explosive is usually blown up in the inner cavity of a thick-walled cylinder on which the annular specimen is placed, and the ring is not stretched by the pressure of the powder gases, but only due to the inertial forces of the inner cylinder (the so-called free expansion).

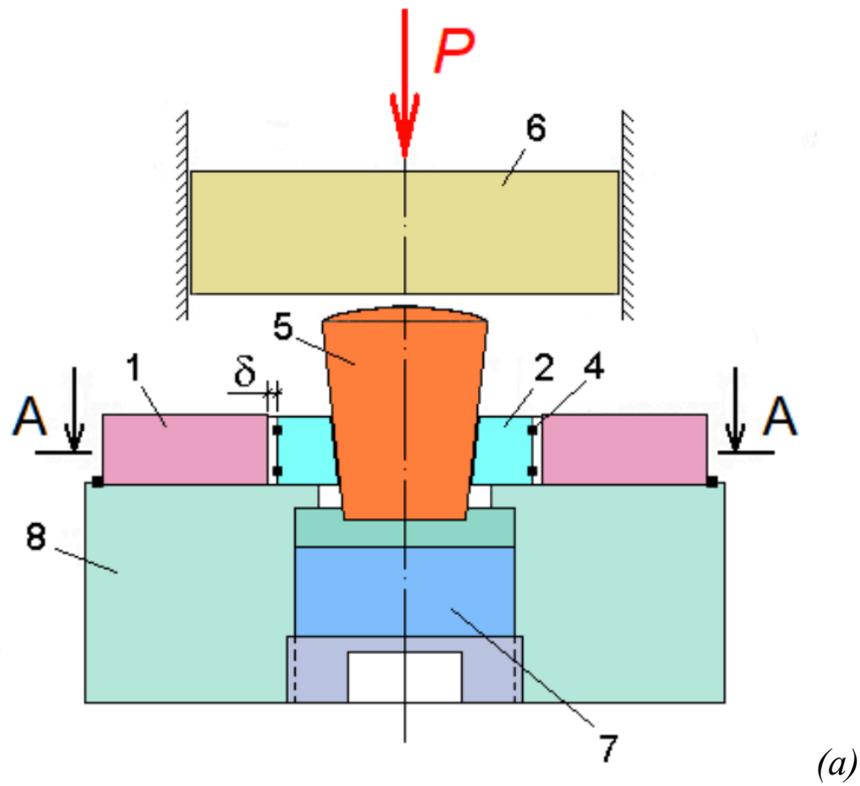

*(a)*

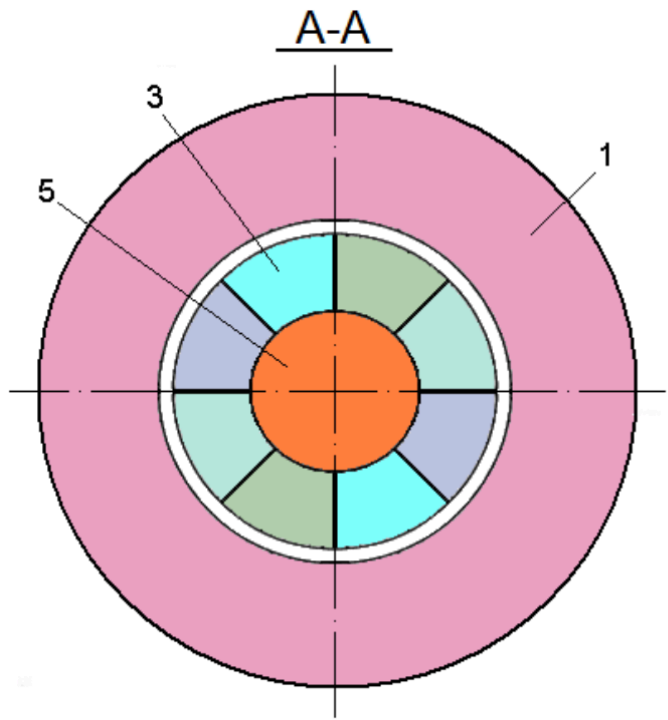

*(b)*

*Figure 1. Schematic diagrams of multi-sector device for loading ring specimens. (a) Side view. (b) Top view. From [27].*

The device works as follows: the annular specimen (labelled 1) is stretched by the action of the expanding mandrel (labelled 2), which consists of several sectors (labelled 3). The mandrel is held together by means of elastic rubber rings (labelled 4). The expansion force is created by the movement of the conical expanding element (labelled 5), which is set in motion as a result of the impact of the striker (labelled 6).

The load on the specimen is determined by measuring the deformation of the sector (labelled 3) of the expanding mandrel (labelled 2) using a strain gauge (not shown in the figure) glued to the side of the sector.

A feature of this technique is the presence of a gap $\delta$, due to which the influence of the rest inertia of the mandrel-expanding element system on the recorded load in the specimen is eliminated and the transition process is completed before loading of the specimen begins. In addition, the elastic vibrations excited in the mandrel (labelled 2) and in the expanding element (labelled 5) upon impact are suppressed by the damping element (labelled 7) mounted in the housing (labelled 8) along the axis of motion of the expanding element (labelled 5).

The method of constructing stress-strain diagrams provides for the consideration of the free delayed expansion of a ring with an initial radial velocity acquired under the action of an internal pressure pulse. For a thin ring, the circumferential stress in the absence of pressure is [36]:

$$\sigma(t) = -\rho R_0 \ddot{W}(t) \tag{1}$$

where $\rho$ is the density, $R_0$ is the initial radius of the ring, $W(t)$ is the radial displacement of the wall of a ring. If an internal force (hydraulic or pneumatic pressure) acts during the test, it must be measured or accurately calculated and then entered into the equation of motion. The deformation of the specimen is determined by measuring the instantaneous diameter or radial strain of the ring:

$$\varepsilon(t) = \frac{W(t)}{R_0} \tag{2}$$

As can be seen from the above dependencies, in the case of free deformation of ring specimens, only the radial expansion of the specimen is recorded.

For this purpose, high-speed photography [32], strain gauges [20], or a shadow optical system with a laser light source and a photomultiplier [33] are used. However, to calculate the stresses the experimental curve $W(t)$ needs to be differentiated twice, which may produce significant errors as the procedure involves complex smoothing spline functions in order to approximate the experimental curve. Thus, despite its considerable simplicity, the expanding ring technique has not yet been widely adopted due to it having insufficient accuracy in determining stresses.

The Taylor impact test has been used for more than 70 years as a simple and convenient method for determining the compressive strength properties of materials at high strain rates [19, 37-41]. This method is based on the assumption of one-dimensional propagation of elastic-plastic waves in a short cylindrical rod when it collides with a rigid barrier. Taylor obtained a relation that relates the dynamic yield strength of the rod material, $\sigma_T$, to the initial impact velocity $V_0$, the initial $\lambda_0$ and final $\lambda_1$ lengths of the rod, as well as the length $H$ of the undeformed section after the experiment:

$$\sigma_Y = \frac{\rho V_0^2}{2} \cdot \frac{(\lambda_0 - H)}{(\lambda_0 - \lambda_1)} \cdot \frac{1}{\ln(\lambda_0 / H)} \ . \qquad (3)$$

Note that this relation was obtained for a rigid-plastic material making the assumption that the constitutive relation $\sigma = \sigma(\varepsilon)$ is independent of the strain rate. Later Purtov proposed taking into account the strain rate hardening of the material to obtain a more accurate value of the yield strength [42].

The Taylor test is a simple and convenient technique for the high-rate compressive testing of materials, in which strain rates of $10^4$-$10^5$ s$^{-1}$ can easily be obtained even at relatively low impact velocities. Erlich & Shockey developed an important variant of the Taylor method involving the symmetric collision of two rods, thereby eliminating the effect of friction at the rod ends on the value of the measured dynamic yield strength [43]. In order to obtain a dynamic deformation graph, Kolsky & Douch used a single Hopkinson bar instead of an anvil in order to record the axial force at the impacted end of the specimen [44]. The idea for this came from the war-time investigations of Taylor & Davies into the dynamic mechanical properties of explosives (this study was not published in the open literature until 1958 [40]).

Taylor impact has been used to measure the dynamic mechanical properties of metals and alloys at elevated temperatures [45, 46] and to obtain the dynamic strength properties of polymers [47-52].

The Taylor test is of considerable interest to defence research laboratories because it is easy to implement and provides data on the behaviour of materials in the strain rate range between that of the split Hopkinson pressure bar (SHPB) and plate impact.

At the present time Taylor impact is widely used for verification of equations of state and constitutive models of material behaviour based on the comparison of the shape of the rod after (or even during) a test with the results of numerical simulation [41, 53-63].

Among the methods of dynamic testing we have mentioned, the one that is most widely used is Kolsky's method of using an SHPB due to its good theoretical foundations, its ease of implementation, and its great informativeness, since it permits control of the changes in the strain rate during a test, in addition to generating stress-strain curves.

## 1.1. The Kolsky bar method and its modifications

Bertram Hopkinson's idea of using a bar to measure the shape of loading pulses was ultimately revolutionary for dynamic testing methods [6]. The principle of operation of a Hopkinson bar is the determination of dynamic stresses, strains, or displacements occurring at the end of the bar from data obtained at some distance from the end. A perturbation that occurs at the end of a long elastic bar propagates along it without distortion (apart from the very high frequency components) at the velocity, $c$, of an elastic wave $c = \sqrt{E/\rho}$ (where $E$ is Young's modulus and $\rho$ is the density of the bar material). Therefore, a strain gauge bonded to the bar registers the force at the end of the bar as a function of time, but with some time delay.

The SHPB (or Kolsky bar) technique was developed during the Second World War by Kolsky, Taylor, Volterra and Davies [18, 37, 64, 65], but it only began to be widely used after the 1970s (Figure 2). The technique allows the testing of a wide range of materials (see Figure 2) in the strain rate range $10^2$-$10^4$ s$^{-1}$. Kolsky suggested that high strain rates could be achieved by placing two Hopkinson pressure bars either side of a specimen [18]. It has since become one of the most widely used devices in experimental practice for studying the behaviour of a material at high strain rates (Figure 2). A review paper has been published recently which contains a bibliography outlining the history of the method [6].

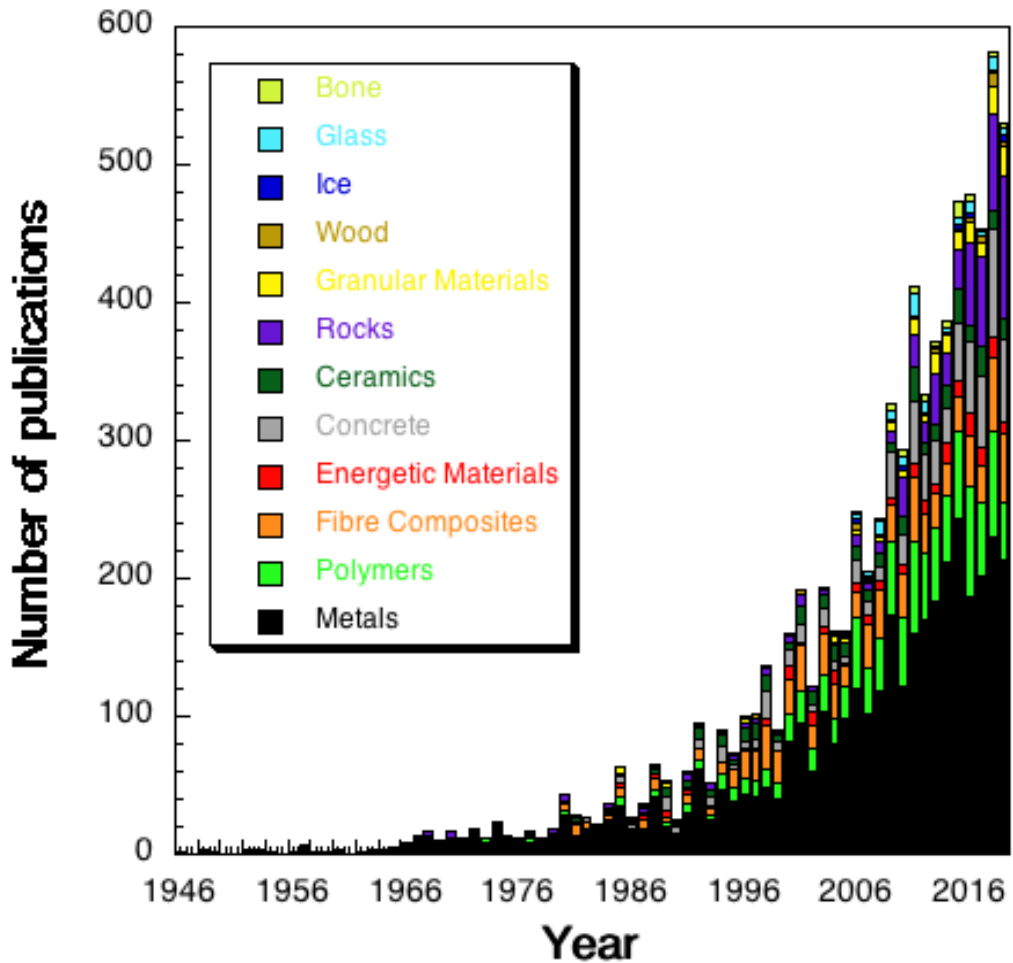

*Figure 2. Number of articles each year since 1946 in which the Kolsky bar method was used to obtain high rate data for various materials. From [6].*

Unfortunately in Russia, the Kolsky bar technique was not widely used until the early 1980s. This was because no critical investigation had been made of the assumptions made in its analysis. After this had been done, the method was used extensively by Stepanov and co-workers [66] at the Institute of Strength Problems in Kiev as part of a vertical test bench. In this configuration, a long measuring bar or tube waveguide was used to measure the stress in a specimen loaded under tension.

The classic SHPB is made of two long thin metallic bars with a high yield strength, between which a small sample of the material under study is positioned. The yield strength of the measuring bars must be higher than the yield strength of the specimen. The bar-specimen system is loaded with an elastic pulse, which is recorded using strain gauges bonded half-way along the bars. A plot of the dynamic specimen deformation can be calculated based on the one-dimensional theory of elastic waves.

The main assumption of the method is that due to the small length of the specimen compared to the length of the loading pulse, a uniaxial stress state is produced in the specimen which means it has a uniform distribution of stresses and strains throughout its thickness. Thus, despite the high strain rate imposed on the specimen, the test can be considered as quasistatic.

The main advantages of the SHPB technique are simplicity of implementation, correct theoretical justification of the phenomena occurring in the system of two long thin elastic bars with a short elastic-plastic specimen placed between them, accurate determination of the significant strain (tens of percent) occurring within the specimen due to the indirect method of measurement and the insignificant mechanical inertia of the strain gauges used to record strain pulses in the bars, and the exclusion of bending of the specimen due to its small length. In addition, the technique allows recording of the history of changes in the strain rate during the entire specimen deformation process.

To date, in addition to the main type of SHPB (for compression tests), other versions have been developed for performing, for example, tension, torsion, or biaxial loading. Descriptions of various versions of the SHPB can be found in the following publications [36, 67-76].

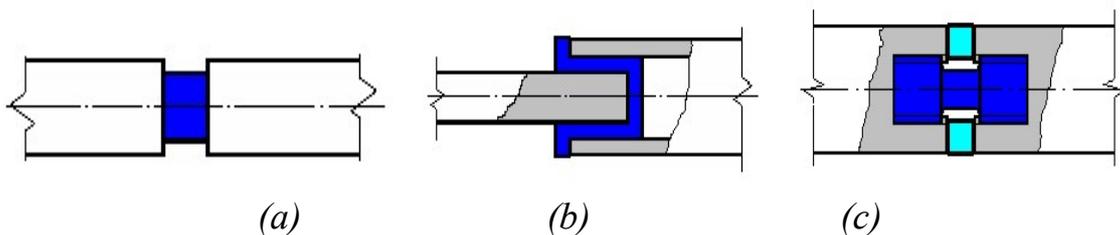

*(a)* *(b)* *(c)*

*Figure 3. Schematic diagrams of the main types of SHPB compression and tension tests. (a) Classic compression. (b) Top-hat tensile. (c) Classic tension.*

An early design for a tensile Kolsky bar system was proposed by Lindholm & Yeakley [77]. In their design, the specimen was hat-shaped and was supported by a thin-walled tube (Figure 3b). The disadvantage of this design is the presence of shear stress and strain components in the specimen, which introduces significant error into the definition of the stress and strain state. In addition, this design is not applicable for testing viscoelastic materials, such as polymers [78].

In 1981 Nicholas proposed another design for a tensile Kolsky bar [72]. In his design, a classic tensile specimen is loaded by a tension wave, which is generated after a compression wave is reflected from the free end of the output bar. To prevent plastic deformation of the specimen during the passage of the compression wave, a split collar was placed around the gauge section between the input and output bars (Figure 3c). Similar designs for tensile Kolsky bars have been implemented by a number of other authors [79-82].

Kishida & Senda designed a tension system that made use of only one Hopkinson bar [83]. In their design, a cylindrical rod of the test material with a length of 600 mm and a diameter of 8 mm was connected using friction welding to a measurement bar of the same diameter. The other end was rigidly fixed to a frame. A tensile wave was excited in the measurement bar, which then loaded the specimen. The strain pulses were recorded using strain gauges bonded both to the measurement bar and also directly to the specimen. This allowed the determination of stresses and strains in the specimen.

A torsional variant of the SHPB was proposed by Duffy $et$ $al.$ [84]. Both the bars and the specimen were thin-walled tubes. To operate it, the input bar was twisted about its axis along part of its length and then released using a mechanical clamp. The amplitude and duration of the shear wave were determined by the amount of spin. Despite the fact that this method allows testing at strain rates up to 30,000 s$^{-1}$, the effect of inertia on the resulting stress-strain curve is insignificant [85].

A biaxial Hopkinson bar system for simultaneous torsion and compression was designed by Lewis & Goldsmith in 1973 [86]. Stiebler $et$ $al.$ later performed tests of an austenitic steel (X2 CrNiMoNNb) under conditions of combined tension-torsion loading, which made it possible to construct yield surfaces at different strain rates [87].

The use of the traditional version of the Kolsky bar for more than one loading cycle in a single test is associated with certain difficulties. First, the correct registration of strain pulses in successive cycles is complicated by interference effects of waves in finite-length rods. Second, for equal length bars, the transmitted strain pulse, $\varepsilon_t$, that passed through the specimen and reflected from the rear end of the output bar in the first loading cycle, will arrive back at the specimen at the same time as the reflected strain pulse, $\varepsilon_r$, that reflected from the specimen and then returned to it again, thereby distorting the shape of the second loading cycle. In order to prevent the return of the transmitted pulse, Lindholm used a momentum trap bar placed in physical contact with the rear end of the output bar [88]. This design permitted only one additional loading cycle of the specimen.

In order to correctly record and analyse several loading cycles, Zhao & Gary developed the necessary mathematical apparatus to analyse a bar system in which strain gauges were placed on each measuring bar at two locations [89]. This allowed the identification of the corresponding strain pulses in both bars for each loading cycle. This design provided for the recording of up to four loading cycles during testing of a foam, which permitted following the behaviour of the material during compaction up to 80% strain.

Bragov & Lomunov used striker bars made from rods of two different materials in order to load specimens with pulses of a single sign but with a variable interval between loading cycles [90, 91]. The amplitude and duration of the loading pulses were varied using a number of different materials and lengths of projectiles. The interval between pulses was changed by varying the gap between the rods. A striker consisting of two components of the same material, but with different cross-sectional areas, produces a similar mode of loading [92].

In contrast to Lindholm [88], where repeated loading was produced by a pulse reflected from the specimen during the first loading cycle, Ogawa achieved alternating loading using a stepped loading bar [93]. In this case, the amplitudes and signs of the loading pulses were set by selecting the cross-sectional areas of the bar. The device is quite versatile, but difficult to implement, since the use of a rod with a variable cross-section requires analysis of the propagation of elastic strain waves in it. A similar technique was devised by Novikov *et al.* to study the dynamic Bauschinger effect [75]. A detailed analysis of the propagation of elastic strain waves in rods having a variable cross-section was performed by Bacon & Lataillade [94].

For testing steels and alloys under conditions close to uniaxial strain and 3D stress, Bhushan & Jahsman proposed placing the specimen in a massive collar to restrict radial expansion [95]. The use of a collar with an effective modulus of 280 GPa allowed the measurement of the dependence of the second invariant of the stress tensor, $J_2$, on the work of plastic deformation, $W$, for a number of aluminium alloys.

For testing materials at elevated temperatures using an SHPB, a coaxial electric heater is usually used, inside which the specimen as well as the ends of the measurement bars are placed. This raises the problem of changes in the impedance $Z = A\sqrt{E\rho}$ (where $A$ is the cross-sectional area of the bars, $E$ is their modulus of elasticity, and $\rho$ is their density) along the axis of the bars with increasing temperature [96]. This problem arises because the forces and displacements in the specimen are measured using strain gauges that are bonded to the bars at a distance from the specimen.

On the basis of solving the one-dimensional heat conduction equation and making the assumption of constant heat flux along an austenitic steel rod (whose coefficient of linear expansion $\beta = 17 \cdot 10^{-6}$ K$^{-1}$) Bacon *et al.* showed that the value of $A\sqrt{\rho}$ does not vary more than 9% in the temperature range 0 to 1000°C [97].

Thus the influence of thermal expansion of the bars can be ignored and it is necessary only to take into account the change in the elastic modulus $E = E_0[1 - \beta(T - T_0)]$, where $E_0$ is the elastic modulus at room temperature.

In addition to compression, tension and torsion testing of metals and alloys, the SHPB technique can be used to study the high rate behaviour of materials with low acoustic impedance, such as low-density polymers, rubber, etc. For this purpose, polymer rods are often used [98]. In this paper Yunoshev & Silvestrov showed that the difference between the results obtained for D16 Duralumin using polymer (PMMA) and steel Hopkinson bars were within experimental error (Figure 4). Siviour *et al.* came to a similar conclusion the same year using metallic Hopkinson bars with a wide range of impedances [99]. The propagation of strain waves in such viscoelastic rods is accompanied by significant dispersion, which necessitates correction of the initial strain pulses when calculating the stress-strain curve. To solve this problem, algorithms based on the solution of the modified wave equation were proposed by Zhao & Gary [89]. A detailed analysis of dispersion effects in finite rods made of various materials has been carried out [94, 100].

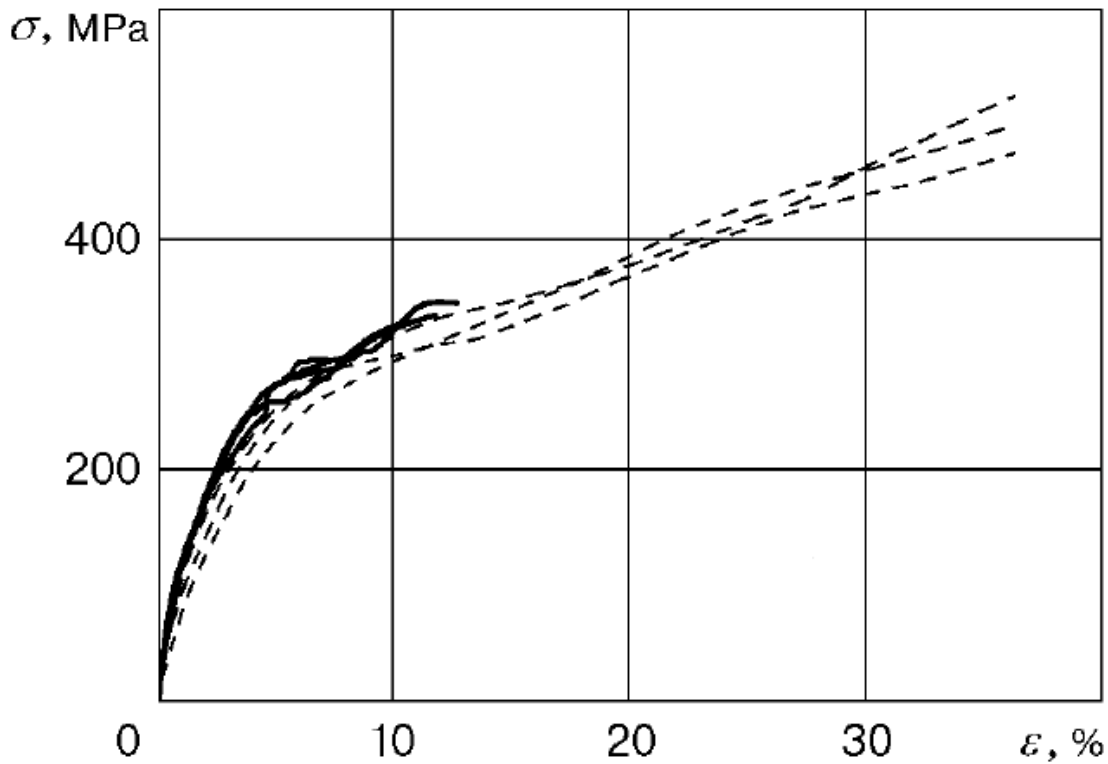

*Figure 4. Dynamic stress-strain curves for D16 Duralumin obtained using Hopkinson pressure bars made from steel (solid lines), strain rate ca. 6000 s⁻¹, and PMMA (dashed lines), strain rate ca. 2200 s⁻¹. From [98].*

As mentioned before, one of the options for creating a complex loading history of specimens in the SHPB is to use rods with variable cross-sections. The possibility of using such rods for testing materials with low acoustic impedance was considered by Bacon & Lataillade [94]. However, it is possible to obtain an analytical solution for the form of the strain pulses (from which the stress-strain curves are calculated) only in a few simple cases [101, 102].

To expand the range of strain rates of the SHPB method, miniature bar systems have been constructed [103-108]. The use of specimens with a diameter of 1 mm and a length of 0.5 mm and measuring bars with a yield strength of more than 2 GPa has allowed testing of a number of metals including titanium, tungsten and copper at strain rates of $\sim 10^5$ s$^{-1}$ [105, 109-111]. However, when testing such thin specimens it is necessary to take into account friction at the interfaces between the bars and the specimen [112, 113].

In SHPB tests, the amplitude of the strain pulse in the input bar is limited by the yield strength of the rod material. This in turn limits the specimen strain rate. For testing at strain rates of $10^4$ s$^{-1}$ and higher, Dharan & Hauser proposed loading the specimen directly with a striker [68]. In subsequent studies, this design was used for testing various steels and alloys both in compression [114] and in shear [115]. Samanta used the direct impact method to test copper and aluminium at elevated temperatures [116]. Gorham, Pope & Field constructed a miniature direct impact system with a 3 mm diameter high-strength tungsten alloy bar which allowed the testing of specimens with a diameter of 1 mm and a length of 0.5 mm so as to be able to neglect inertial effects at strain rates up to $10^5$ s$^{-1}$ [103, 105, 117].

In tests using the direct impact method, the stresses in the specimen are determined in the same way as in the Kolsky method based on the recording of the strain pulse in the output bar. To calculate the displacements in the specimen, assumptions have to be made about the constancy of the velocity of the impacted end of the specimen and the rigidity of the impactor during the test. These assumptions can only be accepted in the case of testing of metals with a low yield strength. When testing high-strength steels, the direct impact method gives significant errors in determining the specimen strain. To avoid this disadvantage, Klepackzo used an optical displacement gauge to measure displacements in a directly impacted specimen [115]. Gorham and co-workers used high-speed photography for the same purpose [103, 105].

The classic Kolsky bar and variants of it are being used to determine the dynamic strength and deformation properties of a wide range of brittle materials, including rocks, ceramics, and concretes, both at normal and elevated temperatures [118-121] [122-124]. Rodríguez *et al.* proposed a method for determining the dynamic fracture strength of brittle materials using the indirect tension test, which is also called the splitting or Brazilian test [125]. In normal compression tests on cylindrical specimens, the load is applied along the longitudinal axis. However, in the Brazilian test, in order to determine the tensile strength, the load is applied across the diameter (Figure 5). This test was originally developed by a Brazilian researcher (hence the name) to determine the quasistatic tensile strength of concrete [126, 127]. An ASTM standard exists for this test method [128].

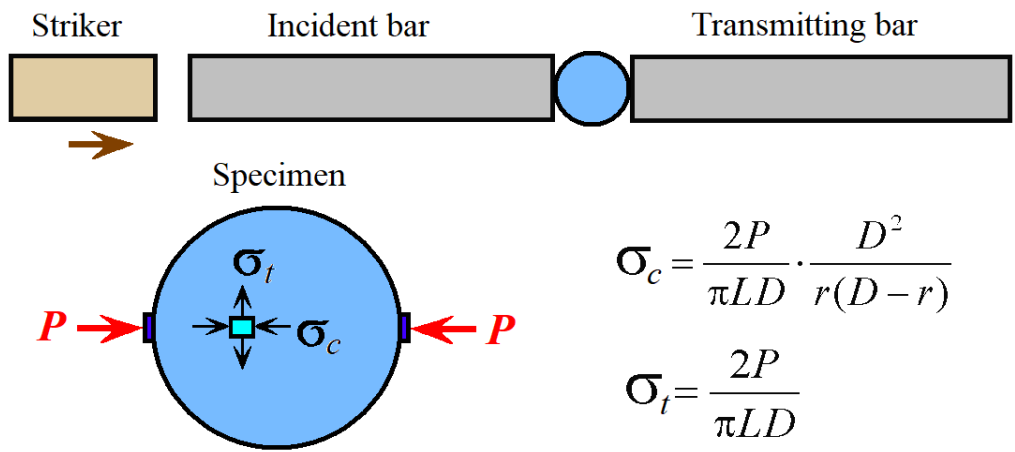

*Figure 5. Schematic diagram of the indirect tension, splitting or Brazilian test implemented in a Kolsky bar.*

The experimental setup and the formulas for calculating the compressive ($\sigma_c$) and tensile ($\sigma_t$) stresses are shown in Figure 5. Here $L$ is the length of the cylindrical specimen, $D$ is its diameter, and $r$ is the coordinate measured along the specimen radius. The contact force $P = A_b E_b \varepsilon^T(t)$, where $A_b$ is the cross-sectional area of the bars, $E_b$ is the Young's modulus of the bar material, and $\varepsilon^T(t)$ is the strain pulse that passed through the specimen to the output bar. The expressions for the compression and splitting stresses are obtained from the elastic solution of the Hertz contact problem [129](pages 107-108). Similar dynamic tests using the Kolsky bar method were performed by Ross *et al.* [130] and numerically analyzed by Tedesco & Ross [131].

In the Laboratory for the Dynamic Testing of Materials of the Research Institute of Mechanics at the Nizhny Novgorod State University (NNSU), the SHPB technique has been developed since the mid-1970s. Up to the present time, various modifications of the method (including some that were developed in-house) have been mastered and successfully applied, allowing a wide range of studies of the mechanical behaviour of materials of various physical types to be carried out (Table 1).

Analysing the variants of the Hopkinson bar technique described above, the implementation which is the simplest and which also has the best theoretical and experimental justification is the SHPB compression scheme, the highest strain rate can be obtained by the torsion version and also by direct impact compression. The greatest degree of deformation and the smallest errors in determining the stress-strain behaviour is given by the torsion variant.

**Table 1. Schematic diagrams of various Kolsky bar test methods.**

| Compression | | | | |
|---|---|---|---|---|
| | Uniaxial stress | 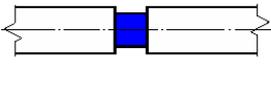 | Bauschinger effect | 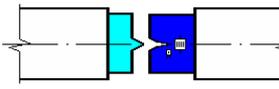 |
| | Uniaxial strain | 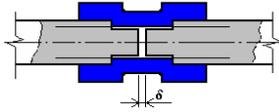 | | 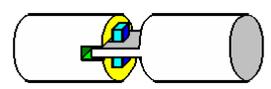 |
| | | 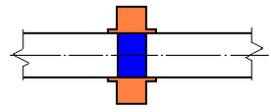 | Indentation | 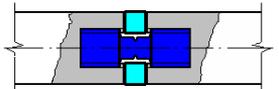 |
| Tension | | 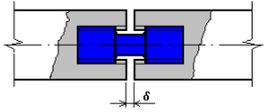 | Brazilian test | 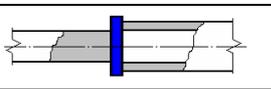 |
| | | 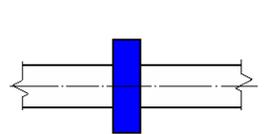 | | 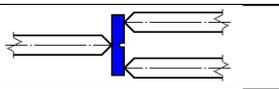 |
| | |  | Crack resistance | 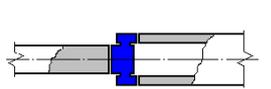 |
| Shear | | 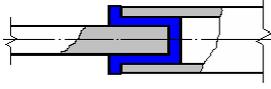 | | 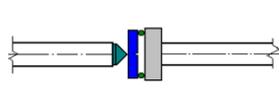 |
| Circular shear | | 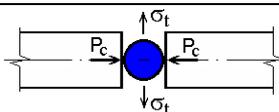 | | 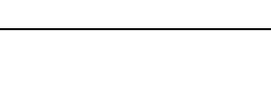 |
| | | 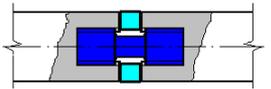 | | 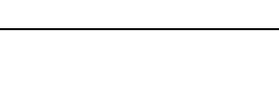 |
| | | 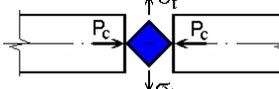 | Direct impact | 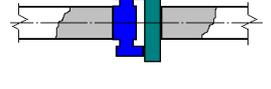 |
| Friction factor | | 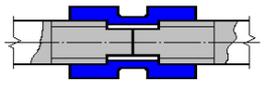 | Punching of a plate | 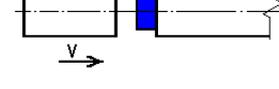 |

The following section gives a derivation of the dependency of the Kolsky bar equations, based on the assumptions of an ideal mathematical model, discusses the differences with the actual physical model of the SHPB, as well as the estimated ranges of possible loading conditions in which these deviations do not cause significant distortion of the results. In addition, an analysis is given of errors that distort the true picture of the deformation along with proposed ways of addressing them. The limits of applicability of the method are identified and a modification of the method is described for implementing alternating loading so as to produce a complex history of strain rate changes. A methodology for testing granular materials such as soil is described.

### 1.1.1. The basic equations of the Kolsky compression bar

The method makes use of the properties of elastic waves propagating within rods, and is based on the exact solution of the wave equation. The mathematical model of the SHPB is a system of three rods: two that are both 'infinitely strong' (very high yield stress) and 'infinitely long' thin rods with a (by comparison) 'soft', very short specimen placed between them. The assumptions of the Kolsky analysis are then as follows:

– due to the very small length of the specimen compared to the length of the loading pulse, a uniaxial stress state is realized in the specimen during the test with a uniform distribution of stresses and strains through its thickness. This is the main premise of the method. Thus, despite the high strain rates within the specimen, the test can be considered to be quasistatic;

– the elastic limit of the measurement bars must be significantly higher than the yield strength of the specimen;

– when elastic waves propagate, there is no dispersion in the bars;

– the distribution of the strain profile within the cross-section of the rod is uniform;

- there are no transverse vibrations within the bars.

A one-dimensional elastic compression wave $\varepsilon^I(t)$ is excited at the end of one of the bars wither by impact or by the detonation of a small amount of explosive. This elastic wave propagates along the bars at a velocity $C$. A picture of the resulting wave propagation in an SHPB is shown in Figure 6 as a Lagrangian $x$-$t$ diagram. Upon reaching the specimen, this wave splits due to the discontinuity in acoustic impedance, $Z$, introduced by the specimen. Part of the wave, $\varepsilon^R(t)$, is reflected back up the input bar and part, $\varepsilon^T(t)$, passes through the specimen into the output bar. Importantly the specimen undergoes elastic-plastic deformation, while the bars deform elastically. The amplitudes and shapes of $\varepsilon^R(t)$ and $\varepsilon^T(t)$ are determined by the ratio of the acoustic impedance of the bar and the specimen materials, as well as by the response of the specimen material to the applied dynamic load. By recording these elastic waves using gauges, it is possible to determine the stress, strain, and strain rate in the specimen using the equations proposed for the first time by Kolsky [18, 132].

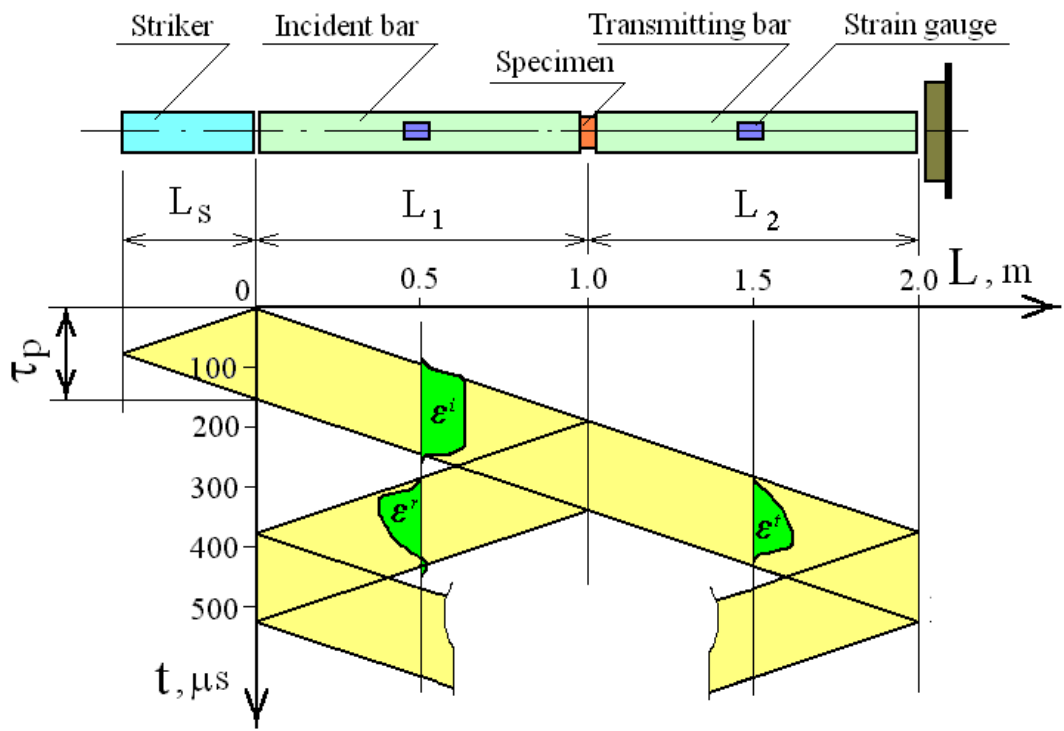

*Figure 6. Diagram of the Kolsky compression bar method and associated Lagrangian x-t diagram.*

When deriving the governing equations of the SHPB method, Kolsky assumed that the specimen has a small length (by 'small' is meant that the time for the wave to pass through the specimen is significantly less than the duration of the loading pulse). This has the consequence that the stress-strain state of the specimen is close to uniform and the elastic-plastic deformation close to quasistatic, even though occurring at a high strain rate. Subject to these conditions, simple the stress, strain, and strain rate in the specimen can be calculated from the elastic strain pulses recorded in the bars [18, 36, 88, 133].

When deriving the Kolsky equations, compressive pulses in the bars are taken as positive. That means that the reflected pulse (which is usually tensile) has a negative sign in the equations.

From the one-dimensional theory of elastic wave propagation in semi-infinite rods, it is known that the strain in the wave is related to the momentum d$U$/d$t$ by the simple ratio [20]:

$$\varepsilon(t) = \frac{1}{C} \cdot \frac{dU}{dt} \ , \qquad\qquad (4)$$

where $U(t)$ is the displacement of particles in the wave:

$$U(t) = C \int_0^t \varepsilon(t) \cdot dt \quad . \tag{5}$$

Figure 7 schematically shows the loading of a specimen by compression pulses in the SHPB system. The aim is to calculate the displacement of the ends of bars 1 and 2 that are in contact with the specimen. The displacement of the left-hand face is $U_1(t)$ which consists of the sum of the displacement $U_1^I(t)$ due to the pulse $\varepsilon^I(t)$ plus the displacement $U_1^R(t)$ due to the pulse $\varepsilon^R(t)$:

$$U_1(t) = C \int_0^t \varepsilon^I(t) \cdot dt + (-C) \int_0^t \varepsilon^R(t) \cdot dt = C \int_0^t \left( \varepsilon^I(t) - \varepsilon^R(t) \right) \cdot dt \tag{6}$$

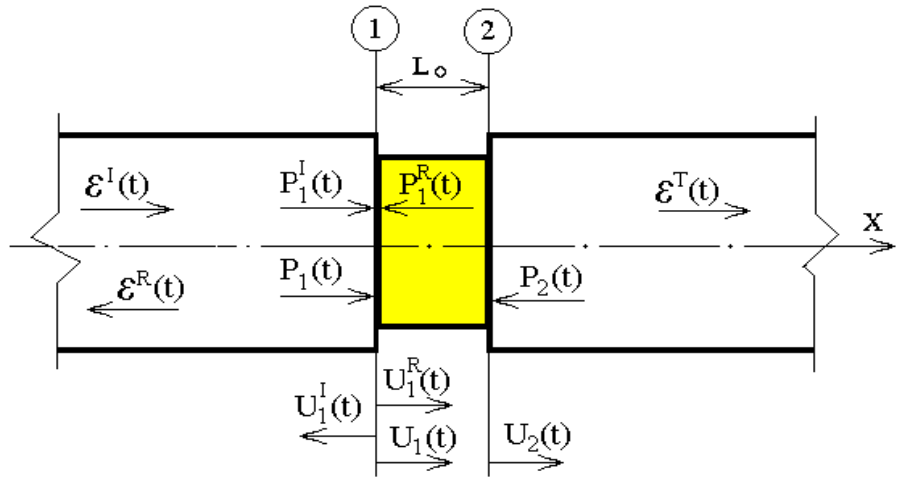

*Figure 7. Diagram showing the displacements of the ends of the specimen due to the dynamic loading produced by the elastic strain pulses in the bars.*

The displacement of the right-hand face $U_2(t)$ is due to the pulse $\varepsilon^T(t)$:

$$U_2(t) = C \int_0^t \varepsilon^T(t) \cdot dt . \tag{7}$$

The average strain of a specimen with length $L_0$ will then be equal to:

$$\varepsilon_s(t) = \frac{U_2(t) - U_1(t)}{L_0} \quad , \tag{8}$$

or, if we express it in terms of the elastic strain pulses in the bars:

$$\varepsilon_s(t) = \frac{C}{L_0} \int_0^t \left[ \varepsilon^I(t) - \varepsilon^R(t) - \varepsilon^T(t) \right] \cdot dt . \qquad (9)$$

Hence the specimen strain rate is given by:

$$\dot{\varepsilon}_s(t) = \frac{C}{L_0} \cdot \left( \varepsilon^I(t) - \varepsilon^R(t) - \varepsilon^T(t) \right) \qquad (10)$$

The stress in the specimen is calculated by considering the forces on the specimen ends. The force at the left-hand end $P_1(t)$ consists of the compressive force $P_1^I(t)$ caused by the pulse $\varepsilon^I(t)$, and the force at the right-hand end is caused by the pulse $\varepsilon^T(t)$. Since the bars have a high elastic limit and hence deform elastically, we may write:

$$P_1(t) = EA \left[ \varepsilon^I(t) + \varepsilon^R(t) \right] ; \qquad (11)$$

$$P_2(t) = EA \, \varepsilon^T(t) , \qquad (12)$$

where $E$ is the Young's modulus and $A$ is the cross-sectional area of the bars. The average force is $P = [P_1(t)+P_2(t)]/2$, hence the average value of the stress in the specimen is given by:

$$\sigma_s(t) = \frac{P}{A_s} = \frac{EA}{2 A_S^0} \left[ \varepsilon^I(t) + \varepsilon^R(t) + \varepsilon^T(t) \right] . \qquad (13)$$

where $A_S^0$ is the initial cross-sectional area of the specimen. As already noted, the stress in the specimen is almost uniform due to its length being short and the incident pulse being of long duration. So it can be assumed with sufficient accuracy that the forces at the ends of the specimen are equal. It follows that:

$$\varepsilon^I(t) + \varepsilon^R(t) = \varepsilon^T(t) . \qquad (14)$$

Substituting this expression in equations (9), (10), and (13), we obtain the following simple expressions for calculating the stress, strain, and strain rate in the specimen, which are most commonly used in practice:

$$\sigma_s(t) = \frac{EA}{A_S^0} \varepsilon^T(t)$$

$$\varepsilon_s(t) = -\frac{2C}{L_0} \int_0^t \varepsilon^R(t) \cdot dt \qquad\qquad (15)$$

$$\dot{\varepsilon}_s(t) = -\frac{2C}{L_0} \cdot \varepsilon^R(t).$$

Rewriting equation (14) in the form of $\varepsilon^R(t) = \varepsilon^T(t) - \varepsilon^I(t)$ and substituting in equations (15), we obtain another set of equations for calculating the stress and strain in the specimen. In a generalized form, these equations can be written as follows [134]:

$$\varepsilon_s(t) = \frac{C}{L_0} \int_0^t \varepsilon_1(t) \cdot dt$$

$$\dot{\varepsilon}_s(t) = \frac{C}{L_0} \cdot \varepsilon_1(t) \qquad\qquad (16)$$

$$\sigma_s(t) = \frac{EA}{2A_S^0} \varepsilon_2(t).$$

One of the three following expressions can be substituted for $\varepsilon_1(t)$ and $\varepsilon_2(t)$:

| to calculate the strain: | to calculate the stress: |
|---|---|
| $\varepsilon_1(t) = \varepsilon^I(t) - \varepsilon^R(t) - \varepsilon^T(t)$ | $\varepsilon_2(t) = \varepsilon^I(t) + \varepsilon^R(t) + \varepsilon^T(t)$ |
| $\varepsilon_1(t) = 2 \cdot \left( \varepsilon^I(t) - \varepsilon^T(t) \right)$ | $\varepsilon_2(t) = 2 \cdot \left( \varepsilon^I(t) + \varepsilon^R(t) \right)$ |
| $\varepsilon_1(t) = -2 \cdot \varepsilon^R(t)$ | $\varepsilon_2(t) = 2 \cdot \varepsilon^T(t)$ |

Thus it is possible to calculate the stresses and strains in the specimen using any two or all three pulses in the bars (a total of 9 sets of equations). On the one hand, this is convenient when for some reason it was not possible to reliably record one of the pulses, and on the other hand, if all three pulses are present, it is possible to compare the deformation diagrams obtained from different sets of equations to assess the accuracy of the main premise of the Kolsky method (equation (14)).

It should be noted that the most accurate are equations (9) and (13), while the most convenient and simple are equations (15), so they are the ones most often used in practice to calculate a stress-strain curve.

Then if time is eliminated from the parameters $\sigma_s(t)$, $\varepsilon_s(t)$ and $\dot{\varepsilon}_s(t)$ that have been obtained, a stress-strain plot may be constructed for the specimen at a known strain rate. This dependence is used to control the change in the strain rate during the deformation process or to evaluate the effect of the history of the change in the strain rate on the resulting stress-strain plot when loading with a complex pulse shape. In addition, the dependence $\dot{\varepsilon}_s \sim \varepsilon_s$ can be used for further construction of a dynamic stress-strain plot for a material at a constant strain rate based on a series of experiments at various strain rates by using a special mathematical apparatus [32] in the case when, for some reason or other, the strain rate during testing was not constant.

The above analysis of the Kolsky compression method is also valid for tensile tests. In this case, the strain pulses $\varepsilon^I(t)$, $\varepsilon^R(t)$ and $\varepsilon^T(t)$ will have opposite signs.

To calculate a true stress true strain plot for large deformations, the stress and strain values need to be corrected. To do this, instead of using the original cross-sectional area of the specimen, $A_s^0$, to calculate the stress, the current value should be used, which is determined using the incompressibility condition applied to the specimen, namely:

$$A_s(t) = \frac{A_s^0}{1 \pm \varepsilon_s(t)} \ . \qquad (17)$$

The minus sign is used in compression, and the plus sign is used in tension.

Then the true stress is calculated as follows:

$$\sigma_s^{tr}(t) = \sigma_s(t) \cdot \left( 1 \pm \varepsilon_s(t) \right) \ . \qquad (18)$$

The true (or logarithmic) strain is calculated using the following expression [135]:

$$\varepsilon_s^{tr}(t) = \ln \left( 1 \pm \varepsilon_s(t) \right). \qquad (19)$$

Thus, when constructing a dynamic diagram, the engineering stress and strain in the specimen are first calculated (for example, using equations (16)), and then these values are modified in accordance with equations (18) and (19) so that a true stress true strain plot is constructed. For small strains (less than 0.15), the true and engineering values of strain are the same to within the accuracy of the test.

## 1.1.2 Analysis of the applicability of the Kolsky bar method

Due to its versatility and simplicity, the Kolsky method has found very widespread usage in the practice of dynamic experiments. Therefore, it has become the object of deep theoretical and experimental critical analysis, aimed at testing the underlying assumptions and identifying the limits of its applicability. As correctly noted by Nicholas [36], none of the known methods of dynamic testing has been subjected to such strict verification as the SHPB method.

The accuracy and reliability of the results obtained using the SHPB method are influenced by the following methodological factors:
- inhomogeneity of the stress-strain state of the specimen due to its finite size (resulting in axial and radial inertia) and the presence of friction at the specimen-bar interfaces;
- dispersion in the propagation of elastic waves in the bars;
- parasitic vibrations produced when the loading pulse is generated;
- influence of the frequency characteristics of the measurement circuitry;
- possible processing errors when constructing the deformation diagram.

The errors associated with these factors were discussed in detail by Lomunov [134]. As noted, when selecting the geometry of specimens in accordance with the Davies & Hunter criterion [136]) inertial corrections do not exceed 5% at strain rates up to $10^4$ s$^{-1}$ (Figures 8 & 9).

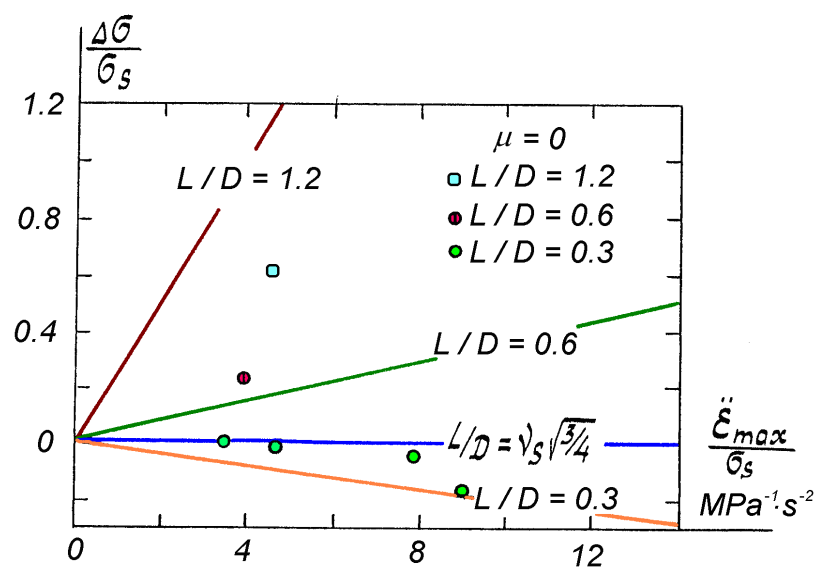

*Figure 8. Effect of specimen geometry (L/D) on the measured stress.*

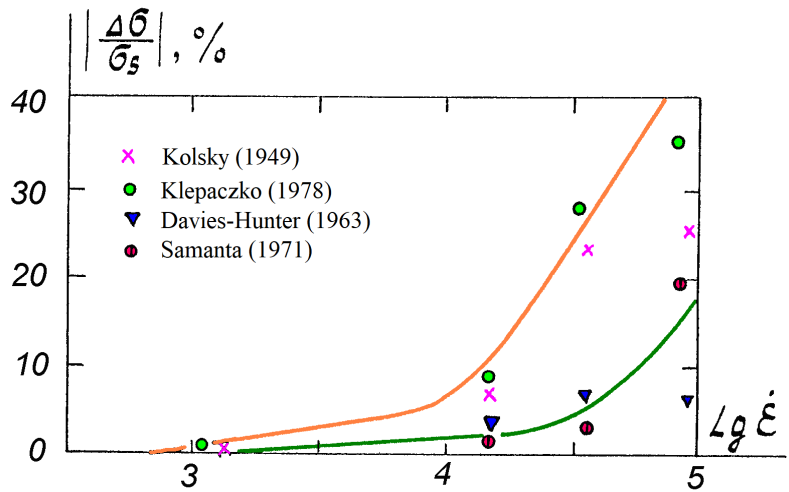

*Figure 9. Inertia corrections proposed by various authors as a function of strain rate.*

The effect of friction on the ends of specimens having the optimal geometry determined by Davies & Hunter [136] can be ignored if the ends of the specimen are carefully polished. Analysis of the effects that introduce errors in the recording of elastic waves in the bars has shown that low-inertia foil strain gauges with a length less than 5 mm, glued at a distance of more than 10 bar diameters from their ends, can reliably be used to record strain pulses propagating in the bars. Dispersion effects that occur for elastic waves propagating in an SHPB system can be minimized using Fourier analysis, or by loading the bar system through a plastically deformable layer. As shown above, the accuracy of synchronization of the initial pulses is the main error in determining the elastic modulus of the corrected diagram. However, the yield stress depends only slightly on the synchronization accuracy. In addition, errors in the experiment, errors in the measuring circuitry, and inaccuracies in data processing contribute to the distortion of the diagram. To build a reliable dynamic plot, you need to minimize the impact of these factors or at least know their magnitude.

***Metrological support for dynamic tests***
The reliability of the results obtained in an SHPB installation is largely determined by the accuracy of the entire measurement chain of elastic strain pulses in finite rods. This chain includes commercially available electronic devices (power supply, recording oscilloscope, amplifiers, generators, etc.), as well as original lab-built components (e.g. power supply circuits for measuring strain gauges with calibration elements). Each electronic device comes with documentation, which indicates the accuracy of its main parameters, guaranteed by the manufacturer. All electronic devices participating in a dynamic experiment are certified by the state to verify that their parameters correspond to those specified in the device documentation. In addition to electronic equipment, all measuring devices (such as micrometers or calipers) involved in measuring the dimensions of the specimen, Hopkinson bars and striker are subject to periodic verification.

### *Experimental errors*

Almost all of the errors that occur in the experiment are random in nature. These include:

- $\delta_{ms}$ - inaccuracies in the manufacture of specimens (roughness, non-flatness and non-parallelism of the end surfaces, hidden internal defects, heterogeneity of the microstructure through the volume due to different heat treatment mode, etc.);

- $\delta_{lc}$ - differences in the specimen loading conditions in repeat tests due to, for example, misaligned insertion of the specimen between the bars, different forces used to press the bars against the specimen, variations in the striker velocity test to test, different conditions for excitation of the elastic pulse in the SHPB, etc.

These errors vary from experiment to experiment and affect the reproducibility of the results of the dynamic test under what is intended to be identical loading conditions. The influence of these errors can be quantified by conducting several experiments under identical conditions and comparing the results obtained.

In 1987, Lomunov considered the case of loading of several specimens made of AMg6M alloy under nominally identical conditions [134]. Three tests were studied. The maximum differences in the amplitudes of the pulses were:

for input pulses $\delta^I = 0.49\%$;

for reflected pulses $\delta^R = 0.78\%$;

for transmitted pulses $\delta^T = 0.66\%$.

As is well-known, the value of the total error is equal to the average quadratic of the partial random errors [73, 137]. Thus the general error in conducting experiments (i.e. the non-reproducibility of the dynamic stress-strain curves obtained under identical conditions) will be made up of errors contributed by each pulse:

$$\delta_{CE} = \sqrt{(\delta^I)^2 + (\delta^R)^2 + (\delta^T)^2} = \sqrt{0.49^2 + 0.78^2 + 0.66^2} = 1.13\% \qquad (20)$$

Errors in the mathematical processing of results on a computer are systematic and are negligible, so they can be ignored.

### *Errors in information retrieval*

Elastic strain pulses in the SHPB are converted by strain gauges into voltage pulses, which are recorded using a digital oscilloscope. A bridge circuit or potential divider can be used for the conversion [70, 138]. The bridge circuit is able to amplify very slow (quasistatic) changes in the gauge resistance. For dynamic tests, a simpler potential divider circuit can be used, allowing all measurement channels to be powered from a single source. Since only the dynamic component of the strain in the rod is recorded during the test, a potential divider was chosen to power the strain gauges. Both groups of strain gauges were powered by a constant current from a standard stabilized power supply (type B5-8).

Small (3 mm) foil strain gauges are used as load cells. To increase the useful signal and compensate for any bending vibrations in the working sections of the bars, four strain gauges are glued at one location evenly distributed around the circumference and connected in series. The frequency response of these strain gauges is linear over a fairly wide range, since according to experimental data [138, 139], their inertia (i.e. their response time to a longitudinal elastic strain pulse) is 0.2-0.5 µs. This makes it possible to consider these sensors as practically inertia-less transducers.

The reliability of the recorded information is affected by many factors, both random and systematic. These include errors related to the gauges themselves:
- influence of the gauge adhesive layer;
- variation of the coefficient of the sensitivity of the strain gauge;
- the linearity of the bridge circuit that transforms strain to an electrical signal;
- the accuracy of calibration of the resistance measurement bridge and sensor calibration;
- the influence of a number of other factors on the measurement circuits such as changes in ambient temperature, magnetostrictive and electromagnetic interference, etc.

In addition, it is necessary to take into account the guaranteed accuracy of all measurement devices involved in the quantitative assessment of the results obtained.

Since a strain gauge has a finite size, it senses the value of strain in the bar not at a point but over its gauge length. The measurement error produced by this depends on the strain gradient within this length. Various proposals have been made in the published literature it is proposed to evaluate this error:

$$\delta'_g = \frac{\ell_g}{2C_0\tau_p} = \frac{3.0}{2\cdot 5.2\cdot 30}\cdot 100\% = 0.97\% \qquad \text{[73] (p. 61)} \qquad (21)$$

$$\delta''_g = \frac{\sin(\pi\ell_g/\lambda)}{\pi\ell_g/\lambda} = \frac{\sin(\pi\cdot 3/50)}{\pi\cdot 3/50}\cdot 100\% = 1.7\% \qquad \text{[70] (p. 38)} \qquad (22)$$

$$\delta_g''' = \frac{\dot{\varepsilon}\,\ell_g}{2C_0\varepsilon_{max}} = \frac{150\cdot10^{-6}\cdot3}{2\cdot5.2\cdot0.003}\cdot100\% = 1.45\% \qquad [20]\ (p.\ 42) \qquad (23)$$

where $\ell_g$ is the length of the strain gauge, $C_0$ is the velocity of elastic waves in the bar, $\lambda$ is the minimum wavelength of elastic waves that can propagate in the bar without distortion, $\dot{\varepsilon}$ is the maximum strain rate at the pulse front, and $\varepsilon_{max}$ is the maximum strain that the strain gauge can record before its response becomes nonlinear. Since the greatest error of gauges occurs when used to record pulse fronts, it is advisable to estimate it using the last of the three formulas given above.

The relative error of the coefficient of the strain sensitivity $k$ (according to the documentation for the batch of gauges that were used) was

$$\delta_{css} = \frac{\Delta k}{k} = \frac{0.02}{2.17}\cdot100\% = 0.92\% \qquad . \qquad (24)$$

According to Kupershlyak-Yuzefovich [140], the error in measuring strain due to the nonlinearity of a bridge circuit when measuring strains up to 1% does not exceed 2%, but when recording actual strain pulses in bars with an amplitude less than 0.2%, this error can be assumed to be equal to $\delta_{lin} = 0.5\%$.

To measure the resistance value of the load cell itself and the calibration resistance, a DC bridge (P4833) was used, which has a guaranteed error within the required ranges of $\delta_{br} < 0.1\%$. To reduce the influence of random errors, measurements were performed 4 to 5 times and the average value of the value was determined. The influence of temperature effects, as well as magnetostrictive and electromagnetic interference on the measurement circuits can be ignored, since the change in temperature of the bridge circuit during measurements (after warming them up for 30 minutes and subsequent calibration) is negligible, and special measures were taken to minimize various external interferences [134].

Thus when performing calibration and determining calibration coefficients, the overall error of this procedure can be estimated as:

$$\delta_{cal} = \sqrt{(\delta_g''')^2 + \delta_{css}^2 + \delta_{lin}^2 + \delta_{br}^2 + \delta_{br}^2} = \sqrt{1.45^2 + 0.92^2 + 0.5^2 + 0.1^2 + 0.1^2} = 1.79\%$$
$$(25)$$

The error of the bridge $\delta_{br}$ is taken into account twice, since the bridge measures the resistances of both the gauges themselves and the calibration resistances.

In addition to the DC bridge, other electronic devices have an influence on the accuracy of the results obtained: the power source, the oscilloscope, and the generator. A B5-8 block with guaranteed output voltage error $\delta_{bp} = 0.2\%$ was used as the power source for the measurement circuits. A digital two-channel memory oscilloscope C9-8 was used as a recording device. Measurement of initial pulses in the bars is performed in stroboscopic mode. The sampling period of the digital oscilloscope (the time interval between two adjacent measurements) when using bars with a length of 1 to 1.5 m is 0.5 microseconds. The number of data points in each channel (governed by the amount of oscilloscope memory) was 1024. The sampling period is set by the oscilloscope's internal quartz oscillator, so the error in determining the absolute time value of each measurement is negligible and can be ignored. The main error in measuring the pulse amplitude (vertical deviation) with a digital oscilloscope C9-8 (according to the documentation) is $\delta_{vd} = 1.5\%$, the error in measuring time intervals (horizontal sweep) is $\delta_{hs} = 0.5\%$.

The magnetostrictive effect in the sensor array causes the strain gauge to generate an electric current during the process of high-rate deformation. However, studies of the most widely used strain gauges, the sensitive element of which is made of constantan, Vigness showed that at strain rates up to $10^3$ s$^{-1}$, the influence of the magnetostrictive effect can be completely ignored [141].

To obtain objective information about the measured strain from the oscilloscope, special measures must be taken to protect the measurement circuit from various sources of electrical interference. Kupershlyak-Yuzefovich recommended that in order to reduce magnetostrictive interference from the bars, the strain gauges should be connected to the bridge and calibration by twisted thin flexible wires less than 10 cm long, oriented perpendicular to the axis of the bar [140]. To see if the absence of magnetostrictive leads in the measurement channels has any effect, an experiment was performed in which signals were recorded from strain gauges with the power off. With the sensitivity of the oscilloscope set to that required for experiments, no interference was observed.

In order to suppress the pickup of the 50 Hz mains frequency from the power source the wire connecting the voltage divider to the recording oscilloscope consists of a two-wire shielded coaxial cable. The screen is grounded at the input of the oscilloscope. In addition, the bodies of all measuring devices, the frames on which the rods and guns are installed, as well as the measurement bars themselves are earthed, and the earthing point is selected to minimize interference.

Taking into account the described measures the total calibration errors on the *x* and *y* axes are given respectively by:

$$\delta_{cal}^{(X)} = \sqrt{\delta_{cal}^2 + \delta_{hs}^2} = \sqrt{1.79^2 + 0.5^2} = 1.86\% \qquad (26)$$

$$\delta_{cal}^{(Y)} = \sqrt{\delta_{cal}^2 + \delta_{vd}^2 + \delta_{bp}^2} = \sqrt{1.79^2 + 1.5^2 + 0.2^2} = 2.34\% \qquad (27)$$

***Accuracy of calculating a dynamic test stress-strain curve***
The measurement of elastic pulses in bars described above assumed the bars are straight. Based on the results of these direct measurements, the stress and strain in the specimen are calculated using the dependencies of the Kolsky method.

Based on the Kolsky formulas (15), the stresses and strains are actually calculated as follows:

$$\sigma_S = E \frac{A}{A_S} \varepsilon^T = E \frac{d^2}{d_S^2} K_Y^T \cdot Y^T \qquad (28)$$

$$\varepsilon_S = \frac{2C}{L} \int_0^t \varepsilon^R \cdot dt = \frac{2C}{L} \cdot K_Y^R \cdot Y^R \cdot K_X^R \cdot X^R \qquad (29)$$

where $K_X^R$, $K_Y^R$, $K_X^T$ are the calibration coefficients on the *x* and *y* axes for reflected and transmitted pulses, $X^R$, $Y^R$, $Y^T$ are the amplitudes of the same pulses measured using the oscilloscope, $d$ and $d_S$ are the diameters of the bars and the specimen, respectively.

As is well-known [137, 142], the relative error of the result when calculating the product of several quantities is equal to the sum of the errors of these quantities, and the error of raising to a power is equal to the error of this value multiplied by the exponent. So the errors of calculating the stress and strain will be equal to:

$$\delta^{\sigma} = \delta_E + 2\delta_d + 2\delta_{ds} + \delta_{cal}^{(Y)} + \delta^T \qquad (30)$$

$$\delta^{\varepsilon} = \delta_C + \delta_L + \delta_{cal}^{(Y)} + \delta^R + \delta_{cal}^{(X)} + \delta^R \qquad (31)$$

In these dependencies, $\delta_E$ represents the error in determining the Young's modulus of the bars. It can be estimated approximately by the value $\delta_E = 3\%$. Errors in measuring the diameters of the bars and specimen, as well as the thickness of the specimen are determined by the accuracy of the micrometer and are equal to $\delta_d = 0.04\%$, $\delta_d = 0.05\%$, $\delta_L = 0.08\%$. The velocity of sound waves in the bar is determined by measuring the time interval between the incident and reflected pulse fronts (no specimen present) recorded by the same gauge. Based on the results of five experiments, the average value and relative error were determined, which was $\delta_C = 0.05\%$. After substituting all the values, you can calculate the final error values for determining the stress and strain in the specimen, which include all the errors described above:

$$\delta^\sigma = 3 + 2\times0.04 + 2\times0.05 + 2.34 + 0.76 = 6.28\% \; ; \qquad (32)$$
$$\delta^\varepsilon = 0.05 + 0.03 + 2.34 + 0.78 + 1.86 + 0.78 = 5.84\% \; . \qquad (33)$$

It should be noted that the calculated errors are determined for the case when all the described errors have the largest value. In fact, with a carefully prepared and executed experiment, these errors are significantly smaller.

## 1.2 Modifications of the Kolsky bar method

In accordance with the objectives of various investigators, novel methods have been proposed, implemented and numerically analysed, which significantly expand the capabilities of the traditional Kolsky bar method. Examples include modifying the SHPB so as to be able to perform tensile and shear tests. To study the effects of strain rate history, specimens have been subjected to alternating loading using loading pulses of one or more different signs. Also methods for performing incremental tests with abrupt changes in strain rate have been implemented. SHPB variants have been developed and tested for the study of dynamic hardness, crack resistance, and dynamic friction, as well as for the study of the dynamic compressibility of materials with low cohesion such as soils. A variant of SHPB has been developed and implemented for multiple loading of low-density specimens and obtaining a compression ratio of more than 70%. In addition, a modification of the traditional SHPB loading design has been proposed to provide a small variation in the strain rate of the specimen during the test.

## 1.2.1 Modifications of the Kolsky bar method for testing materials in tension

A design for performing tensile SHPB tests was first proposed by Lindholm & Yeakley [77]. Their design consisted of in input bar, an output tube and a specimen in the form of a top-hat with a massive base and four parallel grooves in a cylindrical tubular working part (Figure 10). The disadvantage of this design is the presence of shear stress and strain components in the specimen, which results in a significant error in the definition of the deformation diagram.

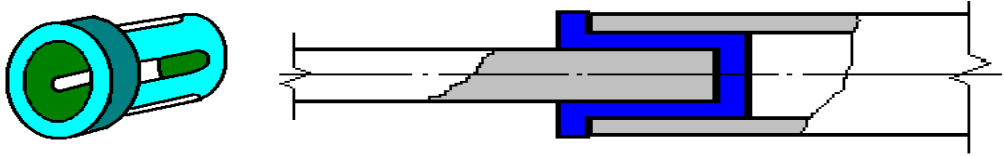

*Figure 10. Schematic diagram of top-hat SHPB tensile test.*

Another variant of the SHPB for tensile testing was proposed by Nicholas [72]. In his design, the specimen is loaded by a tensile wave, which is formed after the compression wave is reflected from the free end of the output bar. To prevent plastic deformation of the specimen during the passage of the compression wave, a split ring was used (Figure 11*a*).

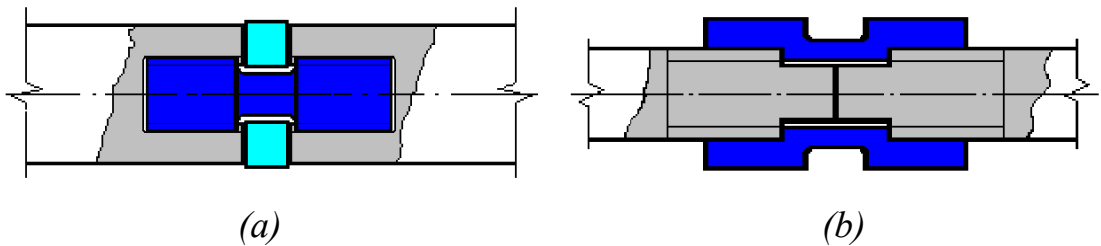

*(a)*                           *(b)*

*Figure 11. Two different tensile SHPB designs.*

Figure 12 presents a diagram of such a test along with the picture of wave propagation in the SHPB system. The main differences to the compression version is that the input bar must be at least twice as long as the output bar (which must have a free rear end) and the specimen is connected to the bars with a thread and the gauge length is surrounded by a split ring.

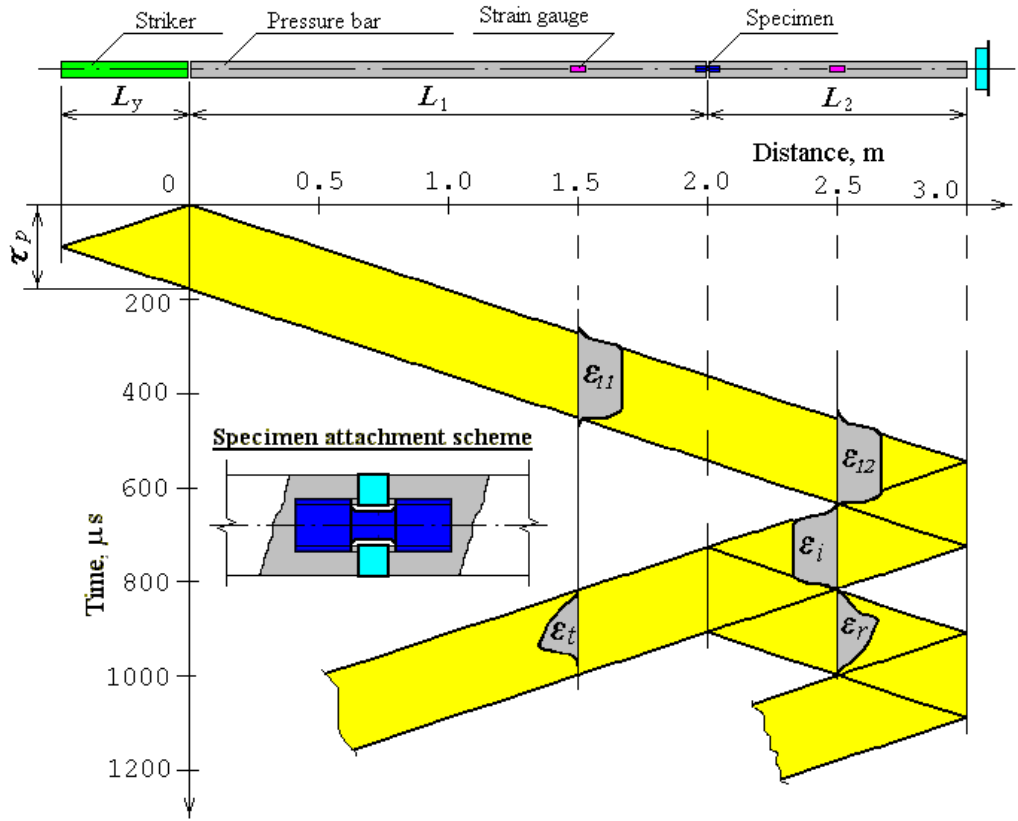

*Figure 12. Lagrangian x-t diagram for a tensile SHPB system.*

The essence of the method is as follows. A longitudinal compression pulse is applied to the left end of the input bar which excites an elastic one-dimensional compression wave $\varepsilon_{11}(t)$ in it. This initial pulse passes freely through the split ring (which has a high yield strength) without causing plastic deformation of the specimen and on into the output bar as a strain pulse $\varepsilon_{12}(t)$. When this strain pulse reaches the free end of this bar, it is reflected as a tensile wave.

The tensile wave is now the initial incident wave for loading the specimen in tension. From the moment of reflection and the formation of the tensile pulse at the free end of the bar and its propagation in the opposite direction along the bar, the experimental design is similar to that of a compression test. The tensile impulse, having reached the specimen, partially passes through it to the input bar and is partially reflected back into the output bar. The specimen then undergoes plastic deformation. The split ring does not experience tension, since it is not attached to the bars. The experimental data processing in this design is performed using the main dependencies of the SHPB method.

Similar variants of the tensile SHPB were implemented by Harding and co-workers [79] and by Staab & Gilat [82]. The design by Nicholas is more technologically advanced in terms of ease of specimen manufacture and is used much more often than the Lindholm design.

In 1995, Bragov & Lomunov implemented a design for dynamic tensile loading similar to that of Nicholas [143]. The specimen is tubular and is screwed onto the ends of measuring rods (Figure 11*b*).

### 1.2.2 Modifications of the Kolsky bar method for testing materials in shear

An option for testing specimens in shear include the use of a specimen in the form of a flat plate or parallelepiped [80]. For this purpose, the measurement bars were made with specially machined ends (Figure 13*a*), between which a flat specimen was placed. In this case, the bars act as a matrix and a punch, i.e. the specimen is subjected to a pure shear strain along two planes (Figure 13*b*).

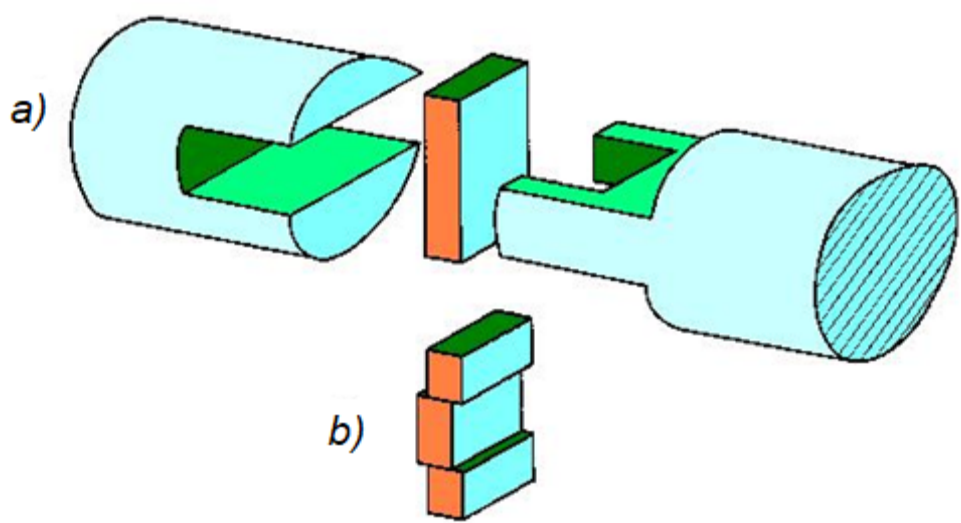

*Figure 13. Schematic diagram of the SHPB shear test proposed by Hauser [80].*

In this case, recording of the elastic compression wave $\varepsilon^T(t)$ in the output bar allows, as in the traditional compression version of the SHPB, the determination of how the tangential stress develops in the specimen:

$$\tau_s(t) = E\frac{A}{A_s}\varepsilon^T(t) \tag{34}$$

where $E$ is the Young's modulus of the output bar material and $A$ its cross-sectional area at the strain gauge location. This is because the output bar in this system acts as a conventional elastic pulse waveguide as in the case of a conventional compression Kolsky bar. However, in contrast to a compression Kolsky bar, the specimen cross-sectional area $A_s$ should be taken not across but along the axis of the SHPB, parallel to the cut planes (see Figure 13*b*), and since the cross-section occurs on two planes, this cross-sectional area must be doubled when substituting in equation (34).

As usual, the pulse reflected from the specimen, $\varepsilon^R(t)$, can be used to calculate the shear strain:

$$\gamma_s(t) = -\frac{2C}{L_0}\int_0^t \varepsilon^R(t)\,dt \qquad (35)$$

since this pulse corresponds to the displacement of the end of the input bar in contact with the specimen. However, the length $L_0$ of the gauge length of the specimen included in equation (35) is determined by the gap between the matrix and the punch. This gap can be accurately determined only before a test by measuring the dimensions of the ends of measurement bars. During the deformation of the specimen, the size of this gap can change in an unknown manner due to possible transverse vibrations of the ends of the bars, thereby 'clogging' the gap with particles of the test specimen, etc. Due to the inability to control the gap (i.e. the gauge length of the specimen $L_0$) the reliability of the latter dependence in equation (35) is doubtful. Since the effective length of the working part of the specimen is determined by the gap between the matrix and the punch, this technique allows us to obtain strain rates of more than $10^4$ s$^{-1}$.

To obtain stress-strain diagrams in ring shear, the SHPB needs to be modified as shown in Figure 14.

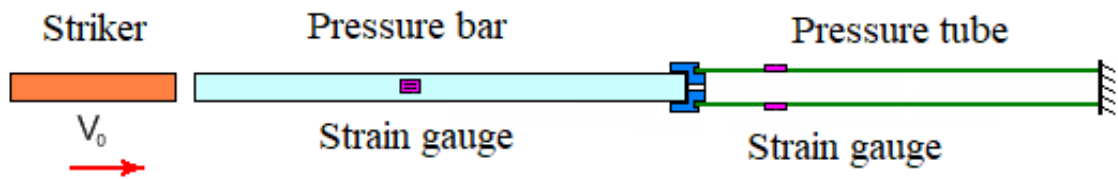

*Figure 14. Schematic diagram of the SHPB ring shear test.*

During a test, the specimen of the design shown in Figure 15 is loaded in the traditional 'input bar – output tube' system.

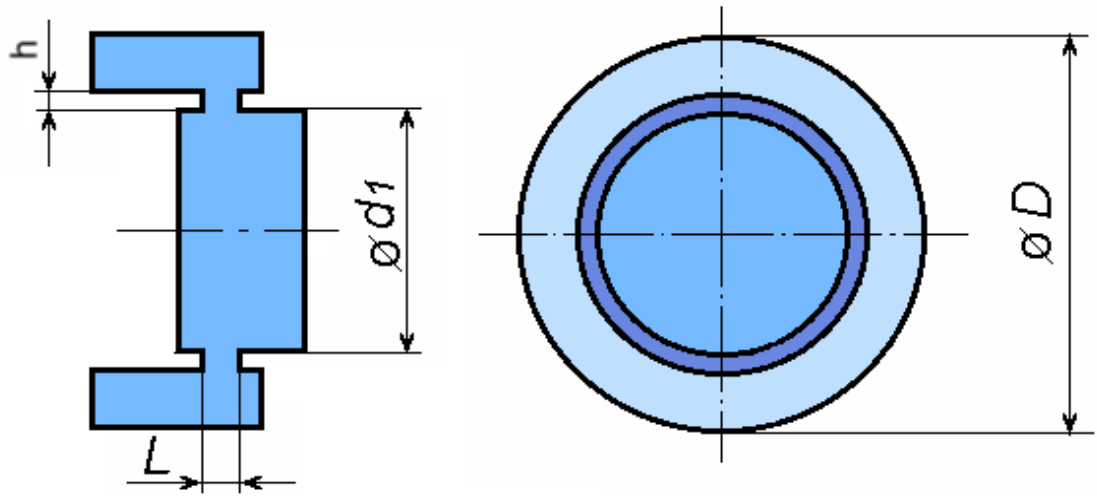

*Figure 15. Design of the specimen used in the ring shear experiment.*

The elastic strain pulses measured in the input and output bars allow the determination of the velocities $V_1$ and $V_2$ of the internal and external parts of the specimen (see Figure 16) as follows:

$$V_1(t) = C_I \cdot \varepsilon^I(t) - C_I \cdot \varepsilon^R(t), \qquad (36)$$

$$V_2(t) = C_T \cdot \varepsilon^T(t) \ , \qquad (37)$$

where $C_I$ is the velocity of sound in the input bar, $C_T$ is the velocity of sound in the output tube, and $\varepsilon^I(t)$, $\varepsilon^R(t)$ and $\varepsilon^T(t)$ are the input, reflected and transmitted strain pulses, respectively.

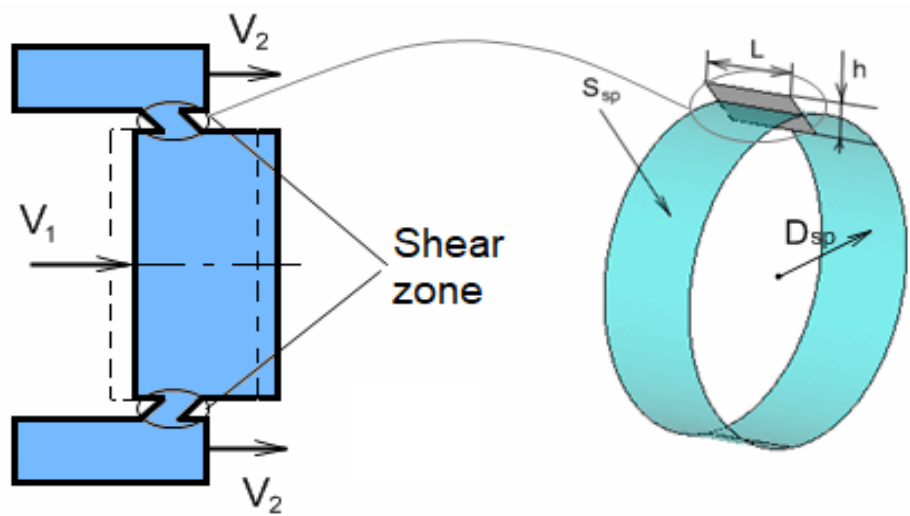

*Figure 16. Schematic diagram showing how shear stresses and strains are applied in a ring shear specimen.*

By integrating these velocity-time relations, we determine the corresponding displacements:

$$U_1(t) = \int_0^t V_1(\tau) \cdot d\tau \ , \tag{38}$$

$$U_2(t) = \int_0^t V_2(\tau) \cdot d\tau \ . \tag{39}$$

The time dependence of the shear strain is then determined by the ratio:

$$\gamma(t) = \frac{U_1(t) - U_2(t)}{h} \ , \tag{40}$$

where $h$ is the thickness of the gauge length of the specimen (see Figure 15).

The force acting on the specimen is calculated from the strain pulse in the output tube:

$$F(t) = E_T \cdot S_T \cdot \varepsilon^T(t) \ , \tag{41}$$

$$S_T = \frac{D_T^2 - D_{T0}^2}{4} \cdot \pi \ , \tag{42}$$

where $E_T$ and $S_T$ are the Young's modulus and the cross-sectional area of the output tube, respectively, and $D_T$ and $D_{T0}$ are the outer and inner diameters of the same tube.

The shear stress is equal to the ratio of the force $F(t)$ acting on the specimen to the area of the annular surface corresponding to the median surface of the part of the specimen that is shearing (see Figure 16):

$$\tau(t) = \frac{F(t)}{S_{sp}} \ , \tag{43}$$

where $S_{sp} = \pi \cdot D_{sp} \cdot L$ ,

$$\tau(t) = \frac{F(t)}{\pi \cdot D_{sp} \cdot L} \ , \tag{44}$$

where $D_{sp} = \dfrac{d_1 + (d_1 + 2 \cdot h)}{2} = d_1 + h$

Eliminating time from the pair of variables $\gamma(t)$ and $\tau(t)$, we obtain a plot of material deformation $\tau(\gamma)$ under ring shear conditions. To compare the plots obtained in this type of test with those obtained under tension or compression, we need to calculate the equivalent values of strain and stress using the von Mises criterion: $\varepsilon_{eqv} = \dfrac{\lambda}{\sqrt{3}}$, $\sigma_{eqv} = \sqrt{3} \cdot \tau$.

### 1.2.3 Ensuring constant strain rate

When testing materials using the Kolsky method, in general, a dynamic diagram of some arbitrary deformation process $\sigma_s = \sigma_s(\varepsilon_s, \dot{\varepsilon}_s)$ is obtained for various, but controlled, values of the strain rate. However, in order to analyse the dynamic behaviour or to determine the parameters of a material model, a set of dynamic material plots is required for a number of fixed rate values. To obtain such a set of plots, a mathematical apparatus can be used for converting the results of a large number of dynamic experiments performed at different strain rates as well as strain rate jump experiments [42].

For the normal operation of this apparatus, the entire stress, strain and strain rate surface under study needs to be populated with experimental data, which requires a large number of dynamic tests. At the same time, by elaborating the experiment, it is possible to significantly reduce the variation of loading rate so that the resulting process diagram can be considered a dynamic diagram for the material.

As can be seen from equations (15), the strain rate during the test is determined by the reflected pulse $\varepsilon^R(t)$ or alternatively, taking equation (14) into account, the difference between the input and transmitted pulses $\dot{\varepsilon}_s(t) = \dfrac{2C}{L_0}\left(\varepsilon^I(t) - \varepsilon^T(t)\right)$ [144]. It follows that the change in the strain rate when a specimen is deformed by the same loading pulse is determined by the transmitted pulse $\varepsilon^T(t)$, which is proportional to the stress in the specimen, i.e. the dynamic behaviour of the material. Since metals generally exhibit significant strain hardening and, in addition, the cross-sectional area of the specimen increases during a test, when the SHPB is loaded using a compression pulse $\varepsilon^I(t)$ of trapezoidal shape of almost constant amplitude (formed when the striker directly strikes the input bar), the specimen strain rate will decrease during the test (Figure 17a). It follows that in order to ensure a constant strain rate during an experiment, it is desirable to have a loading impulse similar in shape to the transmitted pulse, but with a larger amplitude (Figure 17b).

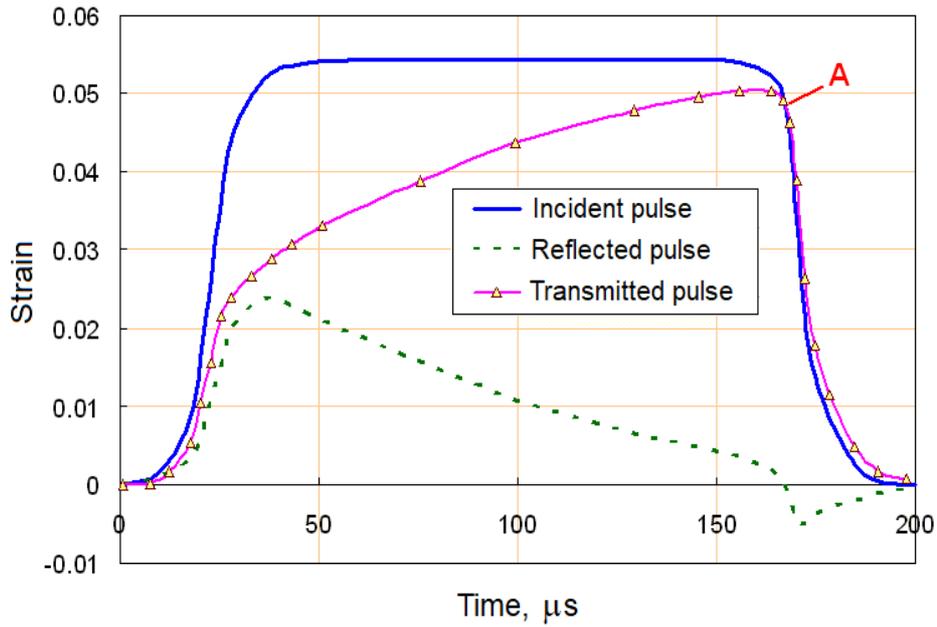

*(a)*

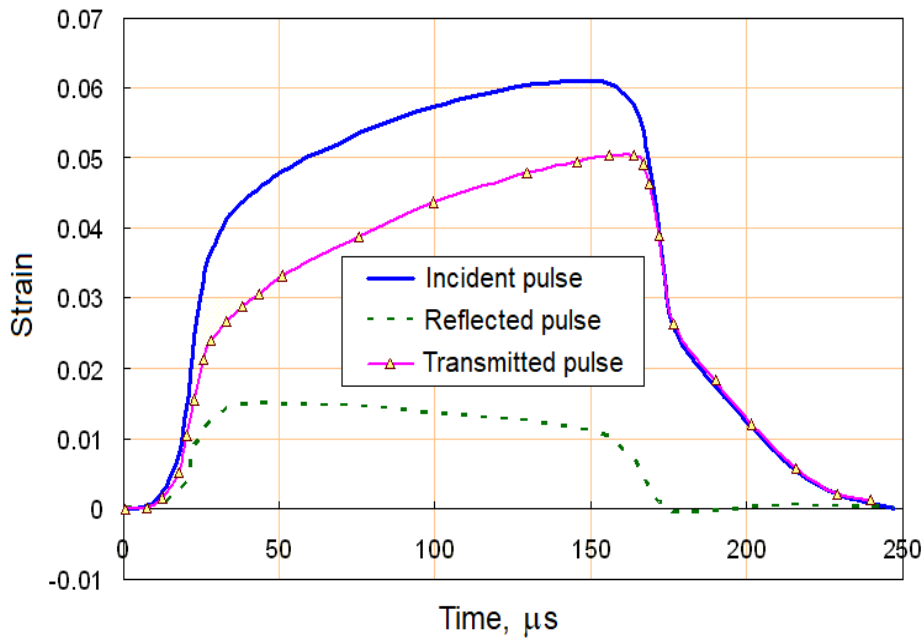

*(b)*

*Figure 17. Plots showing that in order to achieve a constant strain rate in an SHPB experiment, it is necessary to have a loading pulse similar in shape to the transmitted pulse.*

The first studies devoted to solving this problem were carried out about 40 years ago. For these purposes, Ellwood *et al.* proposed adding another bar and an auxiliary specimen to the SHPB system [145]. An alternative proposal by Sato & Takeyama was to load the SHPB with a striker bar of variable cross-section in the form of a truncated cone [146]. In recent years, several types of 'pulse shaper' have been developed, consisting of several components located on the impacted end of the input bar. For example, Frew *et al.* give a description and a detailed analysis of a pulse shaper consisting of two components, annealed copper (α) and tool steel (β), separated by a rigid steel disc that allows both components to expand freely during plastic deformation (Figure 18) [147].

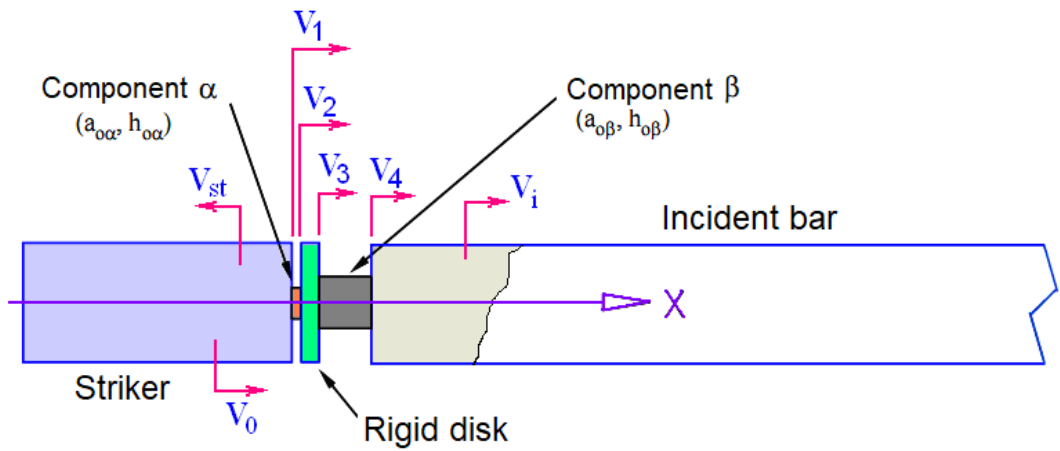

*Figure 18. Loading pulse shaper proposed by Frew et al. [147].*

In the early 1980s, Bragov & Lomunov proposed generating a loading pulse with an increasing amplitude by placing a thin gasket made of a material with significant strain hardening at the impacted end of the input bar [133, 134]. Ideally the gasket should be made of the same material as the specimen under study (Figure 19). The gasket is attached to the end of the input bar using a thin layer of viscous lubricant.

| Parameter / Element | Sectional | Density | Wave velocity | Elastic modulus | Length |
|---|---|---|---|---|---|
| Striker | $A$ | $\rho_y$ | $c_s$ | | $L_s$ |
| Shaper | $a_0$ | $\rho_0$ | | | $h_0$ |
| Loading rod | $A$ | $\rho$ | $c$ | $E$ | - |

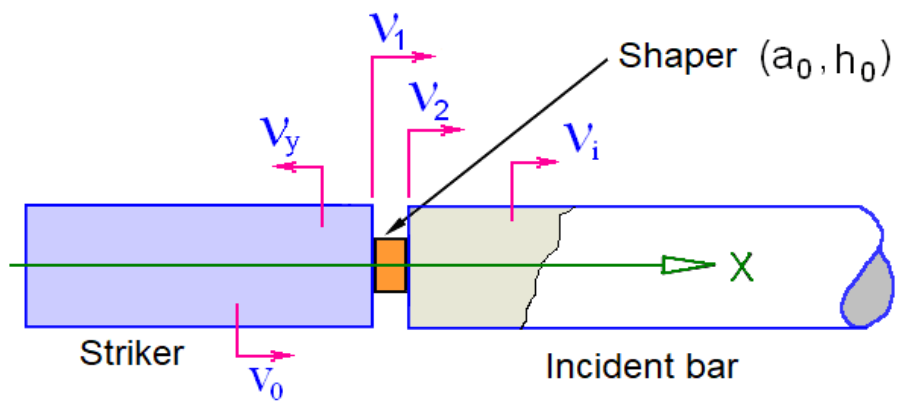

*Figure 19. Loading-pulse shaper.*

When the striker collides with the pulse shaper at a velocity $V_0$, a compressive force is gradually transferred through the pulse shaper into the input bar. An increase in the cross-sectional area of the pulse shaper as it deforms, together with the strain-hardening of the material the pulse shaper is made from, gradually increases its impedance. This causes a monotonic increase in the amplitude of the loading pulse in the input bar as well an increase in its duration.

If the material of the pulse shaper is incompressible, the volume of the pulse shaper remains constant as it deforms, i.e. $a_0 h_0 = a(t) h(t)$, where $a(t)$ and $h(t)$ are the current cross-sectional area and thickness of the pulse driver respectively. The longitudinal strain of the pulse shaper will then be given by $\varepsilon_{ps}(t) = \dfrac{h_0 - h(t)}{h_0} = 1 - \dfrac{h(t)}{h_0}$. From this the current cross-sectional area of the pulse shaper can be expressed as $a(t) = \dfrac{a_0}{1 - \varepsilon_{ps}(t)}$. Since the thickness of the shaper is usually small (about 1 mm), and the impact duration is long, the forces at the ends of the shaper are equal, i.e. $\sigma_s(t) A = \sigma_{ps}(t) a(t) = \sigma_i(t) A$. Hence the stress in the striker and the input bar are equal $\sigma_s(t) = \sigma_i(t) = \dfrac{\sigma_{ps}(t) a(t)}{A}$, if they are made from the same materials.

The stresses in the shaper can be determined from the general relations for the uniaxial stress state $\sigma_{ps} = \sigma_0 f(\varepsilon_{ps})$ where $\sigma_0$ is a constant and $f(\varepsilon_{ps})$ is a function of the axial strain of the striker. Based on the above, the pulse excited in the input bar will be:

$$\sigma_i(t) = \sigma_s(t) = \frac{\sigma_0 a_0}{A} \cdot \frac{f(\varepsilon_{ps})}{(1 - \varepsilon_{ps})} \quad . \tag{45}$$

Since the input bar and the striker remain elastic, the axial strain in the input bar will be given by:

$$\varepsilon_i(t) = \frac{\sigma_0 a_0}{E A} \cdot \frac{f(\varepsilon_{ps})}{(1 - \varepsilon_{ps})} \quad . \tag{46}$$

The velocities $v_1(t)$ and $v_2(t)$ at the ends of the shaper (Figure 19) will be given by:

$$v_1(t) = V_0 - v_s(t) = V_0 - \frac{\sigma_s(t)}{\rho_s c_s} \ ,$$

$$v_2(t) = v_i(t) = \frac{\sigma_i(t)}{\rho c} \ . \tag{47}$$

From these, the strain rate in the pulse shaper can be written as:

$$\dot{\varepsilon}_{ps}(t) = \frac{v_1(t) - v_2(t)}{h_0} \ . \tag{48}$$

In the case of a conventional input bar (without pulse-shaper), a trapezoidal elastic compression impulse reflects into the striker bar at a velocity $c_s$. When this reaches the free end of the striker, it reflects as a tensile pulse. When this tensile wave reaches the boundary with the SHPB, the loading cycle is stopped, and the striker bounces back. In this case, the loading process has a purely elastic wave character. A different picture is seen when an elastic-plastic impact is made on the 'soft' formation. Here there is a combination of wave and inertia effects. The wave excited at the 'striker/shaper' boundary has a profile similar to the deformation diagram of the material of the shaper $\sigma_{ps}(t)$. This wave also moves through the striker to the left, reaches its free end and is reflected as a tensile wave, which when it reaches the junction with the input bar causes additional waves to be reflected and passed into the striker rod.

Let's denote the characteristic time of the double transit time of the elastic wave over the body of the striker $\tau = \frac{2L_s}{c_s}$ . Then the entire process of hitting the shaper can be represented as consisting of several stages, consisting of multiples of the time $\tau$: $0 < t < \tau$, $\tau < t < 2\tau$, $2\tau < t < 3\tau$ etc.

<u>In the first stage at $0 < t < \tau$</u> , when $v_1(t) > v_2(t)$ on the basis of equations (45), (47) and (48)

$$h_0 \dot{\varepsilon}_{ps}(t) = V_0 - \frac{\sigma_s(t)}{\rho_s c_s} - \frac{\sigma_i(t)}{\rho c} = V_0 - \sigma_i(t) \left( \frac{1}{\rho_s c_s} + \frac{1}{\rho c} \right) . \tag{49}$$

Or taking into account equation (45)

$$t = \frac{h_0}{V_0} \int_0^{\varepsilon_{ps}} \left[ 1 - K \left( \frac{1}{\rho_s c_s} + \frac{1}{\rho c} \right) \frac{f(x)}{(1-x)} \right]^{-1} dx \quad \text{where} \quad K = \frac{\sigma_0 a_0}{A V_0} \; . \qquad (50)$$

This equation is valid until the cross-sectional area $a(t)$ of the pulse shaper exceeds the cross-sectional area $A$ of the striker or input bars.

In the second stage $\underline{\tau < t < 2\tau}$ similar processes occur, only with a time shift of $\tau$. In this case, the main pulses in the input and striker bars will be added to the additional pulses reflected back up the striker bar $\sigma_r^1(t-\tau)$ and transmitted into the input bar $\sigma_t^1(t-\tau)$. So the longitudinal forces in the 'striker-pulse shaper-input bar' system will be:

$$\left[ \sigma_s(t) - \sigma_s(t-\tau) + \sigma_r^1(t-\tau) \right] A = \sigma_{ps}(t) a(t) = \left[ \sigma_i(t) + \sigma_t^1(t-\tau) \right] A \qquad (51)$$

The momenta $v_1(t)$ and $v_2(t)$ at the ends of the pulse shaper will be:

$$v_1(t) = V_0 - v_s(t) - v_s(t-\tau) - v_r^1(t-\tau) = V_0 - \frac{\sigma_s(t)}{\rho_s c_s} - \frac{\sigma_s(t-\tau)}{\rho_s c_s} - \frac{\sigma_r^1(t-\tau)}{\rho_s c_s} \qquad (52)$$

$$v_2(t) = v_i(t) + v_t^1(t-\tau) = \frac{\sigma_i(t)}{\rho c} + \frac{\sigma_t^1(t-\tau)}{\rho c} \; . \qquad (53)$$

Similarly, in the second loading stage equation (49) gives the strain rate of the shaper taking into account equation (51) as:

$$\begin{aligned} h_0 \dot{\varepsilon}_{ps}(t) &= V_0 - \frac{\sigma_s(t)}{\rho_s c_s} - \frac{\sigma_s(t-\tau)}{\rho_s c_s} - \frac{\sigma_r^1(t-\tau)}{\rho_s c_s} - \frac{\sigma_i(t)}{\rho c} - \frac{\sigma_t^1(t-\tau)}{\rho c} = \\ &= V_0 - \left( \frac{1}{\rho_s c_s} + \frac{1}{\rho c} \right) \cdot \left[ \sigma_i(t) + \sigma_t^1(t-\tau) \right] - \frac{2\sigma_s(t-\tau)}{\rho_s c_s} \end{aligned} \qquad (54)$$

Subject to equation (45)

$$\sigma_s(t-\tau) = \frac{\sigma_0 a_0}{A_0} \cdot \frac{f[\varepsilon_{ps}(t-\tau)]}{[1-\varepsilon_{ps}(t-\tau)]} \qquad (55)$$

and

$$\sigma_i(t) + \sigma_t^1(t-\tau) = \frac{\sigma_{ps}(t) a(t)}{A} = \frac{\sigma_0 a_0}{A} \cdot \frac{f(\varepsilon_{ps})}{(1-\varepsilon_{ps})} \qquad (56)$$

Then

$$\frac{h_0}{V_0}\dot{\varepsilon}_{ps}(t) = 1 - \left(\frac{1}{\rho_s c_s} + \frac{1}{\rho c}\right)\frac{\sigma_0 a_0}{AV_0} \cdot \frac{f(\varepsilon_{ps})}{(1-\varepsilon_{ps})} - \frac{2}{\rho_s c_s}\frac{\sigma_0 a_0}{A_0 V_0} \cdot \frac{f[\varepsilon_{ps}(t-\tau)]}{[1-\varepsilon_{ps}(t-\tau)]} =$$

$$= 1 - K\left(\frac{1}{\rho_s c_s} + \frac{1}{\rho c}\right)\frac{f(\varepsilon_{ps})}{(1-\varepsilon_{ps})} - \frac{2K}{\rho_s c_s} \cdot \frac{f[\varepsilon_{ps}(t-\tau)]}{[1-\varepsilon_{ps}(t-\tau)]}$$

(57)

The solution of this equation for $\tau < t < 2\tau$ is

$$t = \tau + \frac{h_0}{V_0}\int_{\varepsilon_{ps}^1}^{\varepsilon_{ps}}\left[1 - K\left(\frac{1}{\rho_s c_s} + \frac{1}{\rho c}\right)\frac{f(x)}{(1-x)} - \frac{2K}{\rho_s c_s} \cdot \frac{f[\varepsilon_{ps}(t-\tau)]}{[1-\varepsilon_{ps}(t-\tau)]}\right]^{-1} dx \ , \qquad (58)$$

where also $K = \dfrac{\sigma_0 a_0}{AV_0}$ .

In RIM-NNSU, for simplicity, the pulse shapers that are mainly used consist of a thin disc made of soft metal, or a small-diameter disc made of the material being tested.

As an example, Figure 20 presents a comparison between the waveforms that were obtained when testing the aluminium-magnesium alloy AMg6M using (a) the traditional Kolsky method and (b) when the SHPB was loaded through a pulse shaper made of the same material. A strong difference in the shape of the reflected pulse can be seen in these two cases.

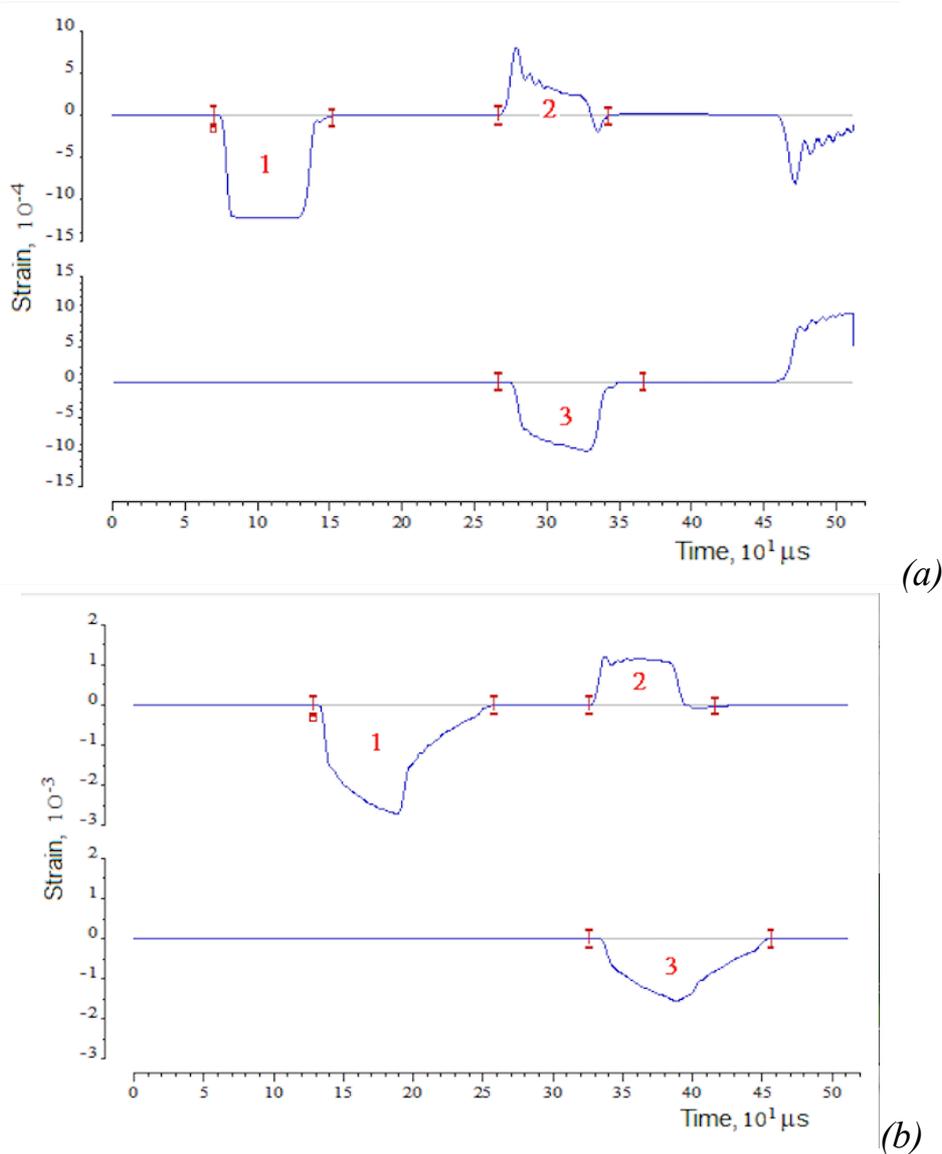

*Figure 20. Generating a constant strain rate in compression experiments.*

The stress-strain diagrams constructed from these waveforms and the corresponding strain-rate deformation plot are shown in Figure 21 (the dotted green line is the quasistatic stress-strain curve). This figure clearly shows the strong influence that a change in the strain rate has on the strength of this alloy. Therefore, a comparative analysis of the dynamic data obtained by different authors must necessarily consider the history of changes in the strain rate during a test.

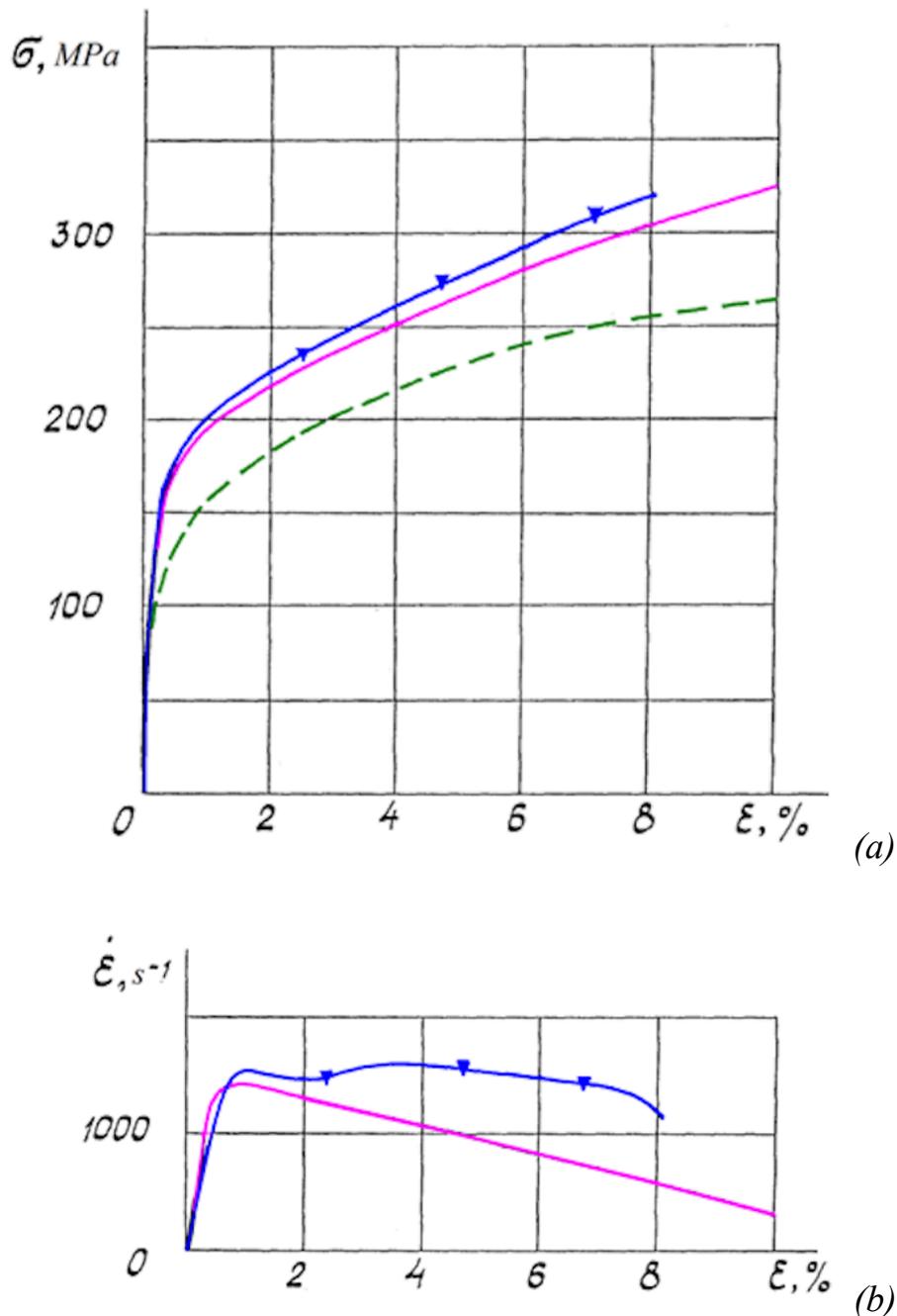

*(a)*

*(b)*

*Figure 21. Stress-strain plots of the AMg6M alloy obtained both dynamically and quasistatically.*

Since the pulse shaper undergoes significant plastic deformation during loading of the SHPB, Bragov *et al.* [148] proposed calculating its dynamic stress-strain curve using the direct impact method [68]. Thus, in a single experiment using an SHPB, a pulse shaper, and the theory of the direct impact bar, it is possible to determine the dynamic deformation plots of two different materials or alternatively the same material but at different strain rates.

As an example, Figure 22*a* presents plots for two different materials obtained in the same experiment: the plot for the steel pulse shaper specimen was calculated by the direct impact method and the plot for the main specimen (an aluminium alloy) was calculated by the traditional Kolsky method. Figure 22*b* shows plots where the pulse shaper and the main specimen were made of the same material (high-strength aluminium alloy D16T), the properties of which are only weakly dependent on the strain rate.

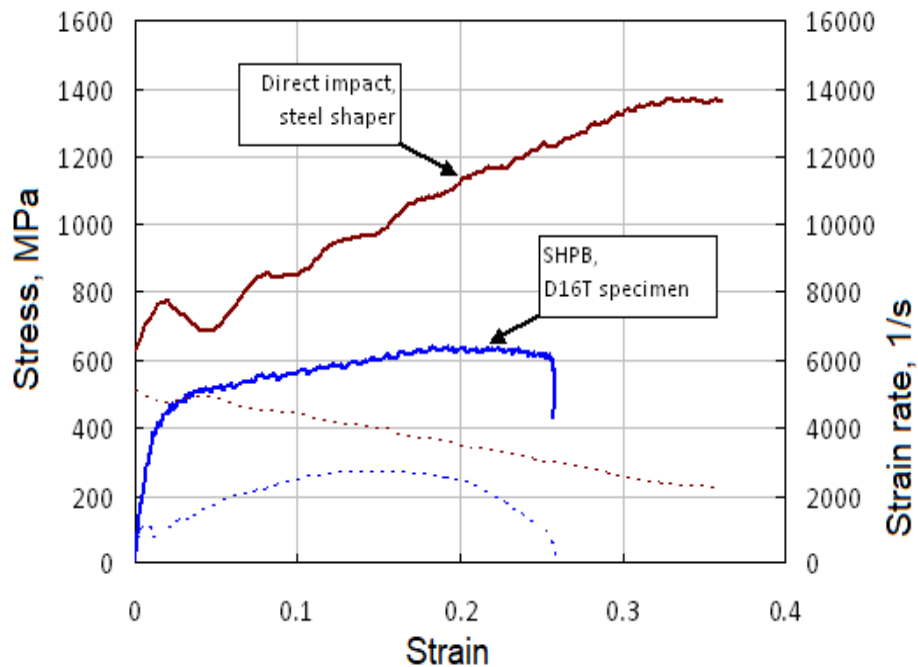

*(a)*

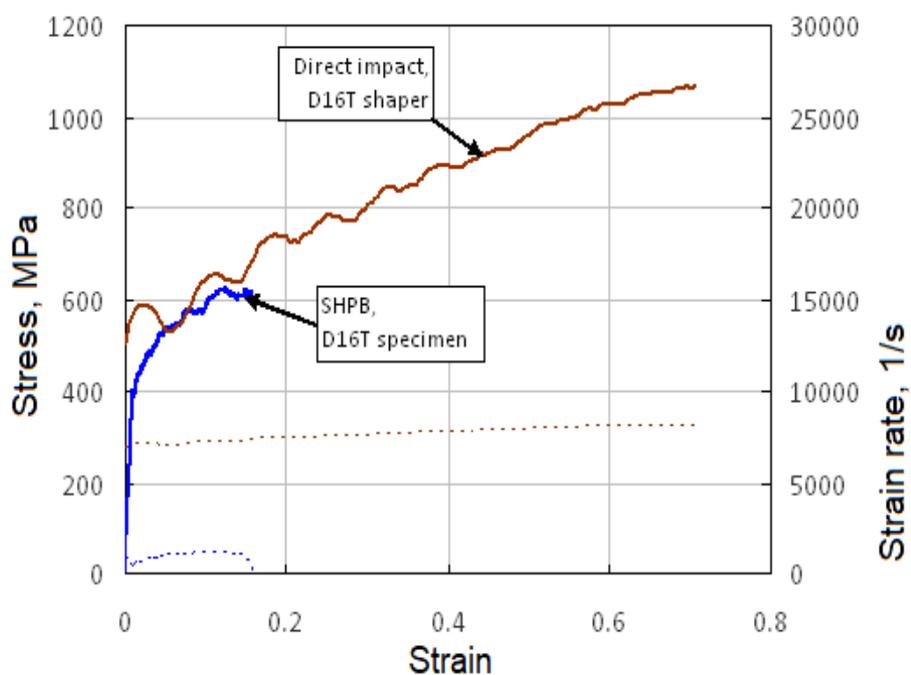

*(b)*

*Figure 22. Stress-strain plots of the main and the pulse-shaper specimens.*

### 1.2.4 Investigation of the effect of strain rate jumps

In order to implement more complex modes of dynamically loading materials, the SHPB can be loaded using striker bars composed of two or more materials with different acoustic impedances, $\rho C$, where $\rho$ is the density and $C$ is the velocity of sound of the striker bar material. Such compound strikers are used to study the influence of strain rate jumps on the behaviour of materials. The rods that make up the compound striker can either be glued together without a gap between them [149, 150], or they can have a gap between of some size $\delta$ implemented by using easily deformable flexible brackets (see Figure 23a, b) [143, 151].

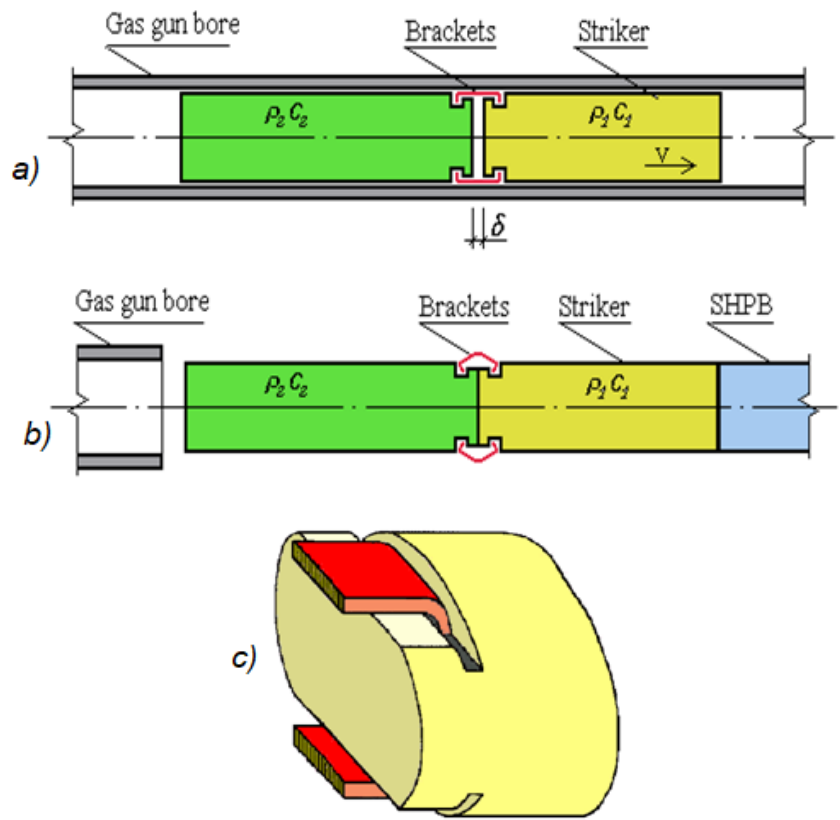

*Figure 23. Compound striker bar with a gap between its components.*
*(a) Within the gun barrel. (b) At the moment of impact.*
*(c) Magnified view of the bracket used to hold the compound striker together.*

In the case of using a striker glued from two or three components with different acoustic impedances (see Figure 24a), the incident pulse is realized with positive (or negative) jumps in amplitude, resulting in corresponding jumps in the strain rate of the specimen (Figure 24a). When using a striker with a gap (Figure 24b), the specimen is loaded with a train of pulses with different amplitudes and a time delay between them (Figure 24b).

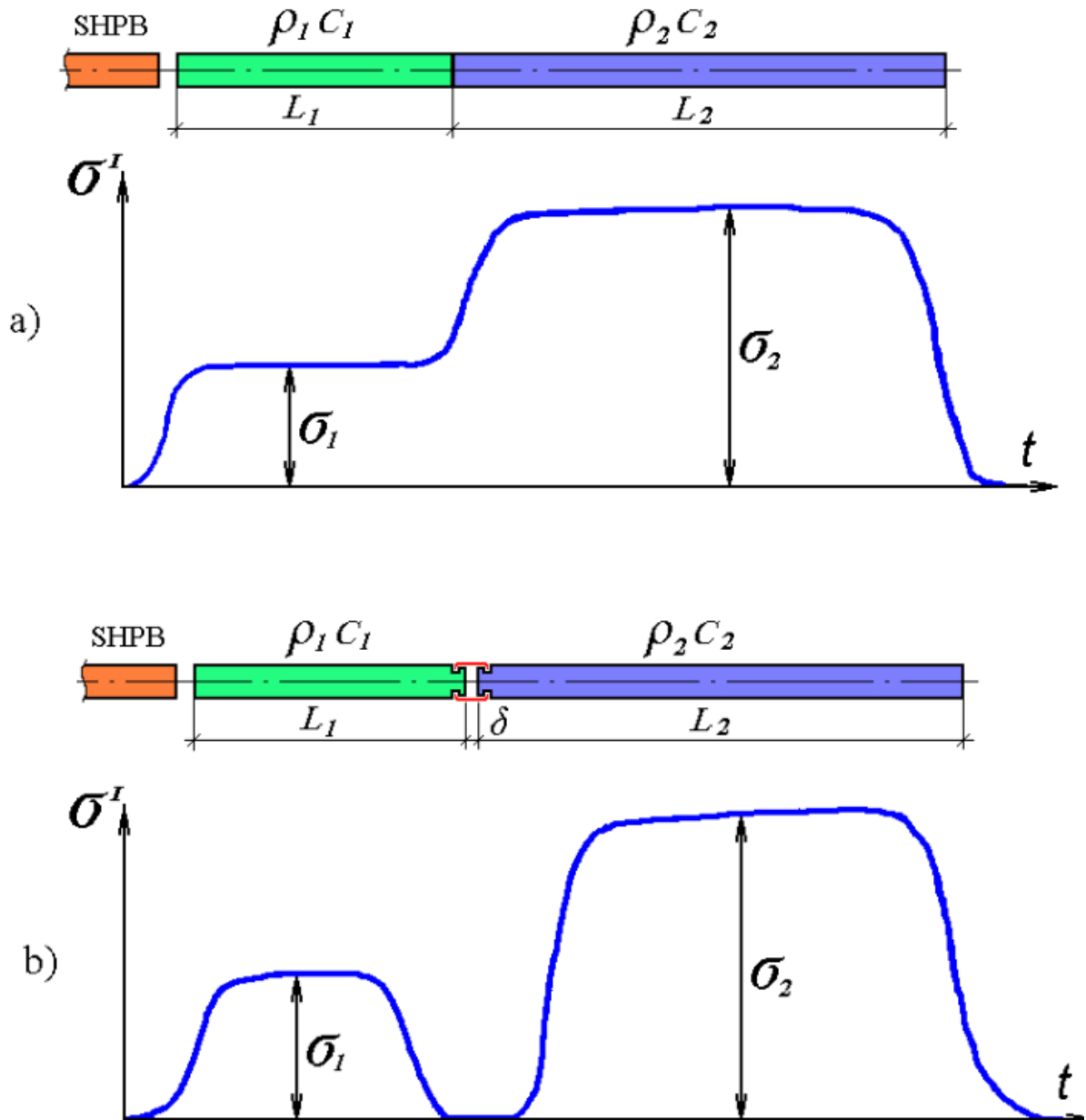

*Figure 24. Compound striker bars for studying the effects of strain rate history: (a) components glued together; (b) components connected by a bracket in order to create a gap.*

Figure 23 shows such a compound striker with a gap $\delta$ at two different times during a test: (a) the striker is still in the barrel of the gas gun and (b) at the moment of impact on the input bar of the SHPB. The brackets are two thin narrow strips of soft aluminium (Figure 23*c*). The ends of the brackets are held in special grooves machined on the ends of the bars that are facing each other. The overall size of the connecting section of the bar-bracket-bar does not exceed the diameter of the rods (Figure 23*c*).

When the striker is accelerating in the barrel of the gas-gun (when inertial forces are acting on it), the brackets cannot bend (see Figure 23a), i.e. the set gap does not change. At the moment of impact on the SHPB, the brackets bend and the second rod generates a pre-loading pulse in the SHPB. The size of the gap is determined by the length of the side of the bracket which can be varied from 0.5 to 5 mm depending on the aims of the experiment. Thus, when using a compound striker with a gap, the specimen will be loaded with a sequence of pulses (according to the number of rods that make up the striker) of the same sign and of different amplitudes with intervals between individual pulses set by the lengths of the brackets. If a compound striker is used without a gap, a compression pulse with a stepwise change in amplitude will be excited in the SHPB.

By varying the ratios of the acoustic impedances of the sections of the striker bars, it is possible to perform alternating loading of specimens in an SHPB both with and without unloading the specimen, or with various changes in the strain rate (such as from a lower level to a higher one, or vice versa), since the strain rate (which is proportional to the amplitude of the reflected pulse $\varepsilon^R$) is determined by the properties of the specimen and the amplitude of the loading pulse.

Taking Figure 25a, as an example, the stress-strain curves (upper chart) and appropriate strain rate-strain curves (lower chart) for aluminium AD1 obtained when using strikers made of three components (dural-steel-dural or vice versa) glued together. Positive or negative jumps in the strain rate when loading the specimen can be seen. The results of using a two-component striker (dural-steel) with a gap are shown in Fig 25b where two consecutive cycles of specimen loading occur at different strain rates with complete unloading between cycles. For comparison, the dashed lines show plots obtained at constant strain rate. It can be seen that aluminium is sensitive to strain rate history (although the effect is small) since the stress reached in experiments where jumps in the strain rate occurred does not reach the value obtained at constant strain rate. This sensitivity is shown to a greater extent when the strain rate falls.

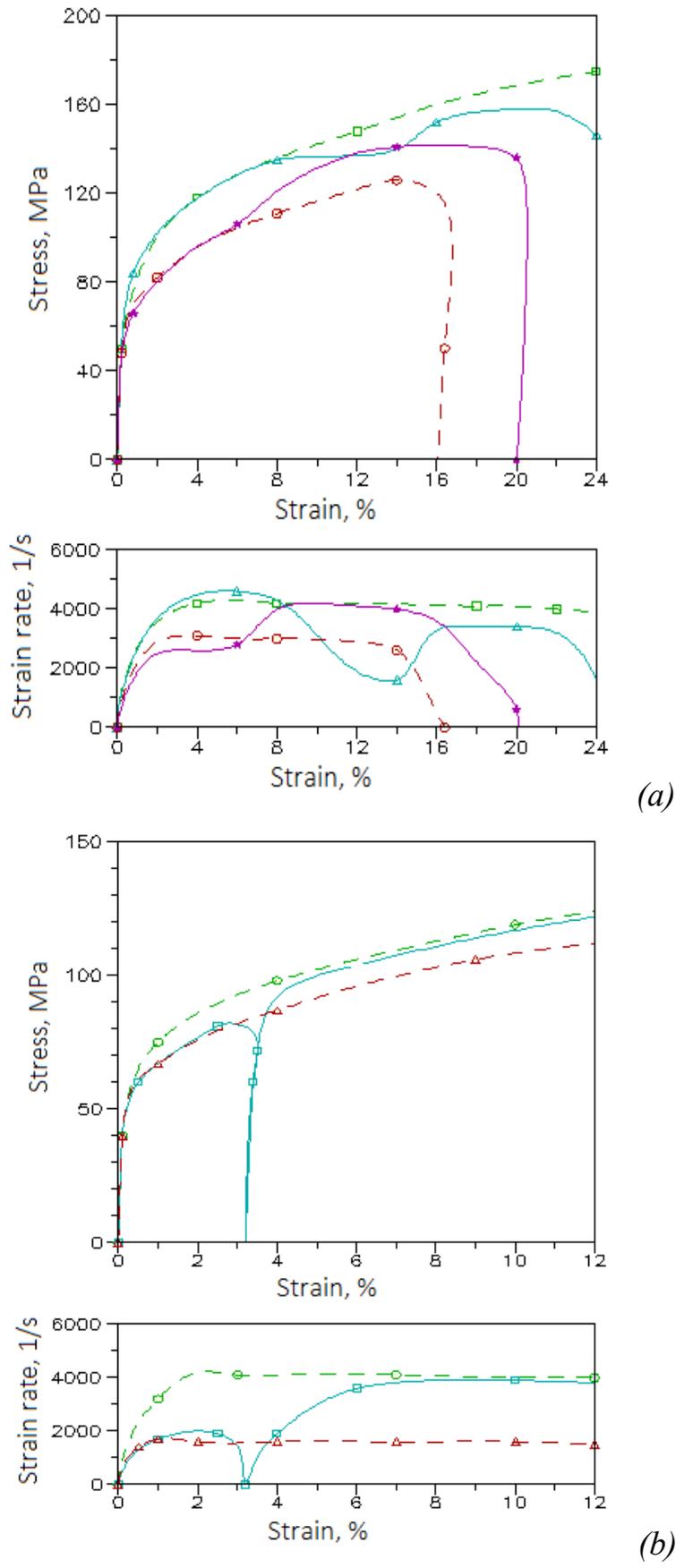

*(a)*

*(b)*

*Figure 25. Deformation diagrams obtained in strain rate jump tests.*

It should be noted that striker bars consisting of two parts made of the same material, but with different cross-sectional areas, can also be used to obtain jumps in the strain rate [152].

### 1.2.5 Alternating loading

The SHPB device described in section 1.1 [93] for studying the behaviour of materials under loading with pulses of opposite sign is difficult to implement. Therefore we designed a simpler device that allows a single cycle of compression-tension loading (Figure 26) [153, 154].

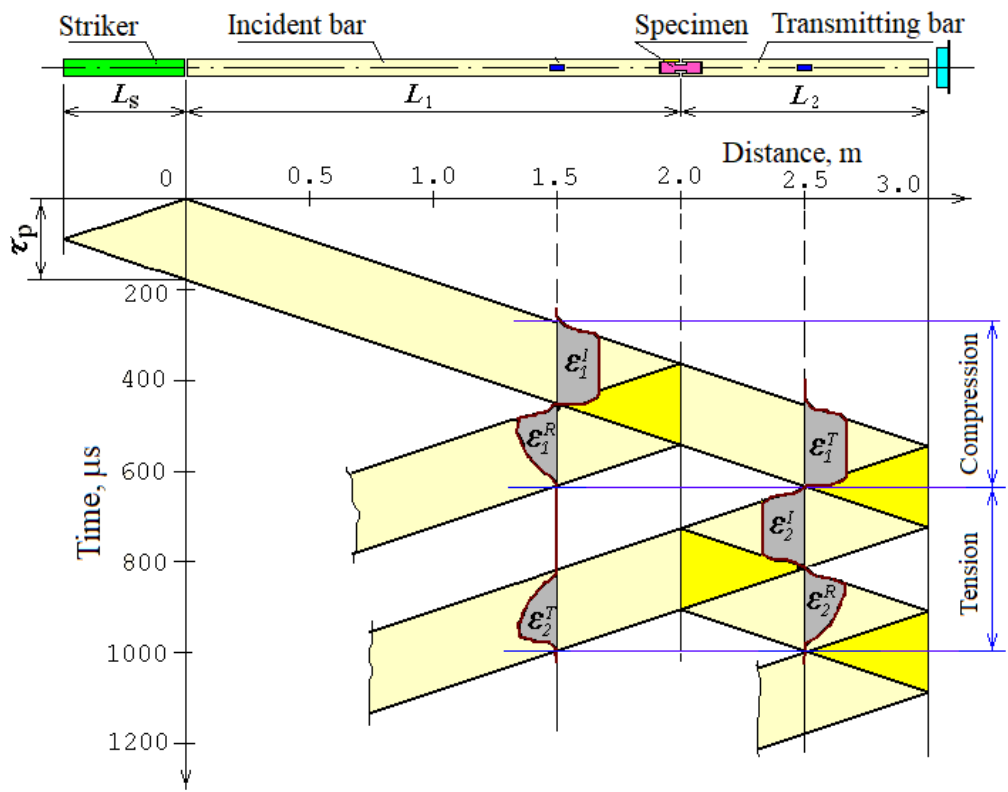

*Figure 26. Lagrangian x-t diagram for alternating compression-tension loading in an SHPB.*

Our design for alternating loading differs from a classic compression bar system in that the input bar is twice as long as the output bar (both bars are made from high-strength steel). Also the specimen is connected to the bars using a threaded connection. At the far end of this design, there is a small gap (3-5 mm) between the output bar and the damping stop in order not to interfere with the reflection of the elastic pulse from the free end of the SHPB during the first compression loading cycle (Figure 26).

We went on to design two variants of the alternating loading SHPB in order to study the Bauschinger effect at high strain rates [143, 153, 154]. These variants differ in the shape of the specimens and the way they are attached to the rods. In the first variant (Figure 27a), a solid cylindrical specimen (labelled 3) is screwed into the ends of the bars (labelled 1 and 5). To limit the amount of deformation during the first loading cycle and also to transmit a pulse into the output bar with an amplitude exceeding the stress level reached in the first loading cycle, split tubes (labelled 2) made from high-strength steel surround the specimen. This design is similar to the tensile SHPB proposed by Nicholas [72]). Since the length of these split tubes is slightly less than the distance between the ends of the rods, a small gap (labelled 4) is present, which determines the maximum value of compression strain produced in the specimen in the first loading cycle. It is possible to adjust this gap (and thus the amount of compressive strain achieved) by having a set of split tubes of different lengths.

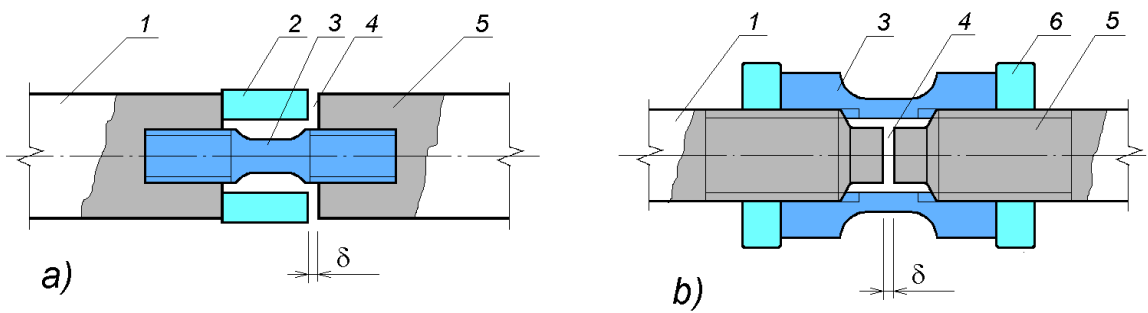

*Figure 27. Two methods for mounting specimens in an alternating compression-tension SHPB.*

In the second variant (Figure 27b) a tubular tensile specimen (labelled 3) is screwed onto the ends of the bars (labelled 1 and 5) and locked using nuts (labelled 6). Part of the thread is removed from the ends of the bars so that a small cylindrical gap is formed between the smooth area in the centre of the inner surface of the specimen and the outer surface of the ends of the bars. This means that the working section of the specimen may freely deform during the passage through it of compression and tensile pulses.

During preparation of the experiment, the bars are screwed into the specimen. Knowing the pitch of the screw thread determines the angle the bars must be twisted through to give the desired size of the gap $\delta$ (labelled 4) which sets a limit to the amount of compressive strain.

Figure 28 presents quasistatic and dynamic plots of alternating loading of 30HGSA steel at room temperature.

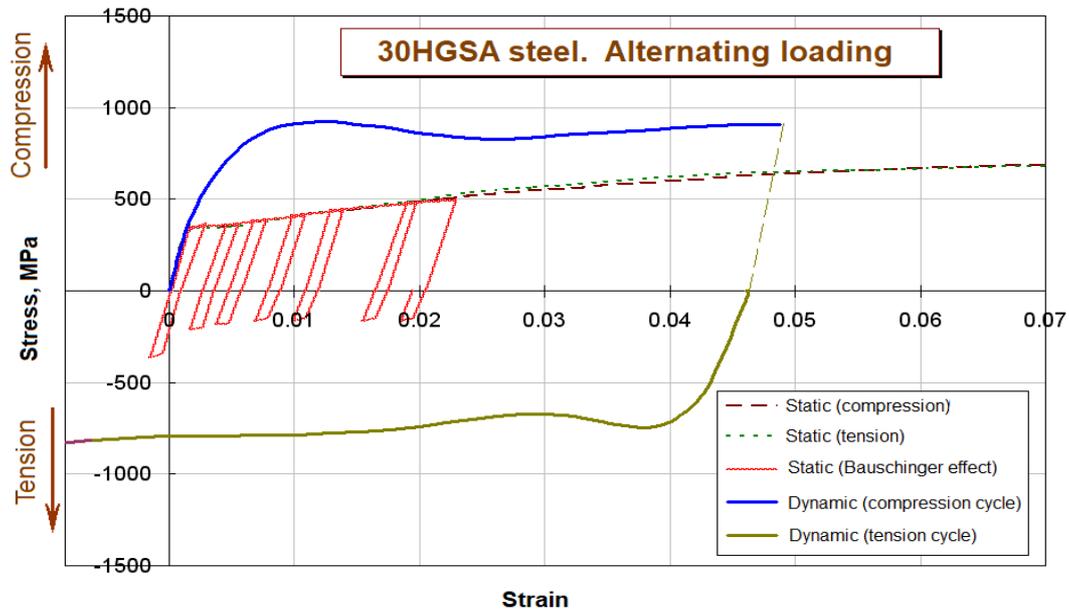

*Figure 28. Alternating compression-tension loading of steel 30HGSA, both quasistatically and dynamically.*

## 1.2.6 Testing materials with poor cohesion (soils)

The Kolsky method can be adapted for the dynamic testing of materials with poor cohesion (such as soils) [155-158]. The test soil sample is located between the ends of the Hopkinson bars in a rigid jacket that restricts its radial expansion (Figure 29).

Since radial strain is prevented by rigid confinement, an axially symmetric volumetric stress state develops in the sample after some time. In this case, the deformed state of the sample can be considered one-dimensional.

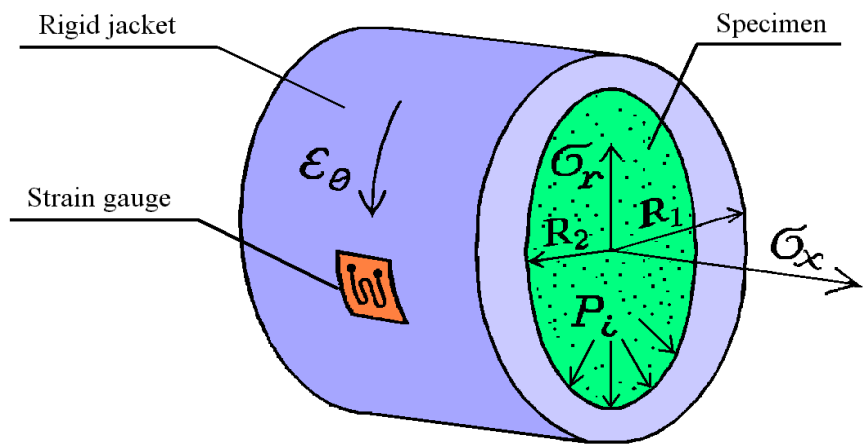

*Figure 29. Stresses and strains in the confinement jacket.*

Then the main components of the stress and strain tensors in the sample will have the form:

$$\sigma_1 = \sigma_x; \quad \sigma_2 = \sigma_3 = \sigma_r; \quad \varepsilon_1 = \varepsilon_x; \quad \varepsilon_2 = \varepsilon_3 = 0, \tag{59}$$

where $\sigma_x$ and $\varepsilon_x$ are longitudinal stresses and strains and $\sigma_r$ are radial stresses.

The axial components of the stress $\sigma_x(t)$, strain $\varepsilon_x(t)$, and strain rate $\dot{\varepsilon}(t)$ in the sample are determined by recording the strain pulses in the Hopkinson bars and analysed using the traditional SHPB equations (15).

The value of the radial component of the stress tensor can be obtained from the solution of the problem of the elastic deformation of a thick-walled tube under the action of internal pressure [129]. The relationship between the internal pressure $P_i$ and the circumferential strain of the jacket $\varepsilon_\theta$ is as follows:

$$P_i(t) = \frac{1}{2R_2^2} \left[ E \left( R_1^2 - R_2^2 \right) \varepsilon_\theta(t) \right] \quad , \tag{60}$$

where $E$ is the Young's modulus of the confining jacket material and $R_1$ and $R_2$ are the outer and inner radii of the tube, respectively. The internal pressure $P_i$, under which the confinement undergoes small elastic strains, is the desired radial stress $\sigma_r$. Thus the radial component of the stresses $\sigma_r(t)$ in the sample can be determined from the outputs of the strain gauges on the outer surface of the confinement $\varepsilon_\theta(t)$:

$$\sigma_r(t) = \frac{1}{2R_2^2} \left[ E \left( R_1^2 - R_2^2 \right) \varepsilon_\theta(t) \right] \quad . \tag{61}$$

Then if time is eliminated from the calculated variables $\sigma_x(t)$, $\varepsilon_x(t)$, $\dot{\varepsilon}_x(t)$ and $\sigma_r(t)$ after their mutual synchronization, a plot of the uniaxial stress-strain of the sample as well as the history of changes in the strain rate can be constructed.

The combination of two stress components in the sample $\sigma_x(t)$ and $\sigma_r(t)$ allows the calculation of a wide range of properties of the tested material.

The maximum tangential stresses (shear resistance) will be on planes located at an angle of 45° to the $x$-axis, and their values on these planes will be equal to:

$$\tau(t) = \frac{\sigma_x(t) - \sigma_r(t)}{2} \quad . \tag{62}$$

The pressure $P(t)$ in the sample is determined through the main stresses as follows:

$$P(t) = \frac{\sigma_x(t) + 2\sigma_r(t)}{3} \quad . \tag{63}$$

The volumetric strain will be equal to:

$$\theta(t) = \varepsilon_x(t). \tag{64}$$

Then (after synchronization) you can plot the curves of the compressibility $P \sim \theta$ of the soil and its shear resistance $\tau \sim P$.

The magnitude of stresses in the volumetric stress state is determined by $\sigma_i = \frac{1}{\sqrt{2}}\sqrt{(\sigma_1 - \sigma_2)^2 + (\sigma_2 - \sigma_3)^2 + (\sigma_1 - \sigma_3)^2}$ or taking into account the ratios (59): $\sigma_i(t) = \sigma_x(t) - \sigma_r(t)$.

The magnitude of strain is given by:

$$\varepsilon_i(t) = \frac{\varepsilon_x(t)}{(1+\nu)} \approx \frac{2}{3}\varepsilon_x(t) \quad . \tag{65}$$

In simple axisymmetric compression of samples that are able to expand laterally, the ratio between the transverse (radial) and longitudinal strains $\frac{\varepsilon_r}{\varepsilon_x}$ is called the Poisson ratio, $\nu$. In contrast, when testing soil samples in uniaxial compression without the possibility of lateral expansion ($\varepsilon_r = 0$), the ratio between the transverse (lateral) $\sigma_r$ and axial $\sigma_x$ stresses $\frac{\sigma_r}{\sigma_x}$ is called the lateral pressure coefficient, $\xi$.

General expressions of Hooke's law for an axisymmetric volumetric stress-strain state (in the elastic region) are:

$$\begin{aligned} E\varepsilon_r &= \sigma_r - \nu(\sigma_x + \sigma_r) \\ E\varepsilon_x &= \sigma_x - \nu(\sigma_r + \sigma_r) = \sigma_x - 2\nu\sigma_r \end{aligned} \quad . \tag{66}$$

If there is no radial component of the strain ($\varepsilon_r = 0$), the first of these equations becomes $\sigma_r (1 - \nu) = \nu \sigma_x$. Hence, taking into account equation (65) for the elastic behaviour of the material, it is possible to determine the relationship between the coefficient of lateral pressure $\xi$ and the Poisson ratio $\nu$:

$$\xi = \frac{\nu}{1 - \nu} \qquad \text{and} \qquad \nu = \frac{\xi}{1 + \xi} \ . \qquad (67)$$

All the parameters considered change with time, so the relationships between them should be considered throughout the test. The relationship between the axial and radial stress components is given by:

$$\sigma_r(t) = \xi \sigma_x(t) = \frac{\nu}{1 - \nu} \sigma_x(t) \ . \qquad (68)$$

Thus the coefficient of lateral pressure

$$\xi(t) = \frac{\sigma_r(t)}{\sigma_x(t)} \qquad (69)$$

and the coefficient of transverse expansion (the Poisson ratio)

$$\nu(t) = \frac{\sigma_r(t)}{\sigma_x(t) + \sigma_r(t)} \ . \qquad (70)$$

Thus the modified version of the Kolsky method allows the determination of the Poisson ratio, the volume compressibility curve, and the dependence of the shear resistance on pressure of a soil sample, as well as obtaining a plot of its uniaxial compression under one-dimensional deformed and volumetric stress conditions. All these parameters can be calculated using equations (61-64) using the Diagrammer® software package.

A numerical analysis of the applicability of the method for testing soils at high strain rates was performed [159]. The analysis of dynamic soil deformation was performed on the basis of Grigoryan's model of a plastic compressible medium [160]. The validity of the main prerequisites of the Kolsky method for obtaining reliable characteristics of the bulk and shear strains of soils was evaluated by checking the uniformity of the stress-strain state in the sample, the influence of friction forces and the effect of the deformability of the confinement on wave processes in the system.

Mathematical modelling of the processes of high-rate deformation of soft soils in the confining jacket was performed in the axisymmetric geometry consistent with the conditions of the experiment (Figure 30). The soil sample (labelled 7) is contained in a steel tube (labelled 4) placed between the ends of the input bar (labelled 8) and the output bar (labelled 6). The steel striker (labelled 1), accelerated in the barrel of a gas gun, strikes the input bar at a velocity $V_0$, exciting a flat one-dimensional elastic compression wave in it. The strain gauge (labelled 2), which is positioned at the middle of the input bar, records the longitudinal strain pulses (incident and reflected). The strain gauge (labelled 5), located at the middle of the output bar, records the strain pulse that passes through the sample. The strain gauge located on the outer surface of the confining jacket (labelled 3) records the circumferential strain from which the radial stress can be determined.

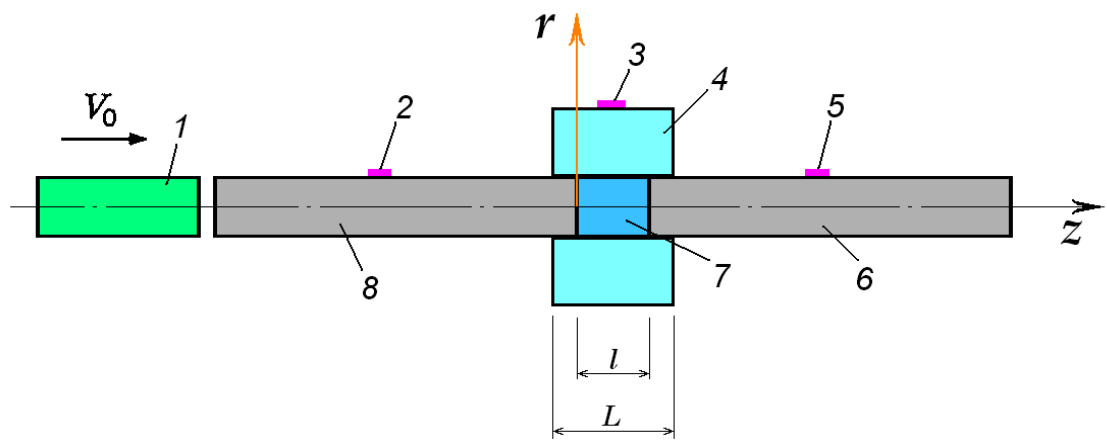

*Figure 30. Schematic diagram for setting the calculation for a confined soil sample.*

To describe the shock-wave process of system deformation, the variational-difference method is used [161] which is based on the dynamical relations of elastic-plastic materials. The original system of equations is written in a cylindrical coordinate system $rOz$, where the axis of symmetry $Oz$ coincides with the axes of rotation of the bars, and the axis $Or$ ($r \geq 0$) is perpendicular to $Oz$ and lies along the contact boundary of the input bar with the soil. The variational equation of motion is formulated in Lagrangian variables based on the principle of possible displacements in the Jourdain form:

$$\iint_{\Omega} \left( \sigma_{rr} \delta \dot{e}_{rr} + \sigma_{\theta\theta} \delta \dot{e}_{\theta\theta} + \sigma_{zz} \delta \dot{e}_{zz} + 2\sigma_{rz} \delta \dot{e}_{rz} \right) r \, d\Omega -$$
$$- \int_{G} \left( p_r \delta \dot{u}_r + p_z \delta \dot{u}_z + q_r \delta \dot{u}_r + q_z \delta \dot{u}_z \right) r \, ds + \iint_{\Omega} \rho \left( \ddot{u}_r \delta \dot{u}_r + \ddot{u}_z \delta \dot{u}_z \right) r \, d\Omega = 0 \quad (71)$$

here $\sigma_{ij}$ are the components of the stress tensor ($i, j = r, z, \theta$), $p_\alpha$, $q_\alpha$ ($\alpha = r, z$) are the components of the surface load and contact pressure, and $\rho$ is the density. The relationship of the strain rate tensor with the displacement rates is constructed in the current state metric, which allows the description of large displacements during step-by-step reconstruction of the geometry (coordinates $r$, $z$):

$$\dot{e}_{rr} = \dot{u}_{r,r}, \ \dot{e}_{zz} = \dot{u}_{z,z}, \ \dot{e}_{\theta\theta} = \dot{u}_{\theta,\theta}, \ \dot{e}_{\theta\theta} = \dot{u}_r / r, \ \dot{e}_{rz} = 0.5\left(\dot{u}_{z,r} + \dot{u}_{r,z}\right), \qquad (72)$$

where the subscript after the comma denotes differentiation with respect to the corresponding variable. The values of the strain tensor components are determined by time integration of the corresponding strain rate tensor components in equation (72). The components of the stress and strain tensors in the elastic bars and in the confining jacket are connected by Hooke's law. When describing the dynamic deformation of a plastically compressible soil, the strain tensor is represented as a superposition of volumetric and deviatoric components.

To solve the nonlinear wave problem formulated above, an explicit variation-difference scheme was used of the cross type, accurate to second order. The scheme was implemented within the framework of the Dynamics-2 package [162]. Two variants of the confinement geometry were considered: 1) the length of the hoop exceeds that of the sample by 3 mm on each side in order to ensure the alignment of the hoop on the ends of the Hopkinson bars; 2) the length of the clip is 1.7 times the size of the soil sample. As a result, it was found that when using a hoop whose length $L$ exceeds the size $l$ of the soil sample, an error occurs in calculating the radial stress in the soil from the value of the measured circumferential strain of the hoop. Thus for a steel confinement with the ratio of the outer diameter $b$ to the inner diameter $a$ equal to 2, the radial pressure in the soil was overestimated by an amount proportional to the ratio $L/l$.

To overcome this disadvantage, the design of the confinement was changed (Figure 31) so as to centre it on the ends of the measurement bars (labels 1 and 4). Thin guide sleeves (1.5 mm thick) were included, whereas the main body of the confining jacket has a length $L$ equal to the length of the hoop $l$ [163]. A series of tests was performed on the dynamic loading of a sample of sandy soil containing humid air using the proposed confinement (label 3). The length of the sample and the shape of the loading pulse varied. In all experiments, the length of the sample (label 5) reduced with the length of the jacket. The confining jackets were made of aluminium alloy D16T and steel 30HGSA, depending on the amplitudes of the applied loads. The ratio of the outer diameter $b$ to the inner diameter $a$ was equal to 1.5. The high deformability of the aluminium alloy compared to steel allows for more accurate recording of circumferential strains on the outer surface of the confining hoop using a strain gauge (label 2) at low stresses in the soil sample (label 5). Experimental data were processed using the method described by Bazhenov *et al.* [159] in order to obtain dynamic stress-strain plots and also the dependence of shear resistance on pressure.

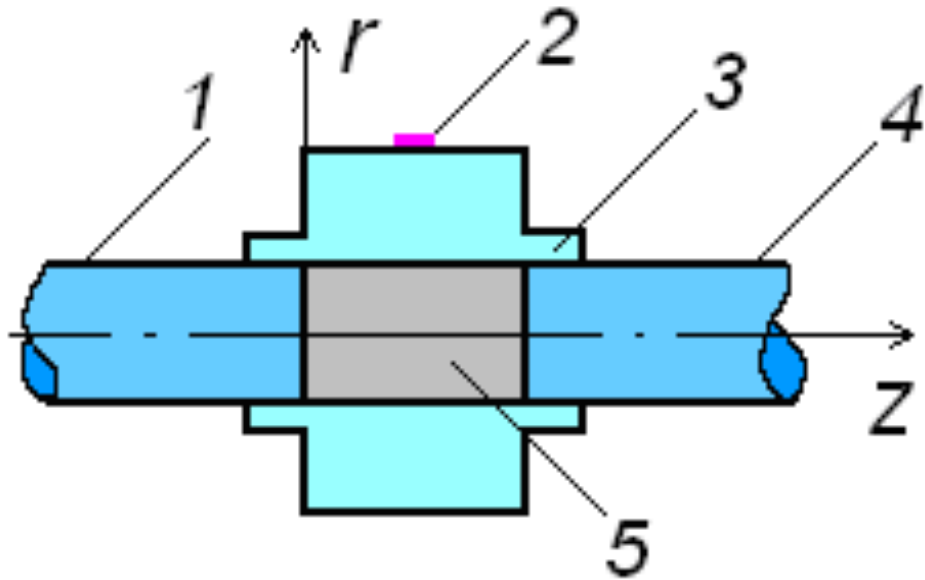

*Figure 31. Schematic diagram of the modified confining hoop.*

To assess the effect of the sample length on the resulting stress-strain diagram, a series of tests were performed of sandy soil containing humid air. In all the experiments, the length $l$ of the sample coincided with the length $L$ of the hoop. Three lengths were used: 6.5, 9.5 and 20 mm. The samples were loaded using trapezoidal strain pulses having a duration of ~175 microseconds with rise and fall times of 15-20 microseconds.

Figure 32 shows the dependence of the longitudinal stress in the soil on the strain for samples of different lengths $l$. The green dotted line is for $l = 9.5$ mm, the solid blue curve is for $l = 6.5$mm, and the dashed red line is $l = 20$ mm. From the data presented, it can be seen that for small strains, the deformation diagram shows fluctuations due to the wave loading of the sample. The amplitudes of these fluctuations are smaller for smaller sample thicknesses.

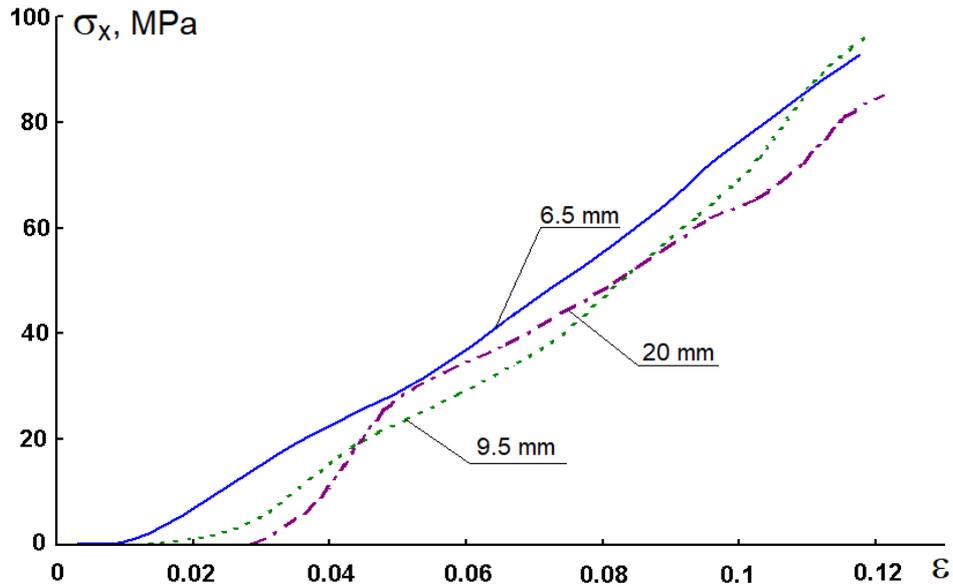

*Figure 32. Effect of sample length on the high rate stress-strain plot for confined soil.*

The minimum length of the sample is determined by the requirement that the sample must be representative of the bulk material. So the sample thickness must be at least an order of magnitude larger than the size of the soil particles. We therefore used samples with a thickness of 9.5 mm in our experiments. Figure 32 shows the section with zero stresses for all three deformation curves, the smallest and largest strains with zero stress being for samples with lengths of 6.5 and 20 mm, respectively.

The average dependence of the sand shear resistance on the pressure was obtained for samples with a length of $l = 9.5$ mm. This relationship is shown in Figure 33 as a solid blue line. The equation of the dotted green line which most closely fits the experimental data is $\tau = P \tan\varphi$, where $\varphi$ is the angle of internal friction in the sand. For humid sand with a density of 1.6 g / cm$^3$, the measured value of the internal friction angle $\varphi = 27.80°$.

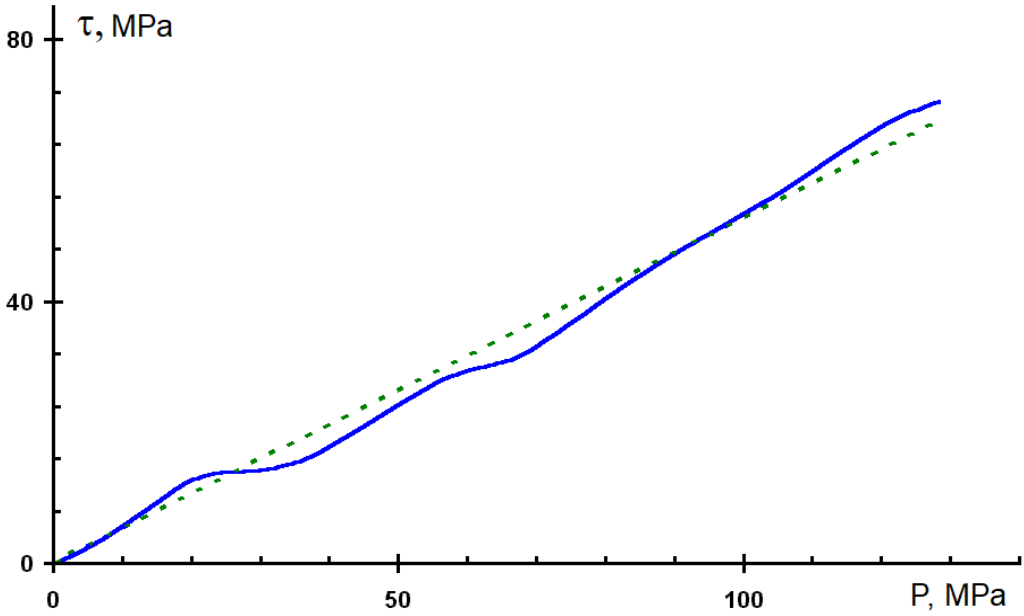

*Figure 33. The curve of high rate shear resistance against pressure for soil.*

To verify the basic premise of the Kolsky method about the homogeneity of the stress-strain state in the soil specimen, a numerical calculation was performed using the 'Dynamics-2' software package using the Grigoryan soil model [160]. In the calculation it is necessary to set the dependence of the pressure $P$ on the strain $\varepsilon$ (both in loading and unloading) as well as the dependence of yield strength on pressure. The parameters of the initial section of the curve $P\sim\varepsilon$ were determined from SHPB experiments [164]. The experimental data are approximated by the power dependence of the form proposed by Rykov [165].

$$P = M\,\theta^n \tag{73}$$

where $\theta = 1-\rho_0/\rho$ .

The constants $M$ and $n$ are determined for each type of soil based on the results of a series of experiments. For pressures more than 200 MPa, the results of plane-wave shock experiments were used [166]. The experimental dependence of the velocity of the shock wave $D$ on the particle velocity $U$ for the linear case $D = A + BU$ can be transformed using conservation laws at the shock wave front into the form

$$\sigma(\theta) = -\rho_0^2\,\theta/(1-B\,\theta)^2. \tag{74}$$

For $P(\theta)$, we apply a similar dependence with constants $A$ and $B$

$$P(\theta) = -\rho_0 A^2\,\theta/(1-B\,\theta)^2. \tag{75}$$

The dynamic dependence of the yield strength $\sigma_T$ on the pressure is assumed to be linear (Figure 33)

$$\sigma_T = \sigma_0 + kP \quad . \qquad\qquad\qquad (76)$$

For non-cohesive soils, in particular dry sands, the value $\sigma_0$, which has the physical meaning of cohesion, is close to zero, $K = 2\tan\varphi$.

Calculations were made for bars and hoops made of D16T alloy. Figure 34 shows the time dependence of the axial stresses in the centre (solid line) and on the side surface of the sample (dotted line) in comparison with the experimental data (markers). There is a good correspondence of results both at the loading stage and during unloading.

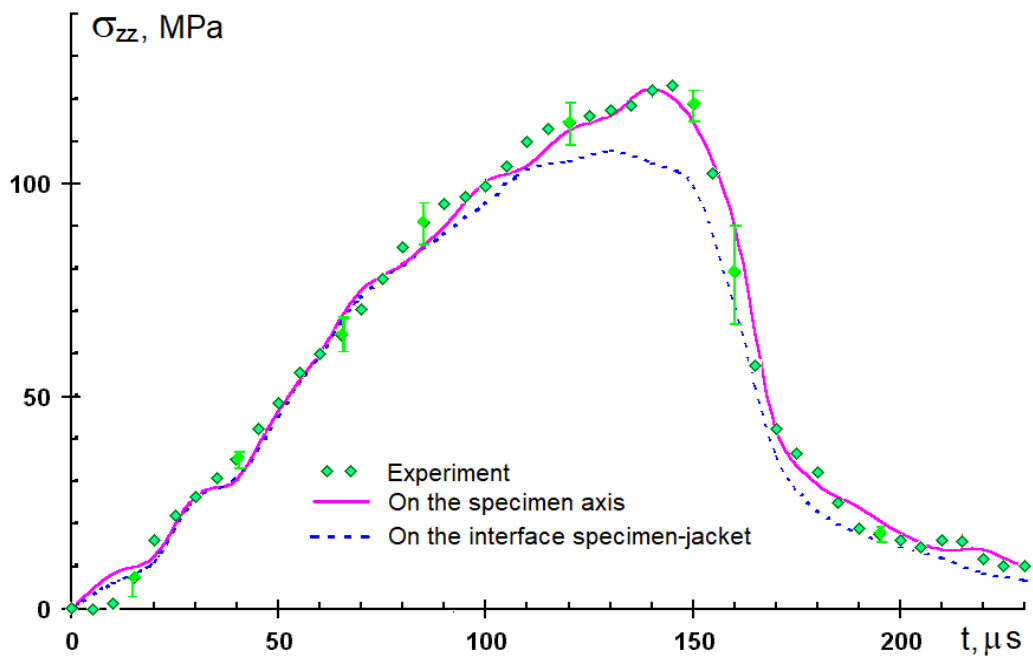

*Figure 34. Plot of axial stress against time for dynamically loaded soil.*

The calculated values of radial stresses in the centre (solid purple line) and on the lateral surface of the sample (dotted blue line) are shown in Figure 35. In the same place, green dots show data obtained from measurements of the circumferential strain of the hoop. The specified confidence limits were determined with a reliability of 0.94 [163]. Thus these results indicate the stress state within the sample was uniform. The differences between the results of the numerical calculations and the experiment were less the experimental error, which indicates that the parameters of the soil model have been correctly determined in the load range considered.

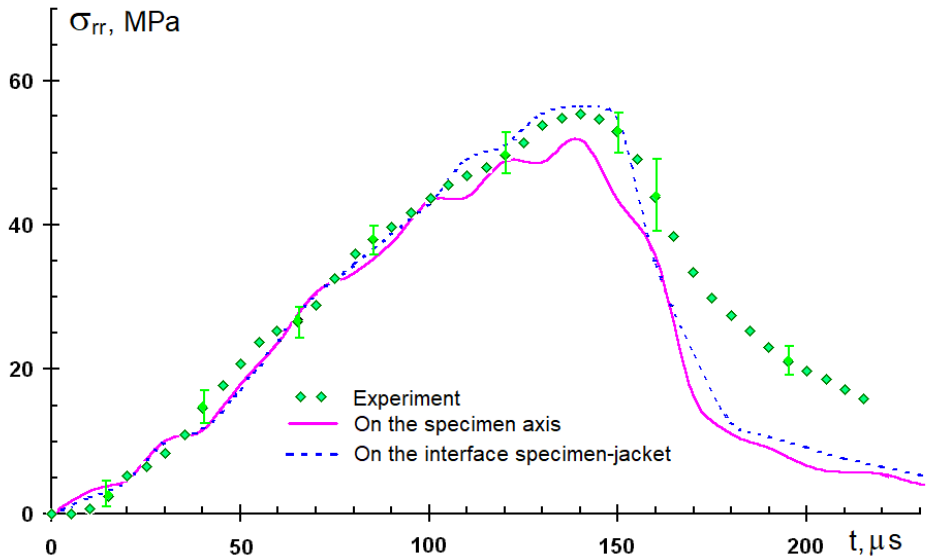

*Figure 35. Plot of radial stress against time for dynamically loaded soil.*

Since soils (especially sand) are quite abrasive materials, the results obtained may be affected by the friction forces that occur on the inner surface of the confinement. To evaluate the effects of friction and the effect of the hardness of the confining material, experiments were performed on loading sand in duralumin and steel jackets that were lubricated on their internal surfaces. In the case of the duralumin jacket, the sample was loaded to an axial stress of ~150 MPa. For the steel jacket, the applied load was ~400 MPa. In Figure 36, the dotted lines show the stress-strain plots obtained using confining jackets with graphite grease applied to the inner surface, and the solid lines are those obtained when no grease was applied.

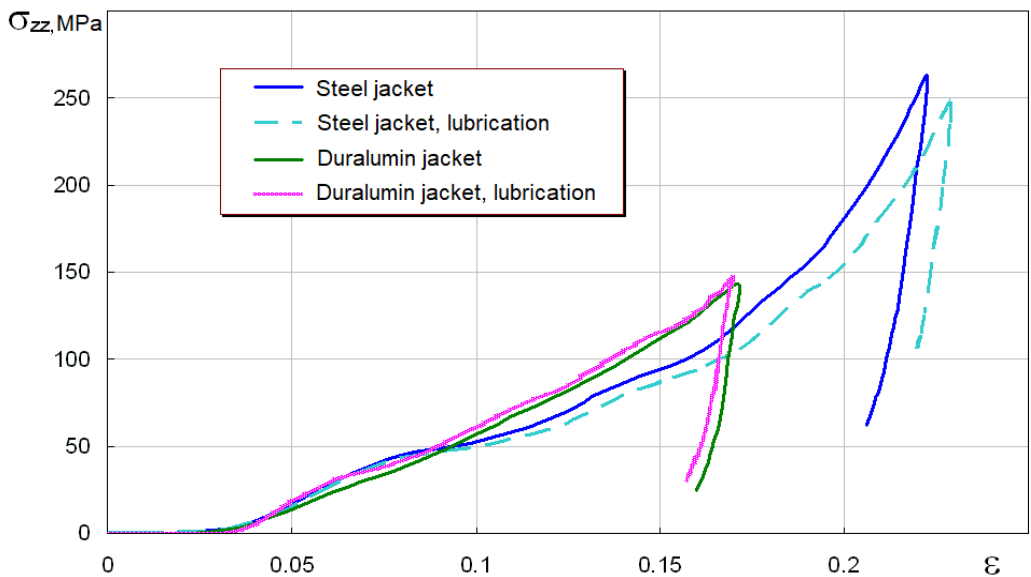

*Figure 36. Influence of the jacket material and the presence or absence of lubrication on the stress-strain curves of dynamically loaded soil.*

It can be seen that for stresses up to about 50 MPa, the plots obtained using both duralumin and steel jackets are almost identical. The effect of friction in both cases is also insignificant. For stresses greater than 50 MPa, the plots of $\sigma_{zz}$ against $\varepsilon$ obtained in steel confinement were noticeably lower. Deformation diagrams obtained using duralumin jackets with and without lubrication are almost identical even at stresses greater than 50 MPa. This can be explained as follows. After the test, noticeable abrasive damage was observed on the inner surface of the duralumin confinement due to the interaction of sand particles and the material of the confinement. The resulting additional resistance forces make a significant contribution to the longitudinal stresses, eliminating the influence of lubrication. When using steel as the jacket material, this effect is significantly less due to the higher hardness of the steel. This is evidenced by the absence of abrasive deformation on the inner surface of the jacket, which is why the plot of $\sigma_{zz}$ against $\varepsilon$ is lower. This is also why the effect of lubrication on the deformation diagram obtained with steel confinement is significant.

Thus, in order to obtain reliable curves of soil deformation over a wide range of pressures, it is possible to use jackets made of D16T aluminium alloy (duralumin) and 30HGSA steel. The use of a duralumin confinement allows the correct determination of longitudinal and radial stresses at low pressures. A steel cage should be used at higher pressures. The values of longitudinal stresses of ~400 MPa obtained in experiments with a steel jacket on steel measurement bars are already close to the data of the plane-wave shock experiment, which allows for direct comparison and construction of deformation diagrams for a wide range of load amplitudes.

Based on the results of this research, a rational design and geometry of the jacket are proposed, which allow reliable data to be obtained for the dynamic compressibility and shear strength of soft soils. The use of jackets made of different materials in the tests allows the study of a fairly wide range of loads, the upper range of which overlaps with the lower range of plane-wave shock experiments. Numerical analysis of the distribution of axial and radial stresses in the samples showed their uniformity, which indicates that the main premise of the Kolsky method is fulfilled.

### 1.2.7 Alternating loading of low-density materials

In compression tests of low-density materials (wood, foams, highly porous ceramics, etc.), the amplitude of the tensile pulse reflected from the specimen is 80-90% of that of the incident wave due to the large difference in the acoustic impedance of the measuring bars and the specimen. This reflected pulse, having reached the front (impacted) end of the input bar, is reflected back as a compression pulse (since the striker is no longer in contact) and, having reached the specimen, loads it a second time. These loading cycles repeat many times until the pulse dissipates. Thus the specimen is loaded many times during a single test, undergoing a strain increment in each cycle. The interval between cycles is equal to the double transit time of the strain pulse along the input bar. To record repeated loads during one experiment, it is necessary in the second and subsequent cycles to exclude the influence on the loading process of the pulse that passed through the specimen and is then reflected from the rear end of the output bar in the form of a tensile wave. This pulse reaches the specimen at the same time as the loading pulse and hence affects the applied strain if the input and output bars are of equal length. The momentum vectors at the two ends of the specimen are directed in the same direction.

In order to prevent the return of the transmitted pulse, Lindholm suggested using a momentum trap bar attached to the rear end of the output bar [88]. In this case, the apparatus provides a record of only one additional loading cycle of the specimen.

In order to correctly record several loading cycles, Zhao & Gary [167] proposed a scheme for placing strain gauges at two locations on each bar. They also developed a mathematical analysis which makes it possible to isolate the corresponding strain pulses in the bars in each loading cycle. This scheme allowed the recording of four loading cycles of foam specimens so that the behaviour of the material under significant compaction (up to 80%) could be monitored. It should be noted, however, that this scheme, while providing accurate recording of several specimen loading cycles, does not prevent distortion of the loading the specimen is subjected to due to the transmitted pulse being reflected from the rear end of the output bar.

In contrast to Lindholm [88] and Zhao & Gary [167], where repeated loading was caused by a pulse reflected from the specimen in the first loading cycle, Ogawa achieved alternating loading specimens by using a stepped bar [93]. In this case, the required combination of loading pulses was achieved by selecting the cross-sectional areas of the bar.

We used a compound striker to load the specimen with pulses of a single sign with a variable interval between them [143, 153]. In this case, the amplitudes and durations of the pulses were varied by using different materials and rod lengths. The interval between the pulses was changed-by varying the gap between the component parts of the rod.

To correctly record the corresponding strain pulses and prevent distortion of the stress-strain state of the specimen, a simple modification of the Kolsky method was proposed, which allows the testing of low-density materials under alternating loads of one sign [168]. To do this, the length of the output bar must be increased in comparison with the length of the input bar by as many times as the number of loading cycles it is desired to record.

Figure 37*a* shows a diagram of the compound split bar and an *x-t* wave diagram for three-cycle loading of a specimen. 1.5 m and 4.5 m long Hopkinson bars were used which allowed the recording of three loading cycles and specimen strains of more than 60%. As an example, an oscillograph is presented in Figure 37*b* of three-cycle loading by pulses of the same sign of a test specimen of Sequoia wood confined in a rigid hoop.

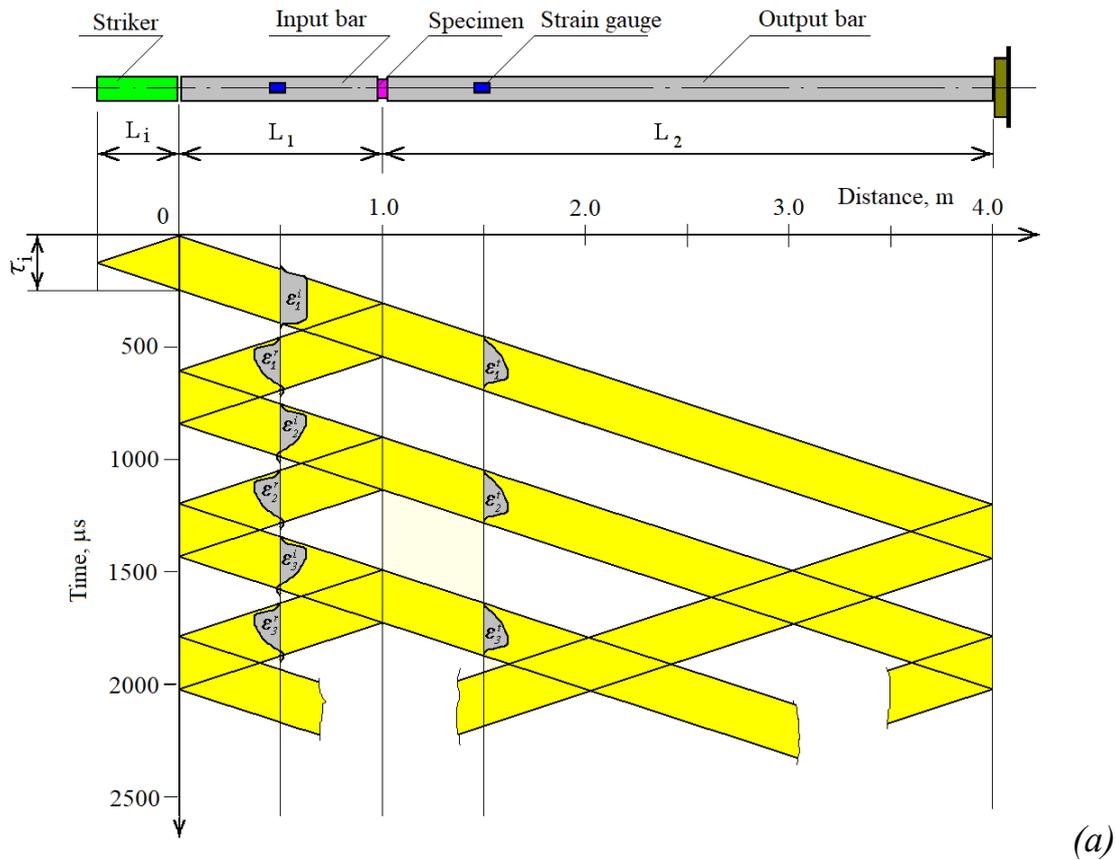

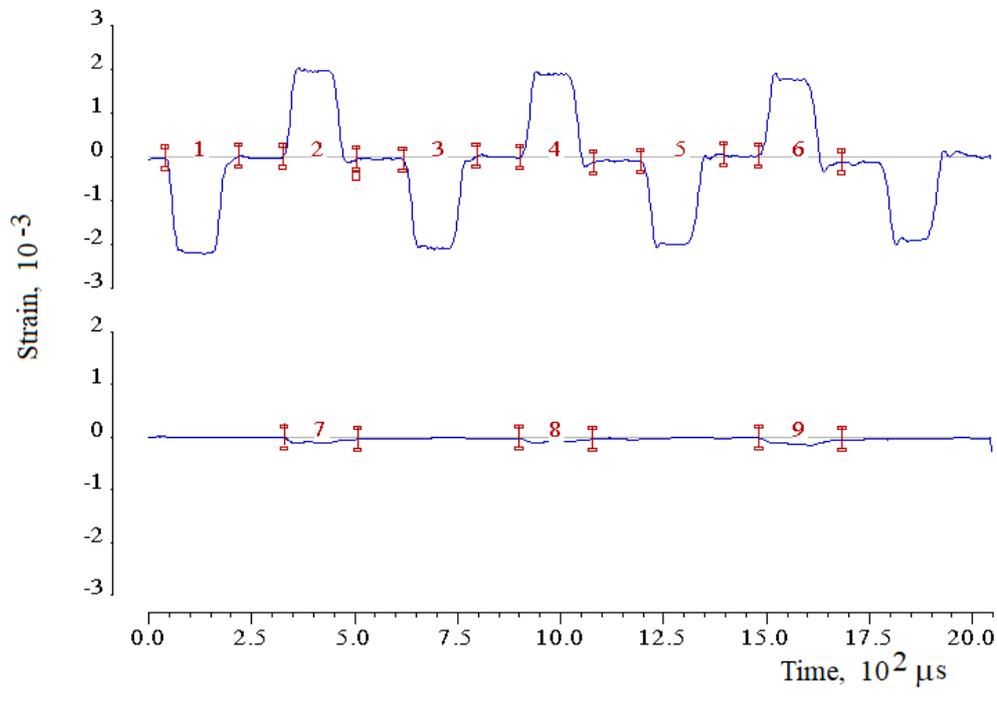

*Figure 37. (a) Lagrangian x-t diagram of wave propagation pattern in an SHPB when recording (b) three loading cycles (typical oscillogram when testing wood).*

The proposed modification can also be successfully used in tensile tests using the Kolsky method, since in this case, due to the significant difference between the cross-sectional areas of the Hopkinson bars and the specimen (12:1), the reflected pulse also has a large amplitude, sufficient to produce plastic deformation of the specimen in the second and even in the third loading cycle.

### 1.2.8 Determining dynamic hardness

Indentation has for a long time been a technological method for evaluating the mechanical characteristics of a material [169, 170]. Hardness, the parameter that is obtained, is the resistance of the material to local plastic deformation either by normal indentation or by scratching. When determining the hardness of a material by pressing in a steel ball or diamond pyramid, it can be estimated by measuring the ratio of the load to the surface area of the dent (Brinell number $H_B$, Vickers number $H_V$) or by the depth that the loaded tip penetrates into the test surface (Rockwell number $H_R$).

Quasistatic indentation methods are well developed and widely used for the indirect determination of a number of fundamental mechanical characteristics of a material, including Young's modulus and yield strength [171]. These characteristics are determined using experimental loading (and/or unloading) curves when using conical or pyramidal indenters with different angles at the vertex [172-174] or spherical indenters [175, 176]. Due to the complexity of the nature of the deformation caused by an indenter, numerical methods play an important role in the development of various experimental methods and procedures for determining the mechanical characteristics of a material [177-179].

Since quasistatic indentation is a relatively simple and well-developed means of determining the properties of a material, the question arises whether this technique can be extended to the dynamic loading of materials while maintaining its simplicity.

Compared to quasistatic studies, there are a relatively small number of published papers on dynamic indentation. Davies & Hunter mounted a conical indenter on a pendulum in order to study the velocity sensitivity of indentation at velocities up to 30 cm/s [136]. Mok & Duffy used a spherical indenter at penetration rates up to 5 m/s [180]. The authors showed that the behaviour of the velocity sensitivity of the dynamic hardness is approximately consistent with the flow stress obtained using an SHPB. Nobre *et al.* used a pendulum impact testing machine equipped with a piezoelectric sensor, which made it possible to measure shock pulses [181]. Tirupataiah & Sundarajan used high-velocity (up to 200 m/s) ball impact to study the relationship between average deflection, average strain rate, and flow stress or dynamic hardness [176]. During these experiments, the time dependence of the signals characterizing the indentation process was not registered. Subhash and co-workers proposed a technique for performing dynamic indentation experiments using the one-dimensional theory of elastic wave propagation widely used in the Kolsky method [182, 183]. They used a single Hopkinson bar to register the time dependence of the force response and to determine the depth of penetration during the experiment, they used a steel strip to which a strain gauge was bonded. The displacement was determined based on the signal from the sensor using a simple theory of the quasistatic bending of beams. This limited the study to a relatively low velocity range for which wave and dispersion effects can be ignored.

Lu *et al.* proposed a direct impact bar technique to allow the measurement of the indentation depth during a test [184]. This was achieved using an optical method (moiré interferometry) to record the movement of the striker bar. The time-depth-force relationship was used to estimate the rate sensitivity of oxygen-free copper. A numerical simulation of the dynamic indentation process was performed in order to verify the proposed method.

Bragov and co-workers modified the split Hopkinson pressure bar method to perform dynamic indentation tests on materials and thereby determine the dynamic Brinell hardness [143, 185]. The main difference from the normal Kolsky compression method is that a replaceable indenter (either conical or hemispherical) is installed between the input bar and the specimen (Figure 38). Solid tungsten-cobalt alloys are used as the indenter material.

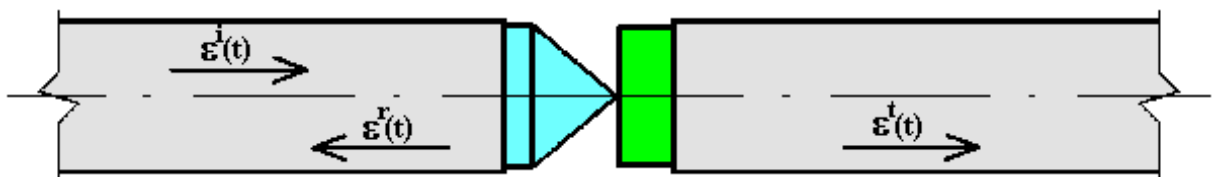

*Figure 38. Schematic diagram of dynamic indentation performed using the Kolsky bar method.*

Recording the incident pulse $\varepsilon^i(t)$ and the reflected pulse $\varepsilon^r(t)$ allows the determination of the depth of the indentation $h(t)$ in the specimen based on the one-dimensional theory of elastic wave propagation:

$$h(t) = c \int_o^t \left[ \varepsilon^i(t) - \varepsilon^r(t) \right] \cdot dt \quad . \tag{77}$$

For a known indenter geometry, this equation allows the calculation of the surface area of the indentation at any time during the application of the load. For a conical indenter with a vertex angle $2\alpha$, the area $S(t)$ of the indentation is given by:

$$S(t) = \pi h^2(t) \frac{\tan^2 \alpha}{\sin \alpha} \ . \tag{78}$$

When using a hemispherical indenter of radius $R$ the area $S(t)$ is calculated using the following formula:

$$S(t) = 2\pi R h(t) \ . \tag{79}$$

The pulse in the output bar allows the determination of how the force $F(t)$ develops with time while the indenter is being pressed into the specimen:

$$F(t) = EA\varepsilon^t(t) \ , \tag{80}$$

where $E$ is the Young's modulus and $A$ is the cross-sectional area of the output bar.

Then defining the dynamic Brinell hardness $H_B(t)$ as the ratio of the force with which the material resists indentation to the area of the dent:

$$H_B(t) = F(t)/S(t) \ . \tag{81}$$

Thus using the relations above, it is possible to calculate the value of $H_B$ at any time during the indentation process.

For reliable calculation of the dynamic hardness from $F(t)$ and $S(t)$, the pulses $\varepsilon^i(t)$, $\varepsilon^r(t)$ and $\varepsilon^t(t)$ must be strictly consistent in time. In order to facilitate this synchronization, the strain gauges are bonded to the Hopkinson bars at the same distance from the specimen.

Figure 39 shows the dependences obtained for the dynamic resistance force to (a) the depth of penetration and (b) the contact area for both conical and hemispherical indenters using a modified Kolsky bar for the aluminium alloy AMg6 at a penetration rate of ~20 m/s. The tangent of the slope of the curve in Plots of the dynamic hardness (Figure 39*c*) can be obtained from the data presented in Figure 39*b*. Figure 39*c* also shows how the dynamic hardness changes throughout the indentation process.

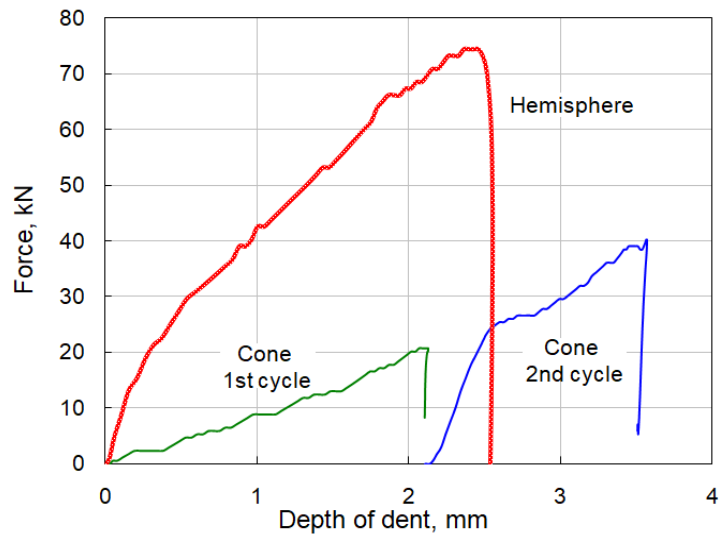

*(a)*

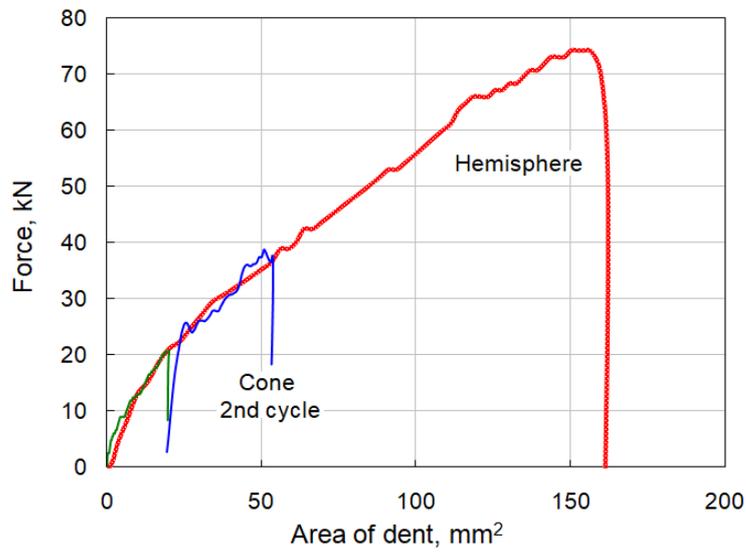

*(b)*

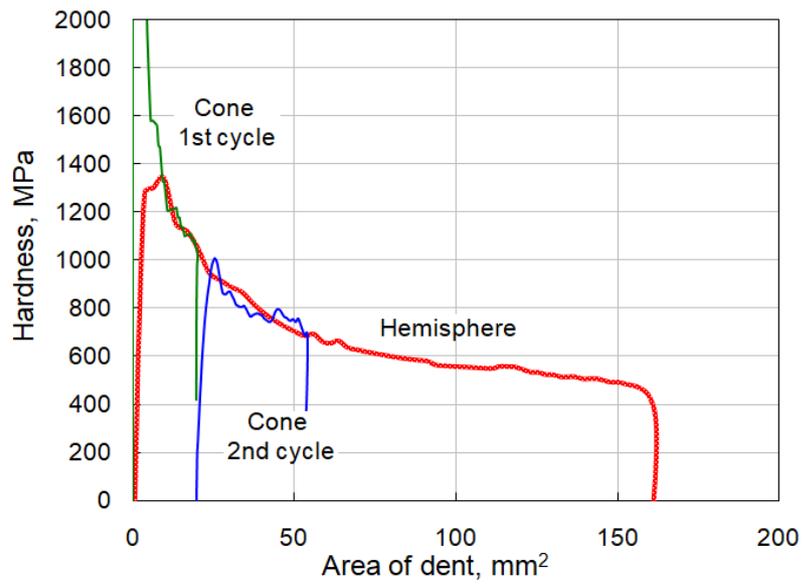

*(c)*

*Figure 39. Results obtained for dynamic indentation experiments performed using conical and hemispherical indenters in the Kolsky bar.*

The standard quasistatic Brinell hardness value for the AMg6 alloy given in the reference books* is 650 MPa, which indicates the reliability of the results obtained. As can be clearly seen, the shape of the indenter has no practical effect on the value of the dynamic hardness.

### 1.2.9 Determination of dynamic crack resistance parameters

Significant progress has been achieved in recent years in determining the fracture resistance of materials due to the development of methods of fracture testing that allows the establishment of a quantitative relationship between the nominal stress, shape and size of defects and the resistance of materials to stable and unstable crack growth. The basic premise of fracture mechanics is that the breaking of structural components is always a consequence of the development of cracks that occurred either during the manufacturing process (e.g. welding, grinding, quenching) or to mechanical loading or chemical corrosion during their use. The development of linear fracture mechanics led to the introduction of new material properties namely the material's resistance to fracture (characterised by the critical stress intensity factors $K_{C(d)}$ and $K_{1C(d)}$). These are considered to be material constants at a given temperature and loading rate and are used for calculations of the plane strain at the crack tip. These calculations are based on a comparison of the calculated stress intensity factor $K_1$ at the vertex of a crack-like defect with the material's characteristic crack resistance $K_{1C(d)}$.

Among the experimental methods for determining crack resistance, the most commonly used tests are Charpy [186, 187], notched rods tests (known as the Izod test [188]), and the SHPB method proposed by Klepaczko [189].

The last method is preferred due to fairly accurate load measurement using strain gauges, simpler analysis and interpretation of the data obtained, a simple test procedure, and meeting the requirements of testing standards. The difference between the method of testing for crack resistance proposed by Klepaczko and the traditional SHPB method is the use of a wedge at the end of the output bar and special compact specimens (Figure 40).

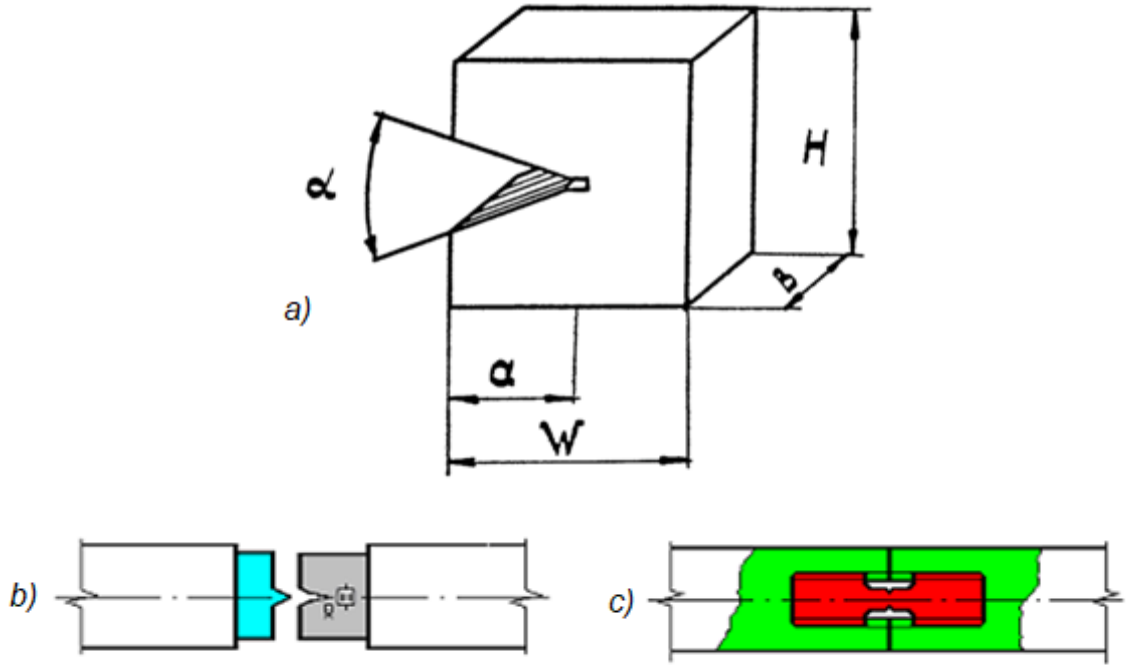

*Figure 40. Options for studying crack resistance using a Kolsky bar.*

By recording the reflected $\varepsilon^R(t)$ transmitted $\varepsilon^T(t)$ waves, the wedge displacement $\delta(t)$ and the total force $P(t)$ can be calculated as a function of time:

$$\delta(t) = -2C \int\limits_0^t \varepsilon^R(V) \cdot dV , \qquad (82)$$

$$P(t) = EA \cdot \varepsilon^T(t) . \qquad (83)$$

This result is important for determining the critical force in the formula for calculating the stress intensity factor. If the critical point $t_c$ on the ray $\varepsilon^T(t)$ is defined, the critical force can be calculated as:

$$P_c(t_c) = EA \cdot \varepsilon^T(t_c) , \qquad (84)$$

where the index $c$ corresponds to the critical state when the crack begins to grow. The relationship between the critical compressive force $P_c$ and the tensile force $Q$ is described by the following expression:

$$Q = \frac{P_c}{2 \tan\left( \dfrac{\alpha}{2} + \beta \right)} , \qquad (85)$$

where $\alpha$ is the wedge angle in radians, $\beta = \tan^{-1} \mu$ is the angle of friction, and $\mu$ is the coefficient of friction, usually measured quasistatically. The small size of the specimens allows you to get a short fragmentation time, in our case about 20-30 μs.

Since the striker bar is at least 10 times longer than the total length of the wedge and the specimen, the process of loading the specimen with an incident wave through the wedge can be considered quasistatic, so it is possible to use the appropriate solution to calculate the stress intensity factor:

$$K_{1C} = \frac{Q}{B\sqrt{W}} \cdot f = \frac{EA\varepsilon^T(t_c)}{2B\sqrt{W} \cdot \tan\left(\dfrac{\alpha}{2} + \tan^{-1}\mu\right)} \cdot f \ , \qquad (86)$$

where $Q$ is the critical force, $a$ is the length of the sharp incision, $B$ is the thickness and $W$ is the width of the specimen (Figure 40a). The function $f = F(a/W)$ is the malleability function, or elastic strain function, calculated by the finite element method and described by the following equation:

$$f = 4.11 - 1.83\left(\frac{a}{W}\right)^{\frac{1}{2}} + 21.13\left(\frac{a}{W}\right)^{\frac{3}{2}} + 11.44\left(\frac{a}{W}\right)^{\frac{5}{2}} + 18.61\left(\frac{a}{W}\right)^{\frac{7}{2}} . \qquad (87)$$

In experiments to determine dynamic crack resistance, usually the time $t_c$ that the crack starts to grow (and its corresponding critical force, which can be used to calculate the stress intensity factor) is determined by the characteristic point on the pulse $\varepsilon^T(t)$. As a rule, this corresponds to either the maximum or the inflection point on the rising part of the pulse passing through the specimen $\varepsilon^T(t)$. More precisely, the time $t_c$ can be determined by recording signals from strain gauges bonded directly onto the specimen. The choice of the location for gluing the gauges onto the specimen relative to the tip of the crack is quite critical. By conducting experiments with different gauge locations and orientations, we found that the most effective way to detect the start of crack straining is to glue a gauge at a small distance (1-2 mm) to the side of the crack's apex. When the crack begins to grow, a release wave begins to propagate from its sides, leading to a sharp drop in the amplitude of the strain pulse recorded by the gauge. The recording of the signal by gauges glued directly onto the specimen makes it possible to determine the (critical) fracture load more precisely.

Our method [143] for obtaining crack resistance characteristics differs little from Klepaczko's. The main difference is that in order to measure the crack propagation rate in a compact specimen, a special sensor is glued where the crack is expected to move. The sensor is a set of individual wires each one with a base of 1 mm, fabricated on a single substrate, with a distance between them of 1.7 mm. All the sensor arrays are connected in parallel and are powered by a standard power supply. When the crack extends, individual wires are broken sequentially, which causes sharp changes in the overall resistance of the sensor and corresponding line jumps on the oscilloscope screen. This makes it possible to determine the rate of crack extension.

In addition, to determine the value of $K_{1C(d)}$, two different Kolsky bar methods have been implemented: (i) the extension of a solid cylindrical specimen with a weakened section (V-shaped notch) and (ii) three-point bending of a beam with a V-neck. Schematic diagrams of the two types of fracture specimens are given in Figure 41.

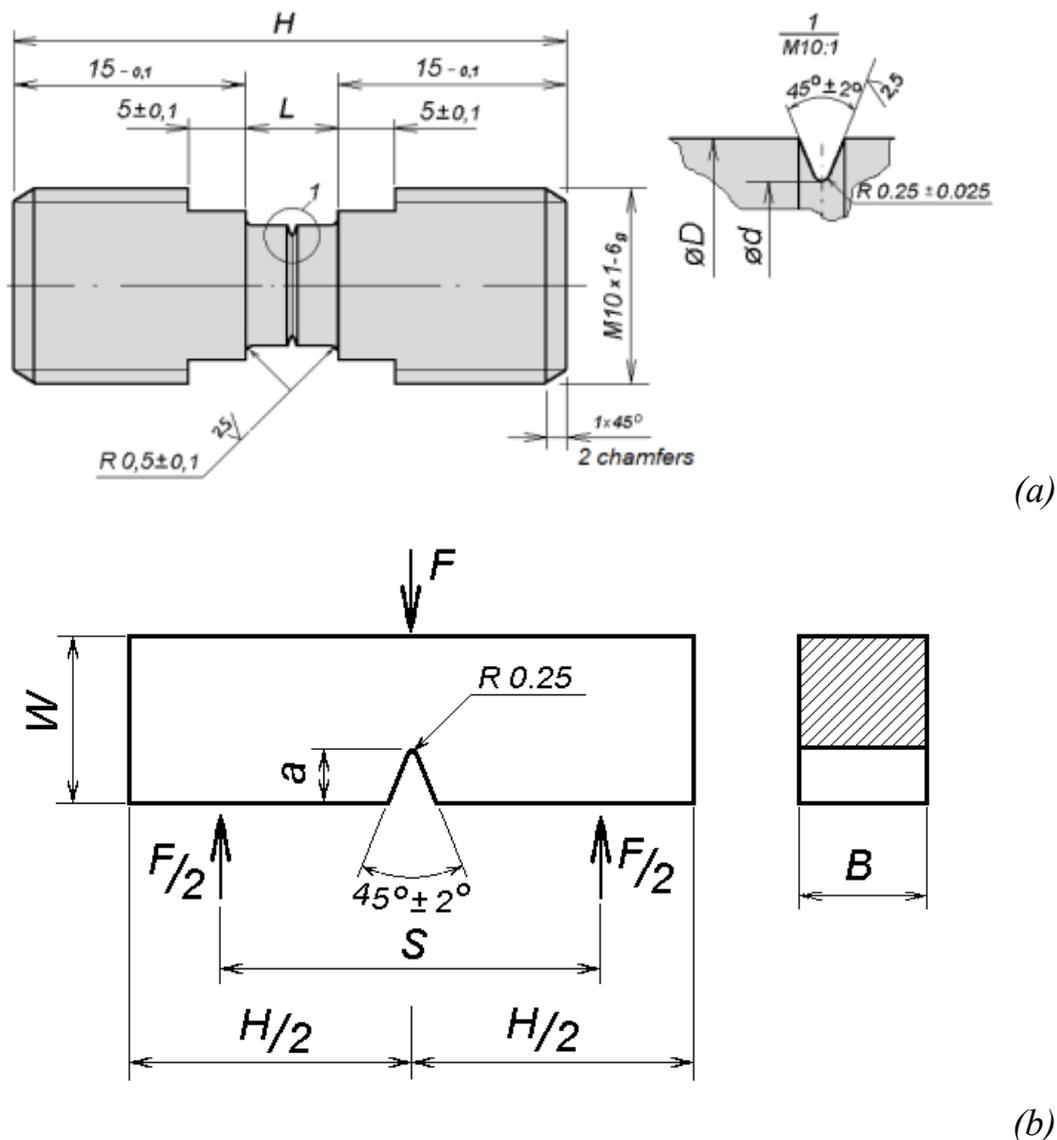

*(a)*

*(b)*

*Figure 41. Two types of specimens for the determination of fracture toughness in a Kolsky bar.*

When loading cylindrical specimens with a V-shaped notch in tension, the stress intensity factor $K_{IC}$ was given by Kogut [74] as:

$$K_{IC} = \frac{P_{max}}{D\sqrt{D}} \cdot f(\lambda) \ , \tag{88}$$

where $P_{max}$ is the maximum force (determined by the strain pulse in the output bar), $D$ is the diameter of the complete cross-section of the gauge section of the specimen, and the shape coefficient $f(\lambda)$ is given by:

$$f(\lambda) = \frac{0.7976\sqrt{1-\lambda}}{\lambda\sqrt{\lambda}\sqrt{1-0.8012\lambda}} \ , \tag{89}$$

where $\lambda = d/D$, $d$ being the diameter of the specimen at the notch.

Currently, standard Charpy tests for impact toughness are performed in our laboratory in accordance with the Russian standard GOST 9454-78 on beam-type specimens with U, V and T type stress concentrators (cracks). The brittleness of materials is better measured using specimens with a V-type stress concentrator as it has the sharpest notch (radius of 0.25 mm). So this type of test is used for the selection and acceptance tests of metals and alloys used in structures needing a very high degree of reliability (e.g. aircraft, vehicles, pipelines, pressure vessels).

When testing a beam with a V-neck, the stress intensity factor $K_{IC}$ is determined based on the experimental value of the maximum (critical) force $F_Q$, at which the specimen begins to break:

$$K_{IC} = \frac{F_Q \cdot S}{B \cdot W^{3/2}} \cdot f\left(a/W\right) \ , \tag{90}$$

where the shape function of the specimens $f(a/W)$, which depends on the length $a$ of the crack (incision) and the length $W$ of the specimen, is determined in accordance with linear fracture mechanics in the following manner [190]:

$$f\left(a/W\right) = 2.9 \cdot \left(\frac{a}{W}\right)^{1/2} - 4.6 \cdot \left(\frac{a}{W}\right)^{3/2} + 21.8 \cdot \left(\frac{a}{W}\right)^{5/2} - 37.6 \cdot \left(\frac{a}{W}\right)^{7/2} + 38.7 \cdot \left(\frac{a}{W}\right)^{9/2}$$
$$\tag{91}$$

In Charpy tests, the energy absorbed per unit area of the fracture section is taken as the characteristic energy of fracture, the so-called impact strength. However, it should be remembered that materials with the same maximum load and work of fracture can be characterized by different ratios between the work of initiation and growth of the crack. So the impact strength indicators obtained as a result of three-point bending tests are mainly used for comparative evaluation of the propensity of materials to brittle fracture under impact loading, and their direct use in calculations is difficult.

To perform dynamic experiments on three-point beam bending experiments, a system of three Hopkinson bars was used: one input and two output (see Table 1.). The ends of the bars in contact with the specimen are ground into 45° wedges to form an edge. Each edge is rounded with a radius of 2 mm.

During the tests, the incident $\varepsilon_i(t)$ (loading) and reflected $\varepsilon_r(t)$ pulses in the input bar are recorded, as well as the pulses transmitted through the specimen $\varepsilon_t(t)$ into the two output bars. Processing these pulses in accordance with the Kolsky analysis allows us to obtain the time dependences of:

(i) the displacement of the wedge (deflection of the specimen)

$$u(t) = C \int_0^t (\varepsilon_i + \varepsilon_r) \, dt \qquad ; \qquad (92)$$

(ii) the velocity of displacement of the wedge

$$v(t) = C(\varepsilon_i + \varepsilon_r) \ , \qquad (93)$$

as well as

(iii) changes in the force in the specimen during loading due to pulses in the input bar

$$P(t) = EA (\varepsilon_i - \varepsilon_r) \ , \qquad (94)$$

or on the basis of the last pulse

$$P(t) = EA \cdot 2\varepsilon_t \ . \qquad (95)$$

As it turned out in the course of tests when comparing the recorded pulses, a more reliable result in determining the fracture forces in the specimen is given by the second dependence (for twice the transmitted pulse).

### 1.2.10 Determining dynamic coefficients of friction

To determine dynamic coefficients of friction, we proposed a simple but effective modification to the Kolsky bar method in which a tube replaces the output bar [76]. In the experiments, the friction specimen consists of two parts: a core bushing surrounded by an enclosing hoop (or ring) under tension to produce a tight fit of the two parts (Figure 42).

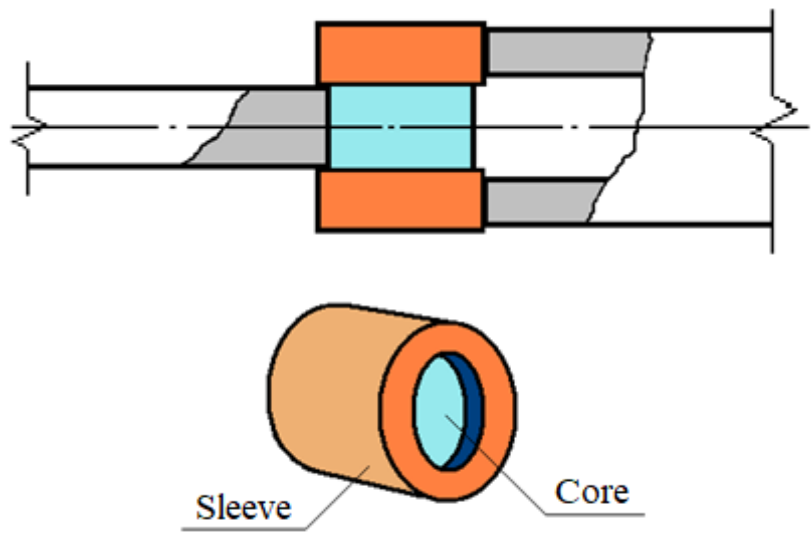

*Figure 42. Schematic diagram of specimen for studying friction using a Kolsky bar.*

The parts of the specimen are in contact with each other on their cylindrical surfaces. The specimen can be made either by mechanical pressing or by thermal shrink-fitting. It is preferable to use the thermal method since in this case the surface finish of the mating elements is not much changed. The thermal shrink-fit method creates a certain controlled tension due to the outer diameter of the core being slightly larger than the inner diameter of the hoop at ambient temperature before assembly. The tension produces a well-defined static normal pressure on the contact surface between the mating parts.

The contact between the mating parts occurs on a cylindrical surface with a diameter $d$ and a length $l$. The amount of tension is determined taking into account the difference in diameters $\delta$ of the hoop and core.

The pressure $p$ on the contact surface for the assembly shown in Figure 43 is related to the tension dependence of $\delta$ [191]:

$$p = \frac{\delta}{\left( \dfrac{c_1}{E_1} + \dfrac{c_2}{E_2} \right) \cdot d} \quad , \qquad\qquad (96)$$

where $c_1 = \dfrac{d^2 + d_1^2}{d^2 - d_1^2} - \nu_1$ , $c_2 = \dfrac{d_2^2 + d^2}{d_2^2 - d^2} + \nu_2$ ,

where $E_1$ and $E_2$ are the moduli of longitudinal elasticity and $\nu_1$ and $\nu_2$ are the values of the Poisson's ratios of the sleeve and core parts respectively. The inner diameter of the sleeve in experiments $d_1 = 0$. In this case, the formula for calculating $p$ is simplified.

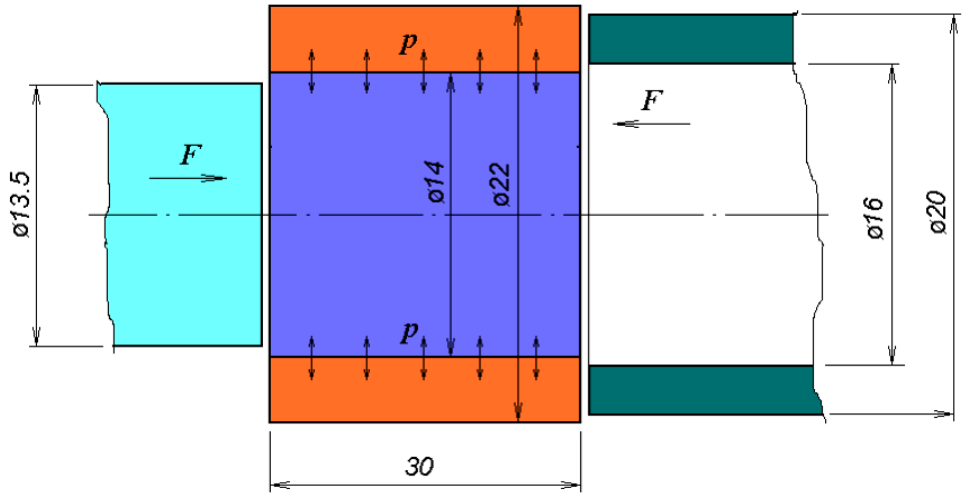

*Figure 43. Pressing-out implementation scheme.*

The axial force $F$ is determined when loading the assembly between the ends of the Hopkinson bars (similar to the pressing process) by the formula:

$$F = E A_b \varepsilon_t(t) \quad , \qquad\qquad (97)$$

where $E$ and $A_b$ are, respectively, the modulus of elasticity and the cross-sectional area of the output tube and $\varepsilon_t(t)$ is the transmitted strain pulse.

The coefficient of friction $\mu$ for an axial force $F$ and pressure $p$ on the contact surface is determined as follows:

$$\mu = \frac{F}{p \pi d l} \quad , \qquad\qquad (98)$$

where $d$ is the diameter of the interface surface and $l$ is its length.

To implement the 'pressing out' of the core in order to determine the coefficient of friction, a steel Hopkinson input bar with a diameter of 13.5 mm and a steel output tube with an internal diameter of 16 mm and a wall thickness of 2 mm were used (Figure 43).

For the correct determination of the friction force and, consequently, the coefficient of friction, it is very important to have an accurate knowledge of the tension in the sleeve as well as the pressure applied by the input bar. Also it is important to maintain a constant length of contact surface between the sleeve and the core during the loading process.

Tests of the method were performed on specimens where the core was made of titanium VT6 and the sleeve was made either of aluminium alloy AK4-1 or of titanium alloy VT6. Based on the dimensions of the available bar and tube, the contact surface diameter was chosen to be 14 mm. There are three types of press-fitting that differ in the amount of tension and, accordingly, the contact pressure:

| The type of press-fitting | $\dfrac{H7}{p6} = \dfrac{\varnothing 14^{+0.018}_{-0}}{\varnothing 14^{+0.029}_{+0.018}}$ | $\dfrac{H7}{r6} = \dfrac{\varnothing 14^{+0.018}_{-0}}{\varnothing 14^{+0.034}_{+0.023}}$ | $\dfrac{H7}{s6} = \dfrac{\varnothing 14^{+0.018}_{-0}}{\varnothing 14^{+0.039}_{+0.028}}$ |
|---|---|---|---|
| Realized tightness, mm | 0.014-0.019 | 0.018-0.022 | 0.024-0.030 |
| Contact pressure Ti-Al, kgf/mm$^2$ | 1.87-2.54 | 2.40-2.94 | 3.20-4.00 |
| Contact pressure Ti-Ti, kgf/mm$^2$ | 2.70-3.67 | 3.48-4.25 | 4.63-5.79 |

The values of elastic constants included in equations (96)-(98) for calculating the coefficient of friction were:

• for titanium $E_1 = 12,500$ kgf / mm$^2$, $\nu_1 = 0.34$,
• for aluminium alloy $E_2 = 7,200$ kgf/mm$^2$, $\nu_2 = 0.33$.

It should be noted that the initial diameters of the components must be carefully measured before assembly in order to accurately determine the amount of realized tension.

For each value of tightness, a series of experiments were conducted under closely similar conditions. based on the results of which the average curves of change in the extrusion force were constructed from the results (Figure 44). The average square deviations of the data were determined to have a relative error of 5%.

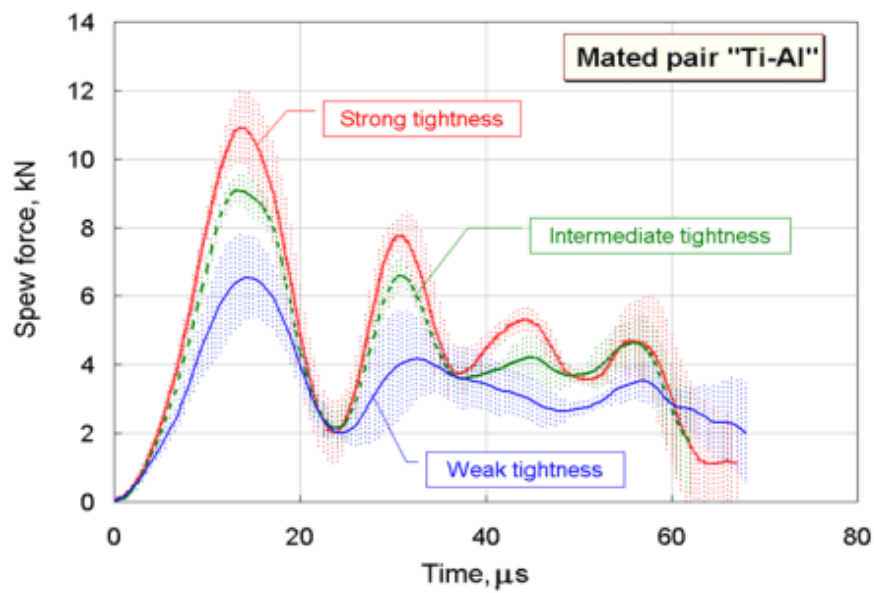

*(a)*

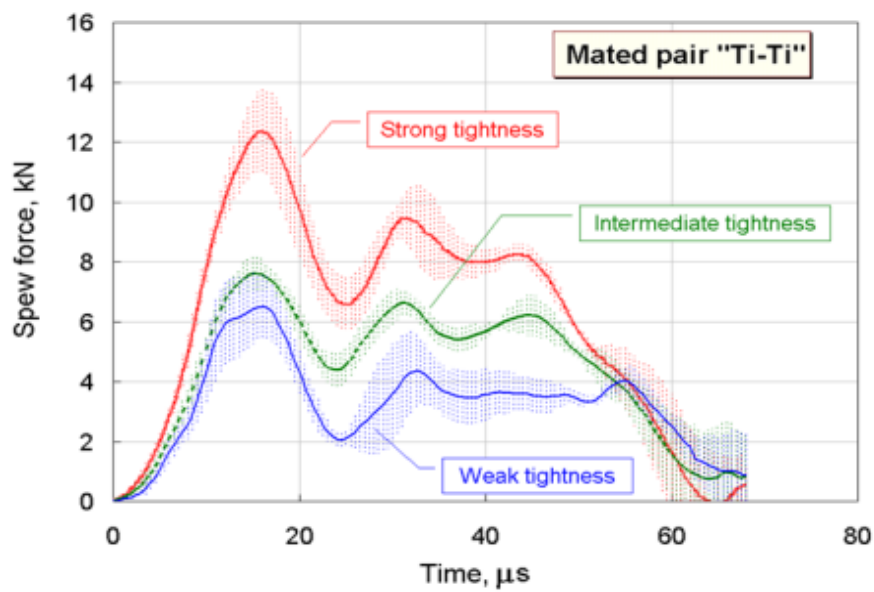

*(b)*

*Figure 44. Changing the core extrusion force from the 'Ti-Al' and Ti-Ti' assemblies.*

The 'oscillatory' nature of the extrusion forces with time should be noted. This fact may indicate a stick-slip motion of the cylindrical surfaces. It should be noted that even under low-velocity loading conditions, it is not possible to perform smooth sliding of the assembled specimen: the movement occurs in jumps. In the calculations of the static friction coefficient, the first maximum on the force profile ($F_1$) was taken; for the calculation of the sliding friction coefficient, the average force value ($F_2$) was selected for the time interval 30-60 microseconds.

Based on the results obtained, the average values of the coefficients of static and sliding friction were determined (Figure 45), which were 0.45 and 0.20 for titanium-aluminium pairs, and 0.28 and 0.22 for titanium-titanium pairs, respectively.

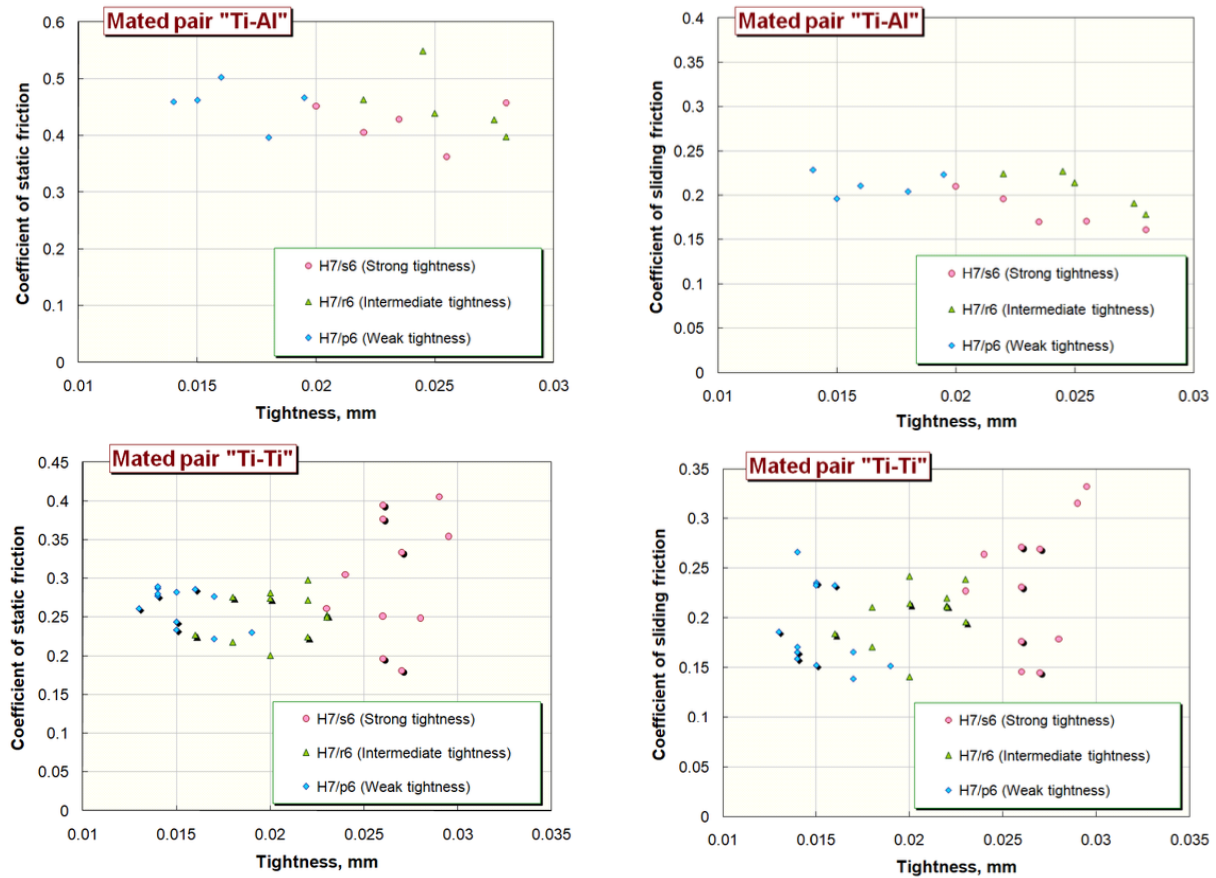

*Figure 45. Values of the coefficients of static and sliding friction for the pairs 'Ti-Al' and 'Ti-Ti'.*

To evaluate the main effects that occur during the test and can affect the result, a numerical analysis of the proposed experimental scheme was performed [192]. The simulation was carried out for a flat axisymmetric setting in two stages: in the first stage, the static stress and strain fields were determined in the specimen before press-fitting, and in the second stage, a dynamic problem was solved corresponding to the full-scale testing in the SHPB system.

The calculation scheme corresponding to the first part of the problem, implemented in ANSYS, is shown in Figure 46. To determine the stress-strain state of the specimen, its geometry was constructed with an initial negative gap $\delta$ corresponding to the real specimen's tension (the difference between the inner diameter of the sleeve and the outer core diameter). At the core-sleeve boundary, a symmetrical node-to-surface contact was set.

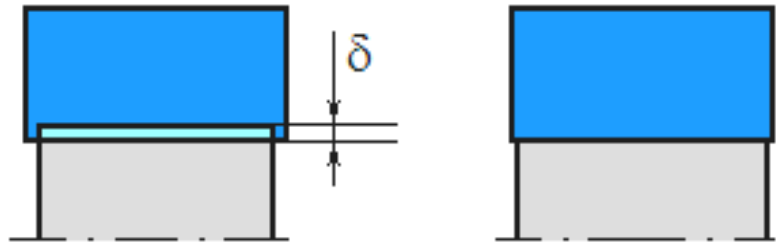

*Figure 46. Setting the task of determining the stress-strain state of a specimen before press-fitting starts.*

Figure 47 shows two sets of contact pairs for processing the core-to-core contact interaction. During the iterative procedure, in which the size of the gap $\delta$ was reduced to zero, the stress-strain state of the specimen was determined. Since the stress in the specimen is significantly lower than the yield strength of its constituent materials, the problem was solved elastically. The components of the stress tensor obtained at this stage for each model element, as well as the deformed configuration of the assembly, were saved to a file in LS-DYNA PP format for dynamic calculation for a prestressed specimen.

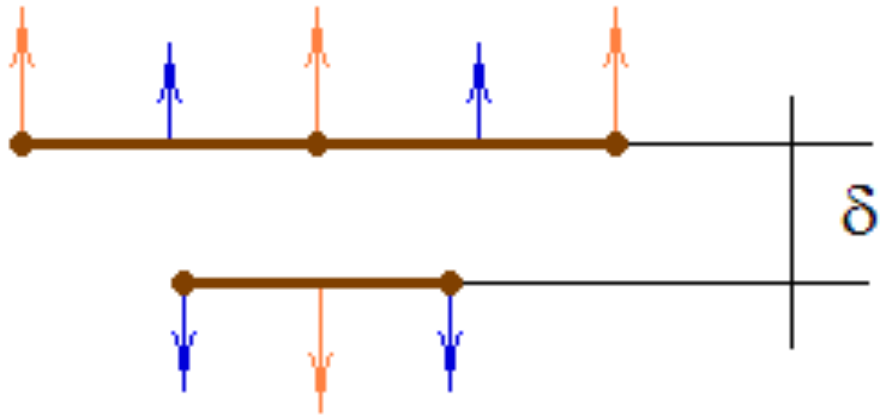

*Figure 47. Symmetrical contact 'node to surface' with initial implementation*

The calculation scheme of the second part of the problem is shown in Figure 48. At the end of the input bar, the load pulse was set, calculated in accordance with the strain pulse recorded in the full-scale test. A sensor on the output tube measured the pressing force.

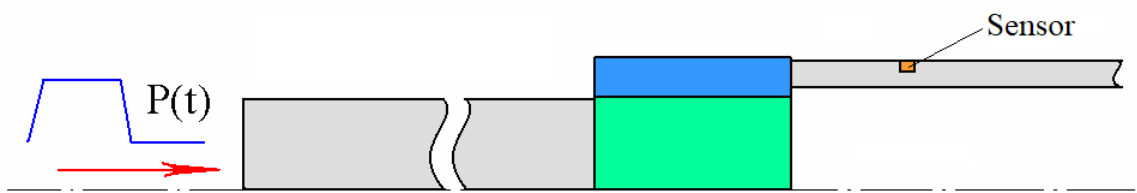

*Figure 48. Design scheme of specimen pressing in the SHPB system.*

To assess the influence of the specimen geometry on the pressure distribution over the contact surface of the core and the sleeve, three configurations were considered having different ratios of the lengths of the core and sleeve (Figure 49). From the point of view of preserving a constant value for the pressure on the contact surface during core pressing, the first scheme (A) having equal lengths is the least preferable, since it requires for the calculation of the coefficient of friction to take into account the change in the area of the contact surface when the core is pushed out of the sleeve.

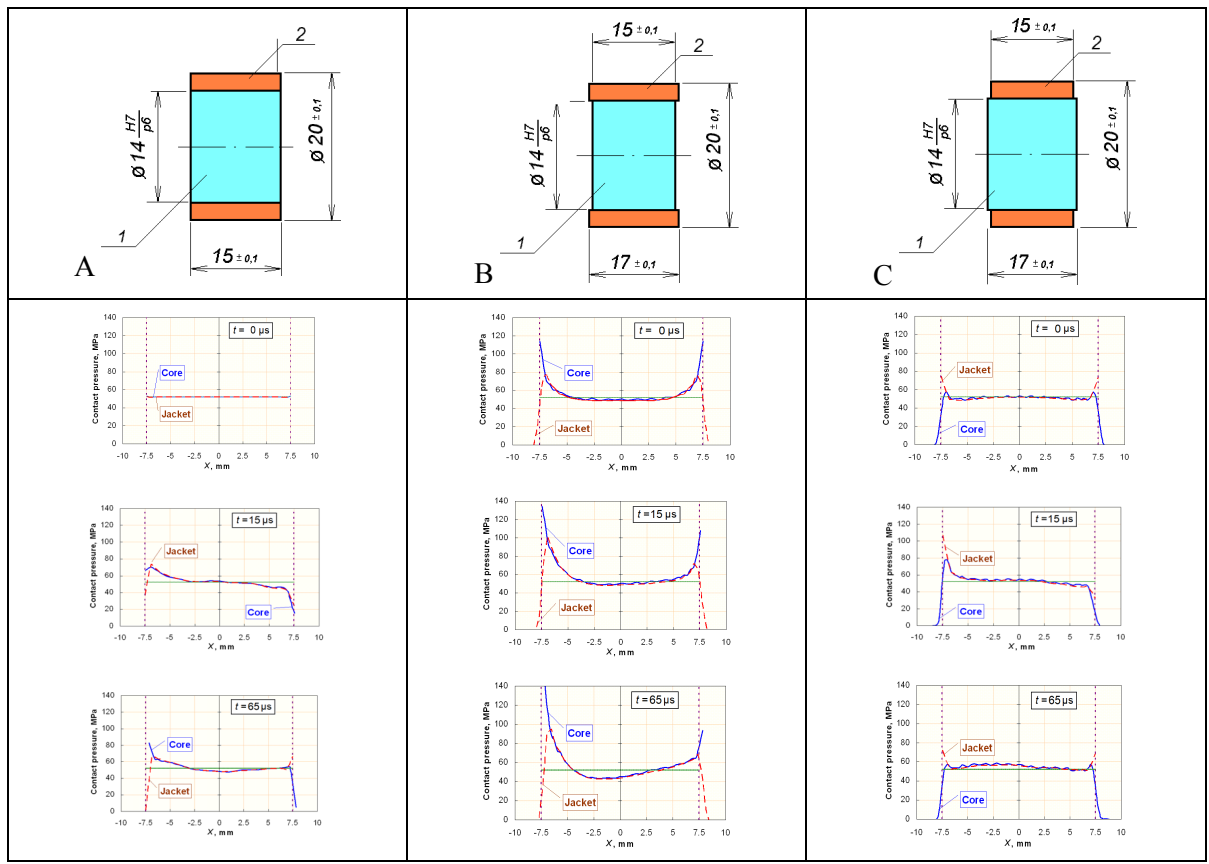

*Figure 49. Specimen configuration options and evolution of the static pressure during the pressing process.*

During dynamic loading of the assembly, radial stresses appear in the core due to the Poisson's ratio, which changes the value of the static load. Modelling of the pressing process allowed us to estimate the values of these additional radial stresses for the specified assembly configurations. Figure 49 shows the profiles of the distribution of the normal pressure value over the core surface and the sleeve corresponding to each configuration, obtained in calculations for VT6 titanium at various times. A solid blue line with markers shows the distribution of contact pressure along the length of the core, and a red dotted line with markers that for the inner surface of the sleeve. The perforated green line shows the analytical value of the normal pressure.

It can be clearly seen that configuration (B) gives the worst result in terms of ensuring uniformity of the normal pressure on the contact surface. Therefore, taking into account the above description of scheme (A), configuration (C) should be considered optimal for the study of the dynamic coefficient of friction. This configuration has maximum contact pressure uniformity on the contact surface. It also preserves the surface area during a test.

Figure 50 shows how the normal force of interaction between the core and the sleeve changes over time. It can be seen that the peak of the contact force occurs soon after the core is first strained. The force is almost constant when the inner part of the assembly is undergoing steady movement.

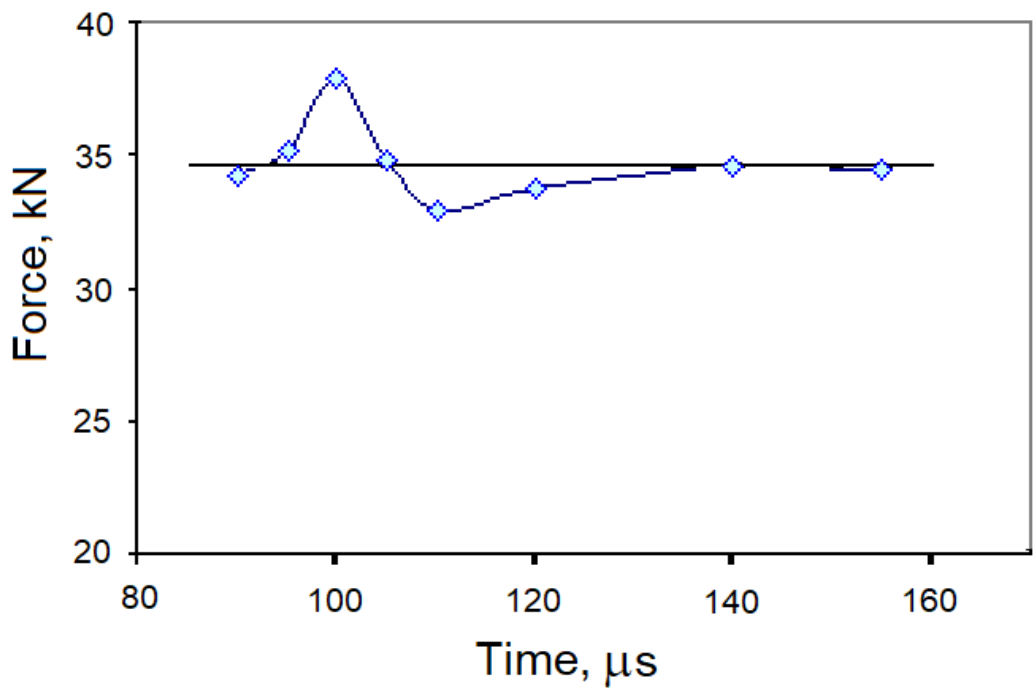

*Figure 50. Changing the contact force during the pressing process.*

Figure 51 shows a comparison of strain pulses in Hopkinson bars: those determined during the experiment (discrete points) and as a result of computer simulation of dynamic compression (solid lines). In the calculation between the core and the sleeve, a constant coefficient of friction was set, determined from the experimental data.

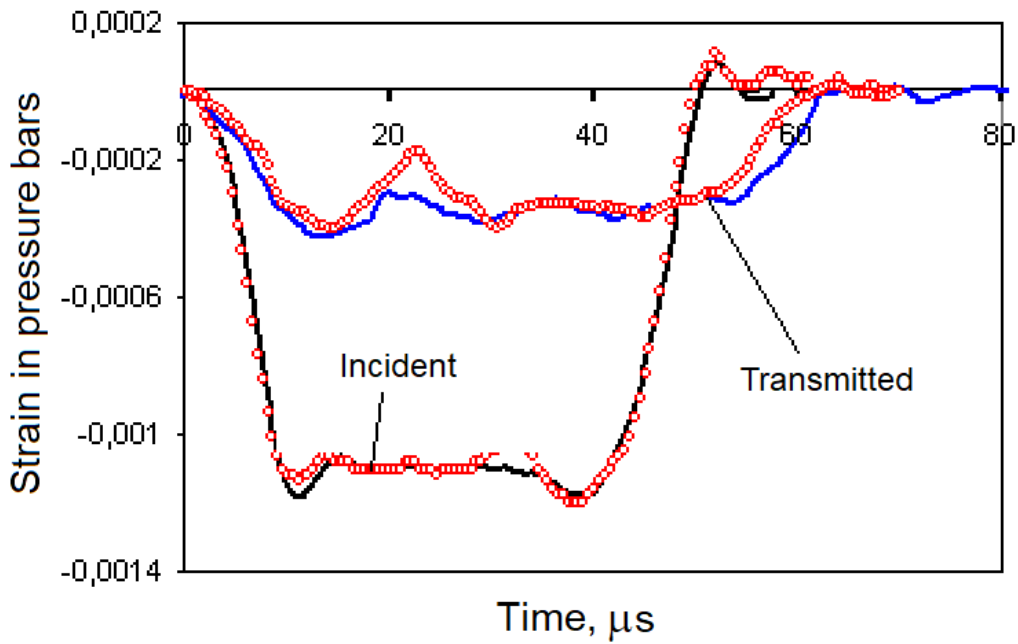

*Figure 51. Comparison of pulses in the Hopkinson bars in a sliding friction experiment.*

## 2. USING THE KOLSKY BAR METHOD FOR TESTING BRITTLE MATERIALS

Interest in the dynamic properties of materials has been growing for a long time, and intensive work in this area is being carried out all over the world. The dynamic properties of metals (which are ductile) have been studied much more extensively than those of brittle materials such as concretes, rocks, and ceramics. This is due to the additional requirements imposed on testing facilities for the study of this class of materials. Currently there are no generally accepted standards for the size of installations, and installations for the study of brittle and especially structurally heterogeneous materials are relatively rare both in Russia and in the rest of the world. A large number of different types of materials fall into the class of brittle and structurally inhomogeneous materials. They differ widely in their chemical and structural compositions. Examples include: ceramics, rocks, concretes, and various types of frozen materials held together by binders that are liquid at room temperature (frozen soils, ice, bitumen, etc.). All these types of materials are still insufficiently studied under dynamic loading, so data on their high rate mechanical properties are rarely found in the scientific literature.

Many scientists study the behaviour of brittle structural materials, such as a variety of concretes, rocks, ceramics, and refractory materials under dynamic loading. The results of these studies have been published in numerous articles by domestic and foreign authors. Yu.M. Bazhenov, A.M. Bragov, A.K. Lomunov, A.I. Sadyrin, Yu.V. Petrov, L.E. Malvern, S.A. Ross, P.H. Bischoff, S.H. Perry,

L.J. Malvar, J.E. Crawford, Q.M. Li, H. Meng, E. Cadoni, A. Brara, J.R. Klepaczko, M.J. Forrestal and many others have made significant contributions to the development of this field of science. However, according to the regularly held international conferences and symposiums on the high strain rate properties of materials, interest in this problem is not weakening. This is due to the fact that on the one hand the results of this research are incomplete and sometimes contradictory, and on the other hand due to the appearance of new materials based on cement (for example, high-strength fibre concrete) the problem is still not completely solved.

In this section, the dynamic mechanical behaviour of rocks under dynamic loads is considered. Dynamic loading is said to occur when increasing the loading rate causes a change in the mechanical properties and leads to fragmentation. The sources of dynamic loads include explosions, shock, and seismic events that produce periods of particle acceleration, velocity, and displacement in short time periods. Understanding the impact of dynamic loads on rock is important for solving underground mining problems, earthquake studies, penetration and explosion events, rock degradation processes, studies of large-amplitude stress waves, and protective structural design. Rock dynamics and geophysical research were the subject of the first book which systematically outlined the fundamental principles and experiments for the study of stress waves in rocks [193].

Rock dynamics has applications in earthquake studies, mining, energy, environmental and civil engineering where dynamic loads must be taken account of. Zhang related the typical problems of rock dynamics to the construction and use of underground caves and caverns [194]. The factors which must be considered include: environmental (for example, limiting pressure, temperature, presence of ground water) as well as intrinsic rock properties (for example, the solidity of the structure, anisotropy of properties, composition and grain size). To date, guidelines and standards in dynamic testing and design are virtually non-existent, and deeper advances in the understanding of dynamic behaviour have largely been achieved due to progress in experimental methods. In recent decades, significant research has been carried out into the development of experimental methods for the characterization of the dynamic mechanical behaviour of materials. The experiments that are of main interest in Zhang's review [194] are those that are designed to develop reliable test methods and critically study the mechanical behaviour of rock materials under laboratory conditions.

There have been a few surveys of the behaviour of brittle materials [195-203]. Xia & Yao surveyed various methods for measuring the dynamic properties of rocks using SHPBs [203]. These include measurements of the dynamic strength of rocks in compression, tension, bending and shear, dynamic fracture (i.e.

dynamic simulation and crack expansion viscosity, dynamic fracture energy and fracture rate) and methods for studying the influence of temperature and pore water.

The fragmentation process depends on the structure and properties of the material such as heterogeneity, creep, deformation characteristics, structural defects, etc. The structure of concrete has, of course, a very large influence on the dependence of its properties on loading rate. First of all, the choice of materials used to make a concrete will be of great importance. The more plastic the cement binder, aggregate and contact interface, the more strongly the secondary stress field and the cracking process will change at different loading rates, and the more noticeable will be the difference in strength between high and low loading rates. The greater the plasticity and creep of the components, the greater the ability of concrete to redistribute stresses and localized sharp concentrations in places of serious accidental structural defects, even at high loading rates, and the stronger the manifestations of the 'lag' deformation process. Consequently, as the plasticity of the components increases, the strength of the concrete increases under dynamic loading.

Numerous experimental data show that nonlinear deformation is characteristic of concrete, since, starting at low stresses, inelastic residual strains develop in it in addition to elastic strains, [204, 205]. The behaviour of different concretes under dynamic impact differ: some collapse as the load increases and others when it decreases; some can withstand a dynamic load that exceeds the quasistatic strength without collapsing.

To study the behaviour of structurally inhomogeneous materials under dynamic compression, the traditional Kolsky method is usually used, but with the use of large-diameter input and output bars [119, 120, 206-209].

In the 1990s, Bischoff & Perry [195] and Fu *et al.* [196] considered experimental methods for investigating the behaviour of concrete under dynamic compressive load, Malvar & Ross [197] presented an overview summarizing the effect of the strain rate on concrete during tensile loading, and Zhao *et al.* [198] considered the achievements that had been made in the study of rock dynamics associated with the development of cavities.

There are several comprehensive reviews of experimental dynamic methods [210-212]. In addition, there are reviews of rock dynamics and applications [194, 213, 214], as well as dynamic experimental methods and results obtained using them [194, 214-216].

The International Society for Rock Mechanics (ISRM) has published recommendations on a large number of low rate methods for testing rocks [217-220]. As for the study of dynamic properties, an ISRM Commission currently proposes the use of only three methods for the dynamic testing of rocks [221] using modifications of the SHPB:

- dynamic compression;
- dynamic Brazilian test;
- dynamic bending of a semi-circular specimen containing an incision (NSCB – notched semicircular bending).

Other well-known methodological developments are still considered possible candidates for future ISRM proposed methods.

Several main types of tests have been developed using a modified SHPB system to study the dynamic properties of rocks [222]:
compression tests for determining the compressive strength under uniaxial stress;
methods for determining the strength under uniaxial strain and volumetric compression under static limiting pressure;
Brazilian disc methods for determining the ultimate tensile strength;
methods for studying the bending strength properties of rocks and to determine the mode I crack resistance and the fracture energy.

The main types of quasistatic and dynamic tests developed at the moment are summarized in Table 2 [194].

**Table 2. Testing methods for brittle materials**

| Type of loading | Test method | Quasistatic property | Dynamic property |
|---|---|---|---|
| Tension | DT – Direct tension | $\sigma_t$ [223] | $\sigma_{td}$ [224, 225] |
| Compression | UC – Uniaxial compression | $\sigma_{uc}$ [218] | $\sigma_{td}$ [221] |
| | TC – Triaxial compression | $\sigma_{tc}$ [219] | $\sigma_{tcd}$ [226] |
| | BD – Brazilian disc- | $\sigma_t$ [217] | $\sigma_{td}$ [221, 227] |
| | FBD – Flattened BD – | $\sigma_t$ [228] | $\sigma_{td}$ [229] |
| | Shear | $\tau$ [230, 231] | $\tau$ [232, 233] |
| | HCFBD – Holed cracked FBD | $K_{IC}$ [234] | $K_{Id}$ [236] |
| | CSTFBD – Cracked straight through FBD | $K_{IC}$, $K_{IIC}$ [235] | $K_{Id}$, $K_{IId}$ [237, 238] |
| | SR – Short rod | $K_{IC}$ [239] | $K_{Id}$ [240, 241] |
| | WLCT – Wedge loaded compact tension | $K_{IC}$ [242] | $K_{Id}$ [242] |
| | HCBD – Hole-cracked BD | $K_{IC}$ [234] | $K_{Id}$ [243] |
| | CCNBD – Cracked chevron notched BD | $K_{IC}$ [244] | $K_{Id}$ [245] |
| Bending | TPB – Three-point bending | $\sigma_t$ [246] | $\sigma_{td}$ [227] |
| | SCB – Semi-circular bending | $\sigma_t$ [247] | $\sigma_{td}$ [248] |
| | SENB – Single edge notched bending | $K_{IC}$ [249] | $K_{Id}$ [198, 250] |
| | CCNSCB – Cracked chevron NSCB | $K_{IC}$ [251] | $K_{Id}$ [252] |
| | NSCB – Notched SCB | $K_{IC}$ [220, 253] | $K_{Id}$ [254] |
| | LGG – Laser gap gauge | | $K_{ID}$ [254] |
| | DIC – Digital image correlation | | $K_{Id}$ [255] |
| | CPG – Crack propagation gauge | | $K_{ID}$ [256] |

## 2.1 In compression

The most popular method for the dynamic testing of brittle materials, including rocks, is the SHPB. The SHPB was originally developed to study the dynamic behaviour of ductile metals, polymers and explosives [6, 18, 37, 40, 64]. When the specimen is a brittle / quasi-brittle material (such as rock), the SHPB test conditions may be unsatisfactory in order to obtain reliable experimental data. Therefore, a number of requirements must be carefully met. Experimental methods for studying the dynamic behaviour of rocks should provide ways to generate a stable reproducible dynamic load, and accurate and reliable methods should be used to record the dynamic parameters of the resistance to deformation of brittle materials.

The principles of dynamic loading techniques and their application to the testing of brittle structural materials, such as concrete, mortar, and ceramics, have been discussed by a number of authors [201, 211, 212, 257, 258]. The results of studies of the behaviour of compressing brittle materials in the strain rate range $10^1$-$10^3$ s$^{-1}$ using the traditional SHPB technique (as well as its modifications) are described in numerous works e.g. [130, 259-268] [269-280].

To assess the influence of the specimen geometry on measurements of the strength properties of rocks, Dai *et al.* performed tests in which cylindrical rock specimens with different *L/D* ratios (0.5, 1.0, 1.5, and 2.0) [281]. Dynamic stress equilibrium was achieved in all tests. Since there was no stress gradient in the specimen, the axial inertia effect was negligible. In addition, to minimize errors in the measured strength due to the presence of friction on the faces of the measuring rod/specimen, the ends of all specimens were thoroughly lubricated with vacuum grease. The values of the measured compressive strength of granite depending on the loading rate are shown in Figure 52.

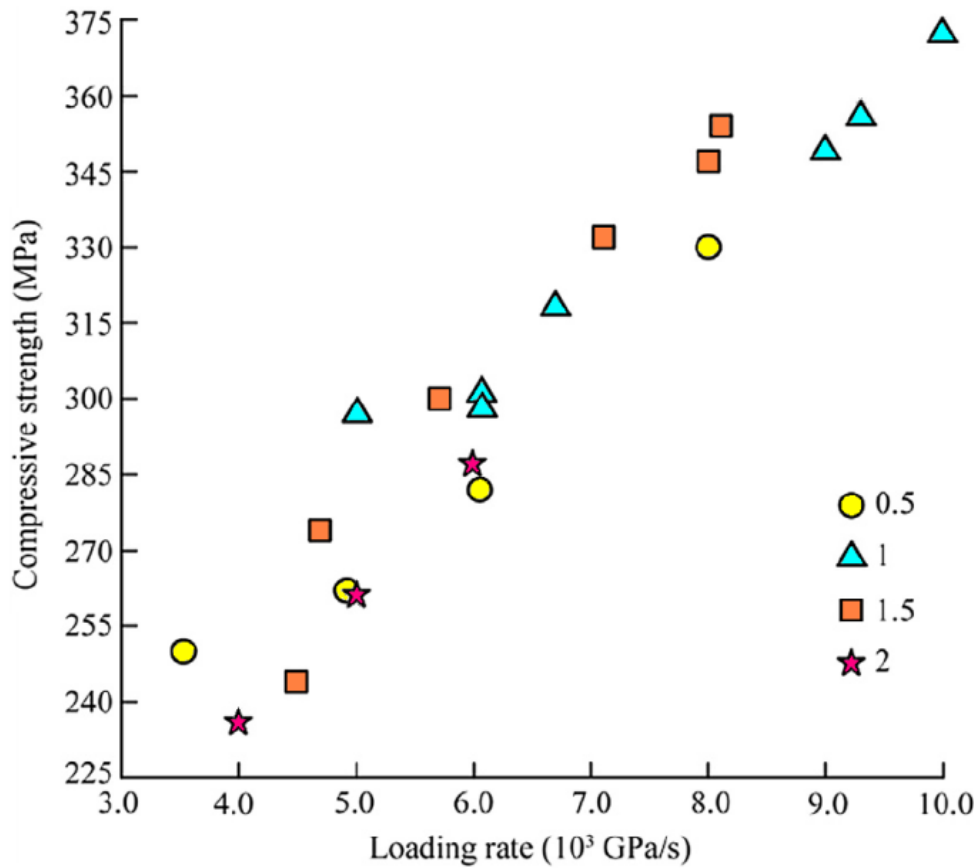

*Figure 52. Dynamic compressive strength values measured on rock specimens with various L/D ratios of 0.5, 1, 1.5, and 2.*

There are no significant differences in the fracture stress for specimens with the selected aspect ratios. Thus for dynamic compression tests of rock specimens with fully lubricated bar/specimen boundaries and, consequently, with reduced axial inertia effects, the aspect ratio has little effect on the test results in the range 0.5-2. However, specimens with $L/D < 0.5$ are very difficult to make, because it is difficult to hold such cylinders during the cutting and polishing processes. In addition, shorter specimens may be partially damaged during processing. On the other hand, for long specimens (such as $L/D = 2$), a very large loading force is required to achieve high loading rates, so data with a strain rate higher than $6 \times 10^3$ s$^{-1}$ for $L/D = 2$ was not obtained, because the Kolsky bars would have been plastically deformed or bent. Dai *et al.* suggested that specimens with an $L/D = 1$ is a reasonable choice for the aspect ratio of rock specimens during compression tests using the Kolsky method [281].

## 2.2 In tension

In dynamic tension, the following modifications of the Kolsky method are mainly used to study the behaviour of brittle materials: direct tension (DT), indirect tension (In-DT), and spalling. For 'direct' tensile loading of brittle materials, the specimen is located in the SHPB and glued to the ends of the input and output bars. An incident tensile wave is then excited in the input bar. The article by Reinhardt *et al.* describes such a device [207], the loading elements of which were a hydraulic jack (with a blocking system) and a drop-weight so as to create a tensile wave. The strain rate they achieved lay in the range 0.05-25 s$^{-1}$. A number of other authors, when implementing a similar loading scheme, excited the incident tensile wave in the input bar after the release of a prestressed high-strength steel rod connected to the input bar (Figure 53) [282-285], while Caverzan *et al.* achieved a strain rate of 150 - 300 s$^{-1}$ when specimens of high-strength reinforced composite were broken [283].

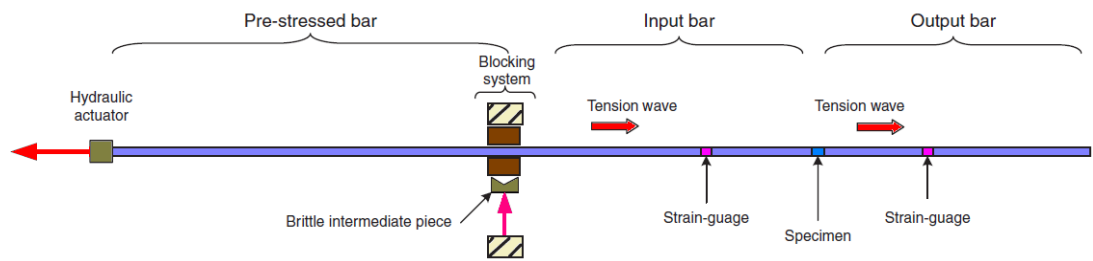

*Figure 53. The test setup for direct tensile SHPB.*

Zhang & Zhao considered variants of dynamic testing of rocks under DT where the specimen is glued to the ends of the input and output bars [194]. The tensile load is generated in the SHPB in various ways (Figure 54). Their ideas for dynamically creating a tensile load were inspired by an article in the ASM Handbook of Mechanical Testing and Evaluation [210] (Figure 54). In the first method (Figure 54*a*), a mass impacts an anvil screwed onto the end of the input bar. In the second method (Figure 54*b*) the anvil is loaded by a compressive wave passing through an additional loading tube. The compressive pulse in the loading tube is generated using a tubular striker. In the third method (Figure 54*c*) the pulse is generated by the detonation of an explosive substance on the surface of the anvil. In the schemes shown in Figures 54*a* & *b*, it is difficult to create a pulse of constant amplitude, whereas in Figure 51*b* the pulse duration is set in the same way as in the compression tests.

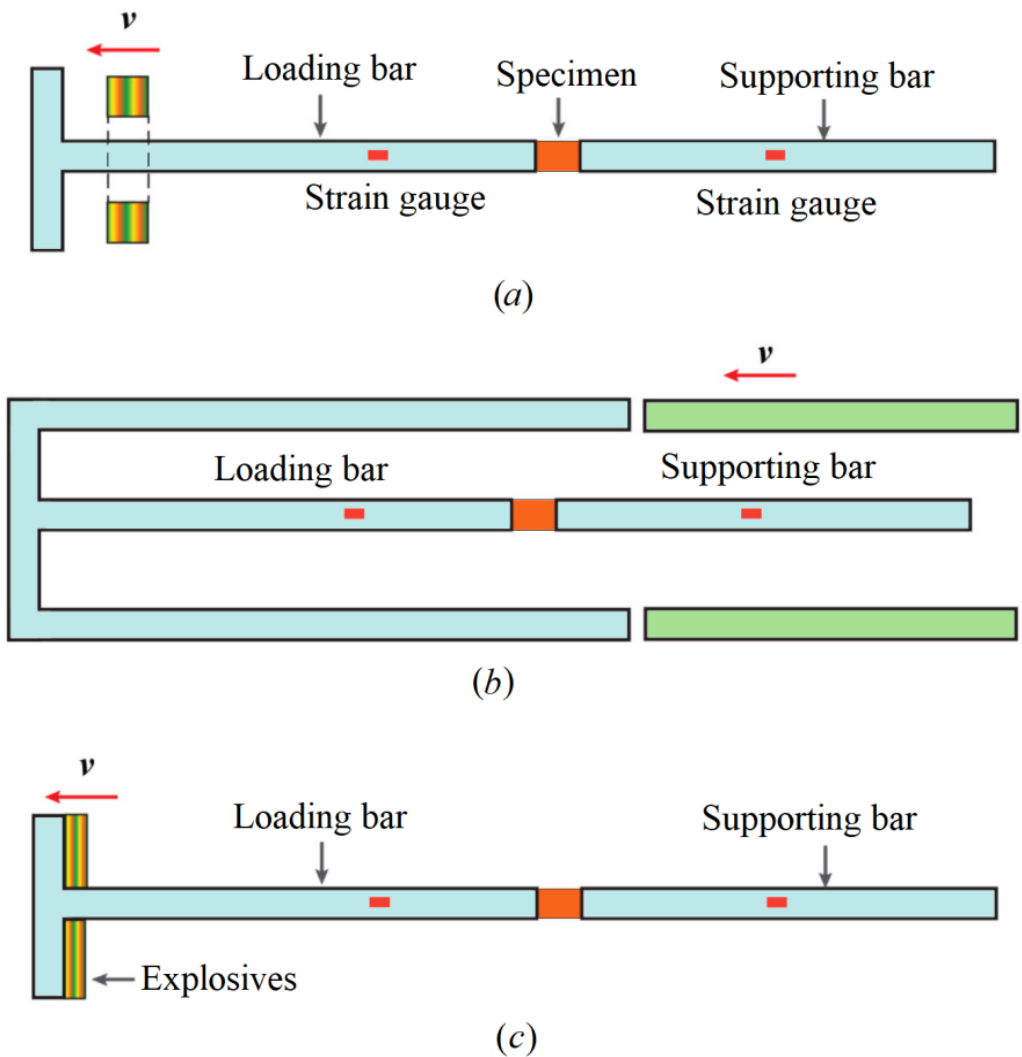

*Figure 54. Schemes of tests for direct dynamic tensile loading of rocks.*

However, compared to the dynamic tensile testing of metals, when a specimen is connected to the Kolsky bars using a threaded connection, the methods used to test brittle materials, such as concrete or rock, are more complex. Rock specimens are used either as parallelepipeds or tablets with flat ends that are glued to the ends of the bars, or as dog-bones or dumbbells that are glued into corresponding sockets in the ends of both bars using high-strength epoxy resin.

The limitations of direct tensile tests have been discussed by Zhang & Zhao [194]: (i) the same limitations as for quasistatic tests; (ii) the complexity of the specimen shape (for example, dog-bones and dumbbells) on the one hand, and the presence of a layer of epoxy resin between the specimen and the bars on the other hand, complicate the experimental set-up leading to a high cost of preparing the experiment and manufacturing the specimens, and may well create undesirable stress concentrations leading to premature specimen failure; (iii) pulse generation methods are difficult to apply, so the stress equilibrium condition may be violated.

Cadoni & Albertini described installations for dynamically testing rocks in DT using a statically pre-loaded input bar [269]. A sketch of the original design is presented in Figure 55. To excite a loading pulse with a smoothly increasing amplitude, the authors use pulse generators based on various mechanical principles.

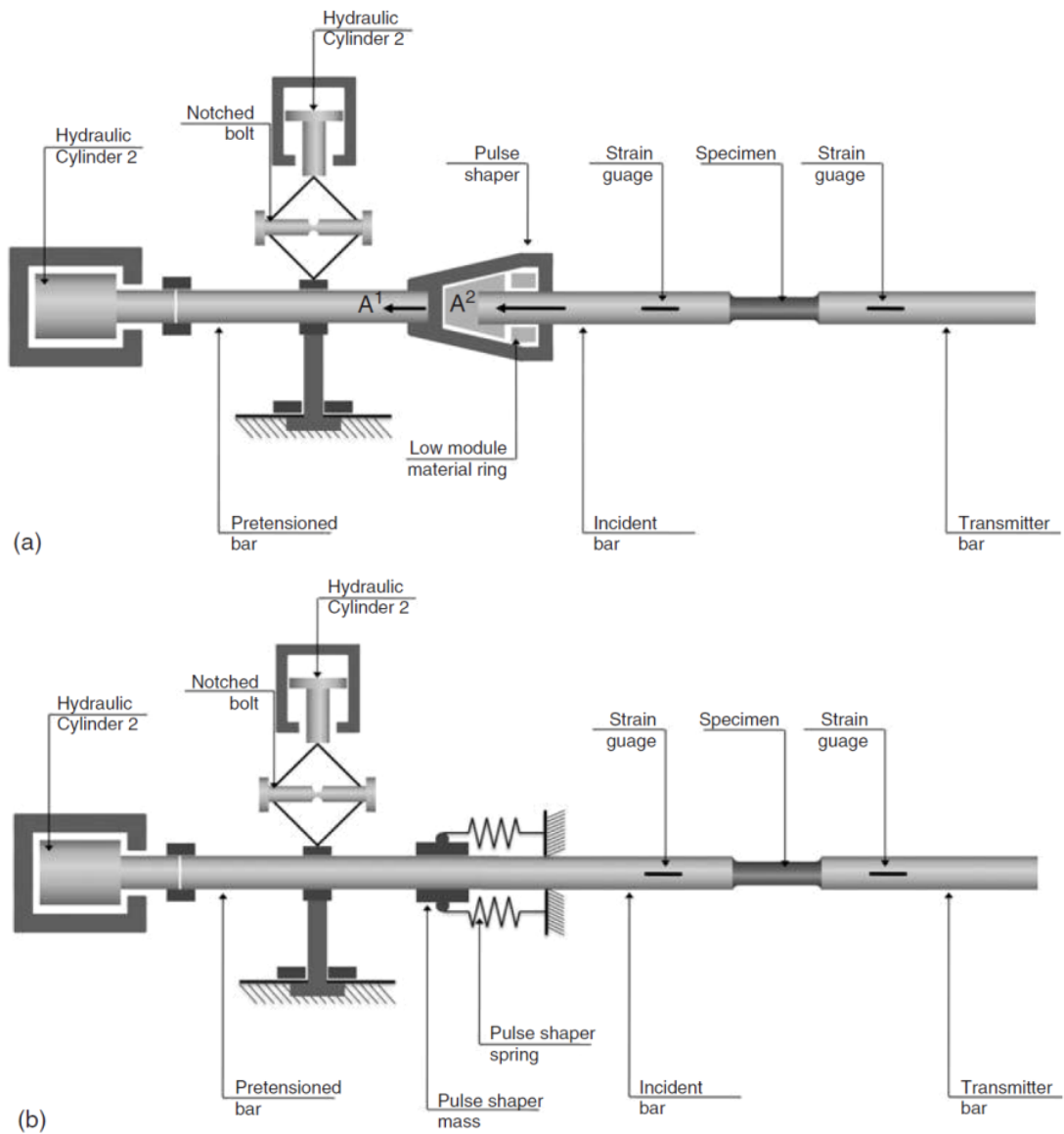

*Figure 55. Modified SHPB schemes for direct tensile testing: (a) with a low-modulus pulse generator; (b) with a spring-mass pulse generator.*

The diameters of the Kolsky bars used are determined by the size of the rock specimen, which should be at least 10 times larger than the size of the average rock grain [221, 245]. To ensure the accuracy of measurement results using simple single-wave analysis, it is necessary to guarantee the actual test conditions using various experimental methods, such as ensuring the balance of forces on the specimen ends, using a specimen geometry that minimizes the radial and axial inertias, and lubricating the ends to minimize the effects of friction.

As for experimental methods of dynamic tensile testing of brittle materials, which are based on the SHPB, it is difficult to find a compromise between the diameter of the Kolsky bars and the specimen size. Both dimensions are important because the phenomenon of wave dispersion in the bars and in the specimen limits the maximum strain rate, which is defined as the ratio of the difference between the velocities of the end surfaces of the specimen to the specimen length.

## 2.3 In indirect tension

To study the properties of brittle materials under indirect tension (In-DT), the following modifications are used: (i) the Brazilian test in which a fully circular disc (or one that has flats at the loading points) is loaded across its diameter and (ii) the bending of a semi-circular specimen (Figure 56).

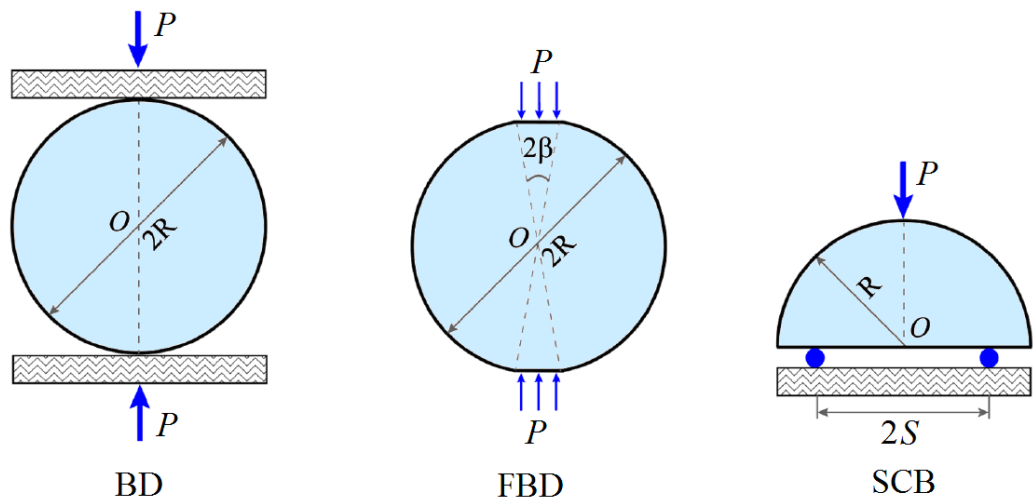

BD          FBD          SCB

*Figure 56. Test schemes for determining the mechanical properties of brittle materials under indirect tension (see Table 2 for designations).*

The 'Brazilian test' can be used to determine the tensile strength of brittle materials when the behaviour of the material is elastic and the specimen is in a state of mechanical equilibrium. In this test, the specimen fails along a diametric plane. This type of test was originally implemented to determine the quasistatic tensile strength of concrete [126, 127]. For this type of test, there is an ASTM standard (ASTM C 496-71) [128]. It is assumed that in this case, the tensile stress is constant for 80% of the specimen diameter. However, the remaining 20% of the diameter at the outer edges are subject to compressive stresses that cause small triangular zones of shear failure in the specimen (Figure 57) [286].

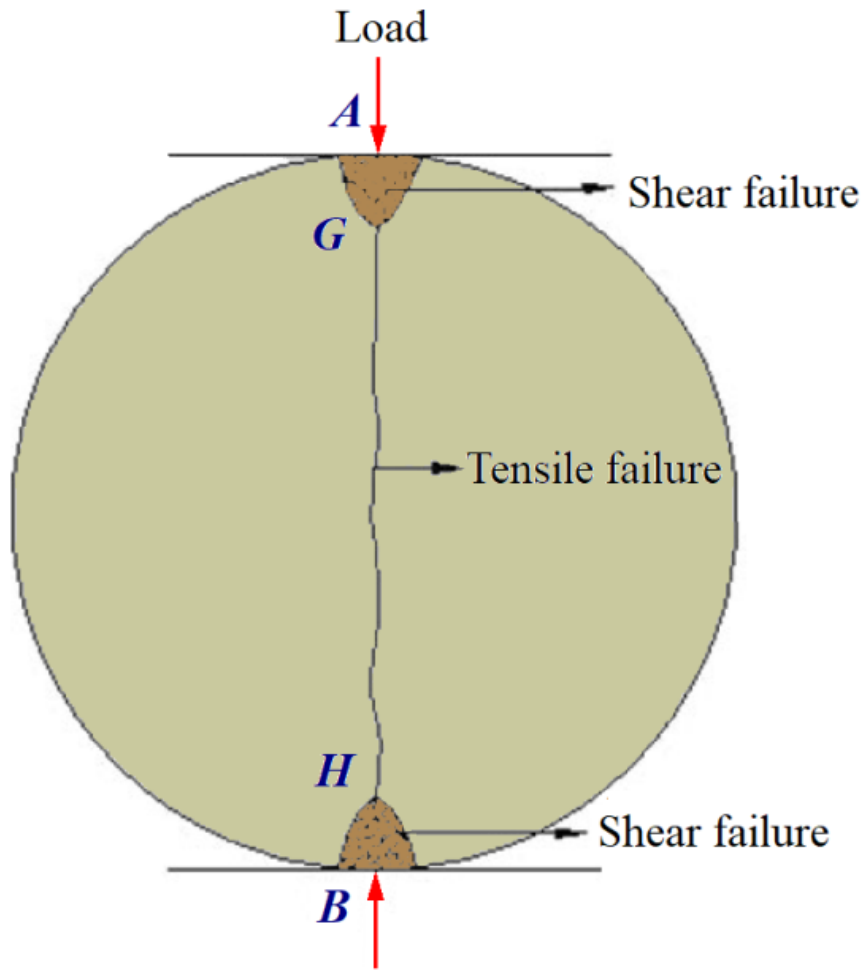

*Figure 57. Tensile and shear fracture zones in a solid circular disc in the Brazilian test.*

The region where the maximum tensile strain occurs is usually described as corresponding to the transition from shear to tensile fracture (Figure 57), as observed for many types of rocks in the Brazilian test [287]. In Figure 57, the central line of the crack GH must be associated with tensile failure. However, it is difficult to judge how the crack initiates during an experiment and how the crack subsequently develops: from the point G (or H) to the centre or from the centre to the point G (or H).

In the Brazilian test, it is assumed that a specimen in the form of a thin disc is loaded with a uniform pressure $P$ that is applied radially to a short zone of the circumference at each end of a diameter. Due to the small contact zone between the loading planes and the specimen, frictional stresses are neglected. Fracture of the Brazilian specimen occurs in accordance with Griffith's criterion.

The four typical loading patterns of the Brazilian disc are shown in Figure 58.

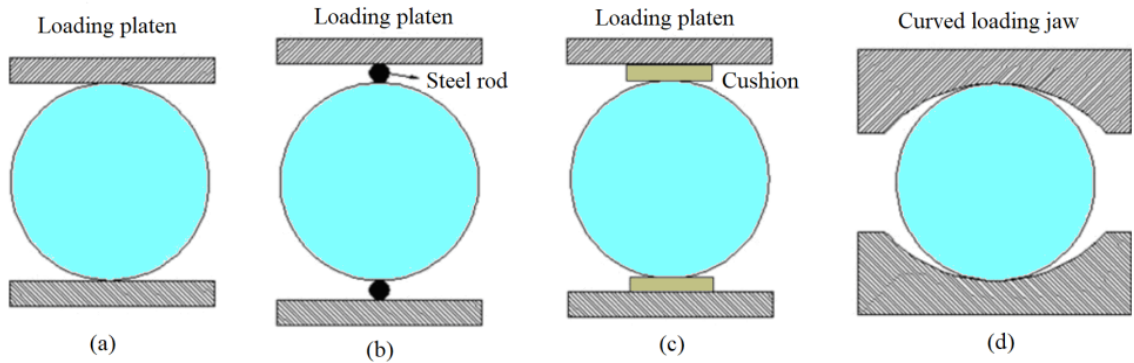

*Figure 58. Test schemes for determining the properties of materials under indirect tension using the Brazilian disc method.*

Wang *et al.* proposed a variation of the Brazilian test in which flats are machined on the sides of the disc at both ends of a diameter so as to create the so-called flattened Brazilian disc (FBD) (Figure 59) [228]. They developed this test for determining the elastic modulus $E$, tensile strength $\sigma_t$, and fracture toughness $K_{IC}$ for brittle materials in one experiment. According to the results of the stress analysis and Griffith's strength criterion, in order to ensure the crack initiates in the centre of the specimen (crucial for the reliability of the tests), the loading angle, $\alpha$, corresponding to the width of the flats must be greater than a critical value, namely 10°.

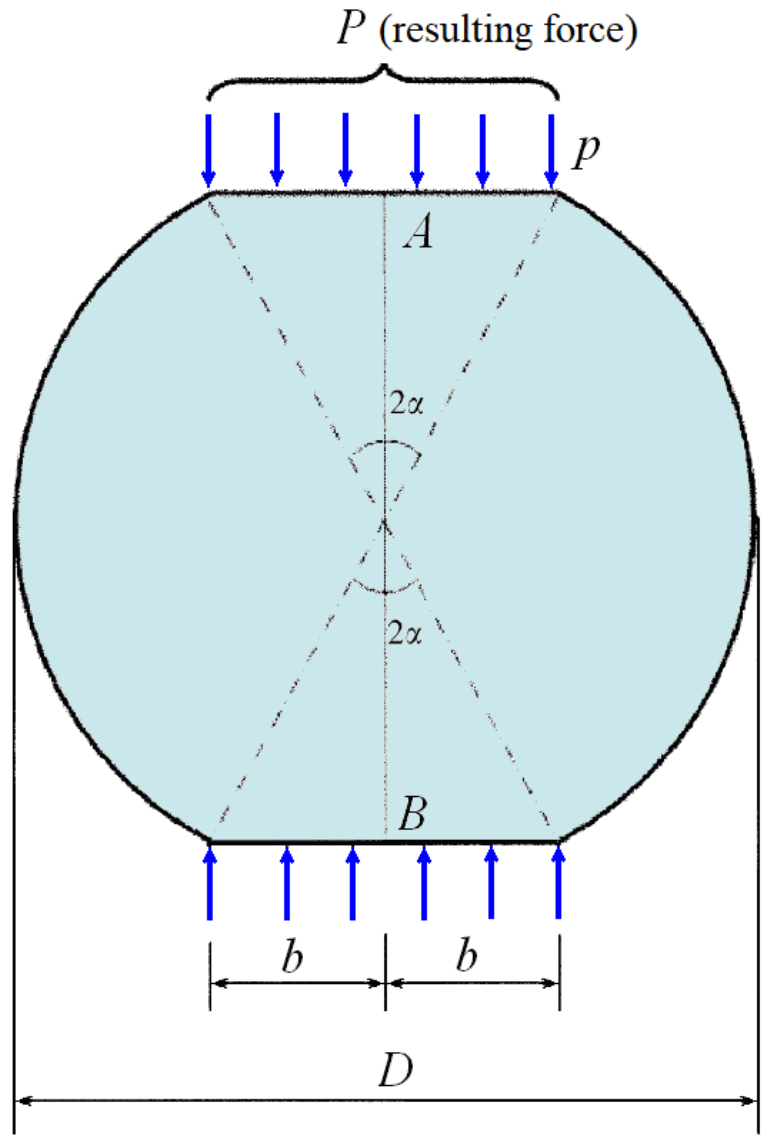

*Figure 59. Test scheme for determining the properties of materials under indirect tension using the FBD method.*

Analysis shows that, based on the recorded full load-displacement curve of the specimen (the curve must include the 'fluctuation' section after the maximum load), it is possible to determine $E$ from the slope of the curve before the maximum load, $\sigma_t$ from the maximum load, and $K_{IC}$ from the local minimum load immediately after the maximum load. The corresponding formulas for calculating $E$, $\sigma_t$, and $K_{IC}$ are obtained, and the key coefficients in these formulas are calibrated by the finite element method. In addition, some approximate closed-form formulas based on elasticity theory are given and their adequacy is shown by comparing them with the results of the finite element method calculation.

## 2.4 In spall

The method of spalling brittle rods allows the evaluation of the tensile strength of the rod material. The test is performed as follows. A compression wave is excited at one end of a long rod specimen. When it reaches the far end, it is reflected back as a tensile wave and, due to the lower tensile strength of brittle materials compared to compression, the tensile wave breaks the rod specimen in one or more places [288]. When using this method, the phenomenon of dispersion in brittle media must be taken into account.

The loading compression wave can be excited by direct impact on one end of the specimen [289] or transmitted to the specimen through a pressure bar (Figure 60) [290-293] [294-298]. The last test is a modification of the Kolsky method, in which the specimen is loaded through a pressure bar by a short compression pulse. The specimen is a cylinder which is attached to one end of the pressure bar. The other end of the specimen is a free surface so that a tensile wave is formed when the compression wave reflects from the far end. The compression wave and the reflected tensile wave interfere. As a result, tensile stresses occur at certain locations along the specimen. If at some point the tensile stress reaches a value equal to the tensile strength of the specimen material, part of the specimen breaks off. The wave then propagates further along the specimen and bar to be recorded by gauges glued both to the specimen and the bar [299-303].

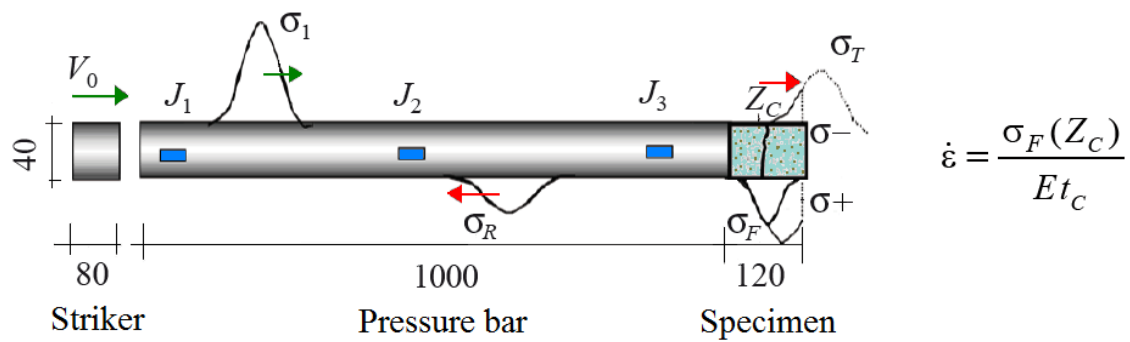

*Figure 60. Schematic diagram of Hopkinson spall test. $J_1$, $J_2$, $J_3$ are strain gauge locations, $Z_C$ is the distance from the spall plane to the free end of the specimen. In the equation for strain rate, $\sigma_F$ is the maximum tensile stress, E is the elastic modulus, and $t_C$ is the time to failure.*

An important limitation of experimental methods for the dynamic tensile loading of brittle or quasi-brittle materials (which are also heterogeneous) is the mechanics of wave propagation and wave dispersion. As for measurement problems, the direct measurement of small strains on the surface of a concrete specimen is not accurate enough. Also the use of resistive strain gauges and displacement transducers to measure the average deflection along the length of a specimen at the moment of spall is has difficulties. For example, linear variable displacement transducers (LVDT) are subject to the effects of inertia and also problems associated with the bandwidth of signal sensors.

Brara & Klepaczko published the results of experiments that they performed on the spall of concrete specimens under water in both saturated and dry states at strain rates in the range 10-120 s$^{-1}$ in a series of papers [290, 291, 293, 294, 296, 298]. They recorded the strain-time history of the specimen using high-speed cameras as well as three gauges bonded to a Hopkinson bar (Figure 60). Figure 60 also includes the formula for strain rate.

Schular & Hansson used an accelerometer attached to the free end of the mounting frame to determine the deflection strength of concrete [297]. In addition, the method of spalling brittle rods has been used to determine the fracture energy of concrete [297, 298].

Gálvez *et al.* recorded the spall process in specimens of aluminium ceramics using a high-speed camera [301], while the tensile and compression impulses were recorded using gauges (Figure 61). They used a light meter to measure the velocity of the striker bar.

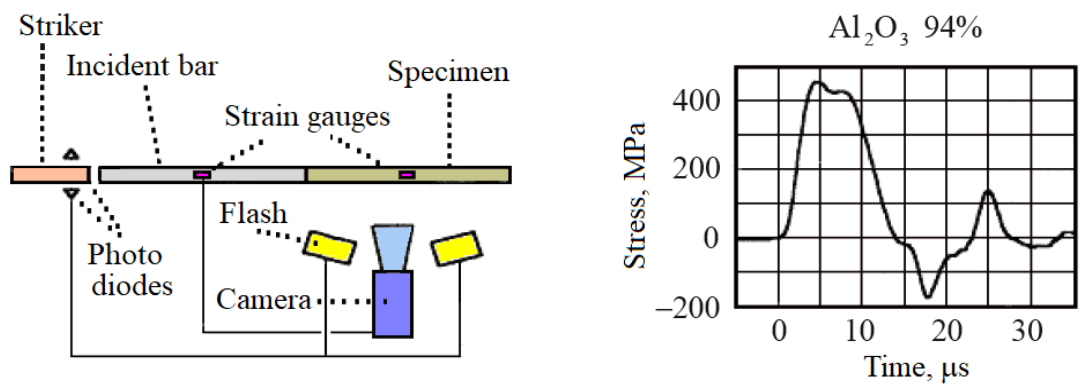

*Figure 61. Schematic diagram of the test set-up and a test result for alumina ceramics. From [301].*

To investigate the dynamic fracture of materials at high temperature, specimens were placed in a high-temperature furnace [303]. To reduce heat outflow, the input bar was connected to the specimen via a transfer bridge (Figure 62), which was also used to record pulses. The connections between the bar, the loading bridge and the specimen were made by means of an adhesive joint. Since the tensile impulse passes through the adhesive joint with an amplitude close to the tensile strength of the specimen, the strength of the adhesive joint must be higher than the breaking strength of the specimen material.

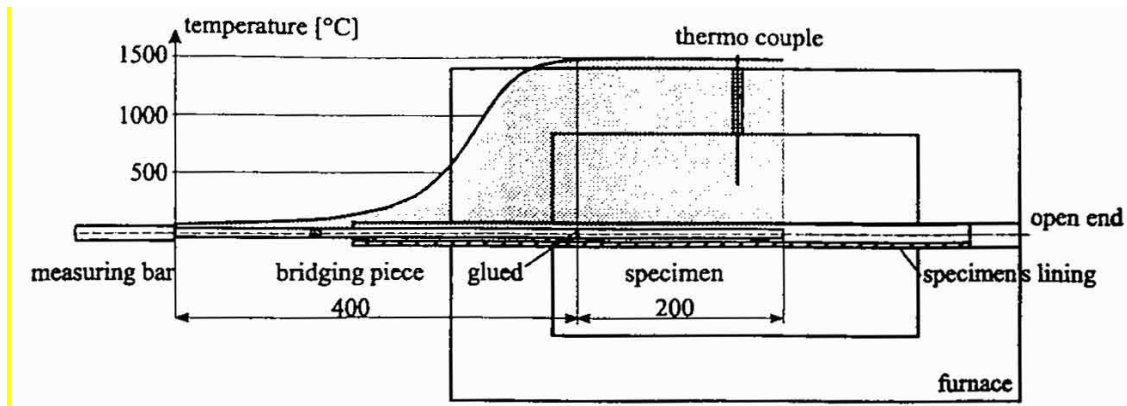

*Figure 62. Experimental set-up and the temperature distribution in the unit bridging piece – specimen. From [303].*

The main advantage of the arrangement, however, is related to its adaptability to dynamic testing of ceramics at elevated temperatures. Applied to specimens located within an open-end furnace, Figure 62, the arrangement includes a bridging piece of the same material and diameter as the specimen itself. It is placed into the thermal transition zone, and serves to bridge the inhomogeneous temperature field at the entrance to the furnace. The specimen itself is positioned in the area of homogeneous temperature distribution. The typical distribution in the unit, Figure 62, has as a consequence a variable wave velocity field in the transition region, due to thermal expansion and the thermal dependence of the elastic moduli of the ceramics.

In order to assess the reliability of the results obtained using these methods and to investigate the influence of inertia, numerical simulations of the three experimental schemes described above were performed by Lu & Li [304]. They concluded that the methods are reasonable, and the results obtained using them (which show an influence of the strain rate on the dynamic behaviour of brittle materials) are reliable.

The spall strength of brittle media can also be studied under conditions of lateral pressure [305]. The experimental set-up is shown schematically in Figure 63. Compression of the specimen was achieved by the pressure of the liquid in the tube surrounding the region where the fracture was predicted to occur. The movement of the spalled rock fragment was measured by interrupting a flat laser beam.

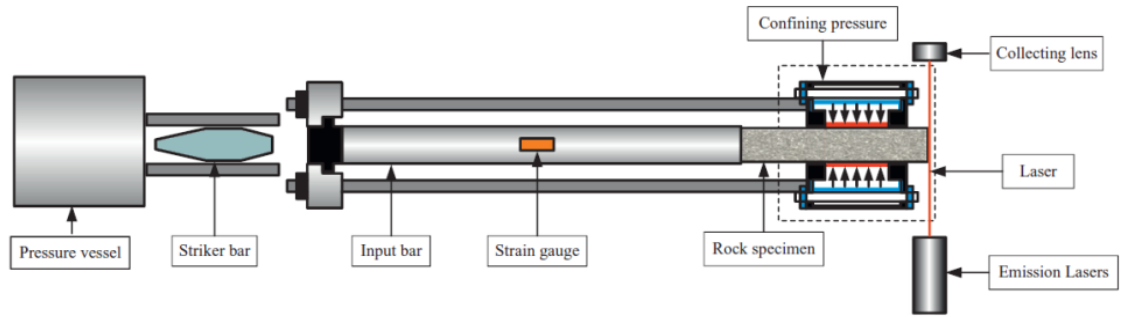

*Figure 63. Schematic diagram of the Kolsky bar apparatus in which lateral pressure was applied hydraulically to the specimen*

In the method for studying spall in long rods, when a stress wave propagates from the input bar into the specimen, there are continuity conditions that determine the energy and velocity at the interface subject to conservation of momentum. However, if a lateral confining pressure is applied to the specimen, momentum is not maintained, and it is difficult to measure the velocity of the spall fragment. Therefore, this study used a modified SHPB device with a system of static pre-compression of the specimen to measure the strength of rock at various compression pressures.

### 2.5 Ensuring uniformity of specimen stress-strain state

Unlike ductile metals, brittle materials have small fracture strains (< 1%). Therefore if the load is applied to a brittle material as fast as it usually is when testing a ductile one, often different parts of the specimen will break at different times, so that, for example, the impacted end of the specimen may fragment while the rear end is still intact. The conditions for constant strain rate and stress equilibrium between the front and rear end of the specimen during most of the test must also be met. To find out to what extent these conditions are met during a test, it is important to record how both strain and strain rate vary with time [306, 307].

The time to reach an almost constant strain rate is mainly determined by the rise time of the input pulse. The time $t_0$ required for the leading edge of the input pulse to pass through the specimen once is determined by the specimen length $L_s$ and the velocity $C_s$ of longitudinal elastic waves in the specimen: $t_0 = L_s/C_s$. It has been suggested by a number of authors that the time $t_{eq}$ needed to achieve stress equilibrium should be some multiple of $t_0$. Lindholm suggested a figure lying between 5 and 10 [308], Davies & Hunter suggested a factor of $\pi$ [136], and Ravichandran & Subhash suggested it should be at least four times [309]. Using Ravichandran & Subhash's suggestion for a specimen for which $L_s = 50$ mm and $C_s = 5000$ m/s, $t_{eq}$ would be about 40 microseconds.

Large-diameter Hopkinson bars should be used to study the properties of brittle heterogeneous materials, such as rocks. However, their use creates certain problems, namely:

(1) a large diameter striker requires a large calibre of gun to accelerate it;

(2) large diameter specimens require a long time to achieve stress equilibrium, which may therefore not be achieved during a test;

(3) since brittle materials fragment at small strains, the effects of friction, inertia, and wave dispersion in a large specimen become more significant.

Therefore, the applicability of the conventional SHPB technique must be carefully studied before reliable interpretation of dynamic experimental data of brittle materials can be performed.

To test brittle materials such as rocks that have an almost linear stress-strain relationship up to failure, a non-dispersive gradually increasing pulse in the input bar is required. Without the correct formation of the incident pulse, it is difficult to achieve dynamic stress equilibrium in such materials, because the specimen can fragment at the end that is in contact with the input bar soon after the loading pulse arrives [310, 311].

The shape of the loading pulse in a traditional SHPB is approximately trapezoidal, accompanied by fluctuations at the transition between the rising section of the pulse and its plateau. Oscillations generated by the rapidly rising section of the incident pulse make it difficult to achieve an equilibrium stress state in the specimen and ensuring the strain rate is constant. However, the stresses must be in balance for a correct test. As just discussed, in order to achieve an equilibrium stress state, it is required that the loading pulse must pass back and forth through the specimen 3-4 times [309]. So in order to achieve accurate measurements in SHPB tests of brittle materials, the dynamic load must be increased slowly enough for the specimen to experience loading in an almost quasistatic manner.

The possibility of using a pulse shaper in the SHPB system was considered by Frantz *et al.* [312]. They found that a slowly increasing incident pulse is the preferred method of loading in order to minimize the effects of dispersion and inertia and thus contribute to the dynamic equilibrium stress state of the specimen.

One of the ways to change the shape of the falling pulse and slow its growth is to change the shape of the striker. For example, Christensen *et al.* [226] used strikers with a truncated cone at the impacting end, and Frantz *et al.* [312] used a striker bar having an impact surface with a large radius of curvature. Li *et al.* [313] and Zhou *et al.* [221] used a cone-shaped or skew-toothed striker to create an approximate semi-sinusoidal load curve that can provide dynamic stress equilibrium and a constant strain rate in a rock specimen.

To achieve a dynamic balance of forces, a cone-shaped striker can be used to create a semi-sinusoidal loading pulse. Figure 64 shows the configuration of a striker bar that provides such a pulse in a 50 mm diameter SHPB system [221].

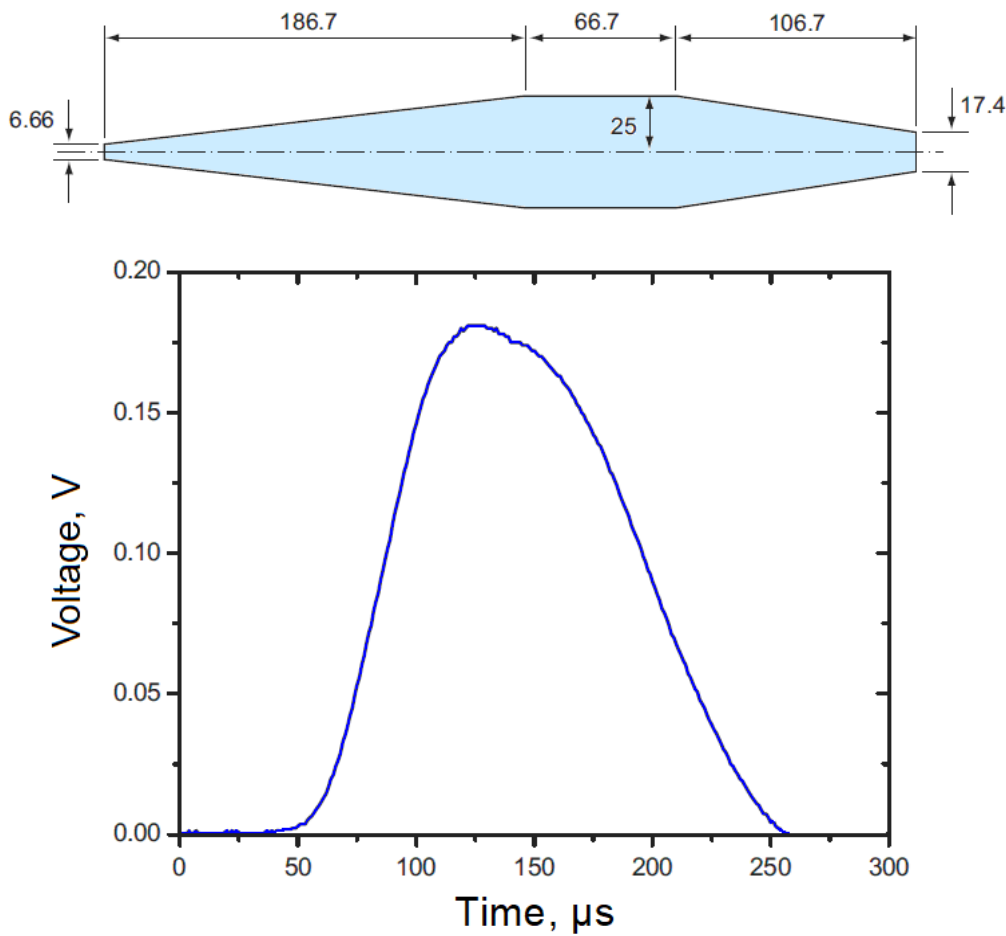

*Figure 64. The geometry of a cone-shaped impactor (dimensions in mm) and the loading pulse it generates.*

The second approach to creating an incident pulse of the required shape is to place a rod of variable shape [314] or an additional (ductile) specimen between the striker and input bars [145].

The third, rather simple and convenient way to generate an incident pulse is to place a small thin disc of soft material between the striker and the loading rod [133, 312]. The disc is called a pulse shaper and can be made of paper, aluminium, copper, brass, or stainless steel with a thickness of 0.1-2.0 mm.

A number of authors [310, 311, 315] have proposed using a pulse shaper in the form of a thin disc of annealed copper on the impacted end of the input bar. During a test, the striker loads the pulse shaper in front of the input bar, thereby generating a non-dispersive pulse that propagates into the input bar (Figure 65). Such pulses have a long rise-time and rather than a force plateau, the load falls. These features contribute to the establishment of a dynamic balance of forces in the specimen [310, 311]. The function of the pulse shaper is to (a) guarantee a constant strain rate during loading and (b) maintain a balance of forces through the specimen.

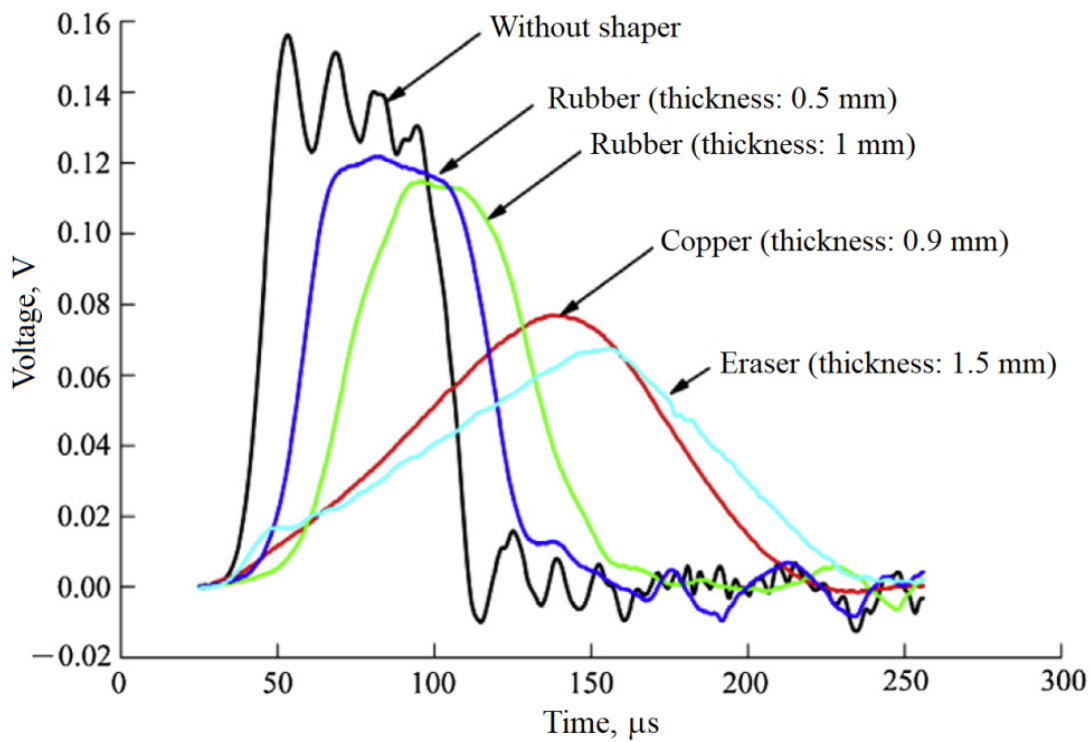

*Figure 65. Shapes of loading pulses when using pulse shapers made of various materials.*

Depending on the materials under study, different load pulse shapes are required. These can be achieved with the correct design of shaper. Experimental results were presented by Frantz *et al.* [312], confirming that a correctly formed load pulse can not only ensure the equilibrium of stresses in the specimen, but can also create an almost constant strain rate in the specimen. Gray & Blumenthal also discussed these issues in their review [316].

Figure 65 shows that a wide variety incident pulse shapes in the input bar can be obtained by changing the material and geometry of the pulse shaper [216]. This figure also shows that when using an appropriate pulse shaper, the loading pulse changes from a rectangular to a trapezoidal shape and the oscillations that are present in experiments performed without a shaper almost completely disappear.

These oscillations in the loading pulse are caused by the non-uniformity of the stress and strain in the input bar near the impacted end. If in addition, a small rubber disc is placed in front of the copper gasket, the slope of the later portion of the pulse can be further reduced to the desired value [216]. In this case, the forces at the two ends of the specimen have no fluctuations and are almost identical until the maximum value is reached. Thus, balance of the dynamic forces at both ends of the specimen can be achieved.

Dai *et al.* showed that with correct procedures, reliable data can be obtained of the dynamic compressive and tensile strength of rocks using the SHPB [245].

## 2.6 In shear

Shear strength is one of the most important mechanical properties of brittle materials. It plays a vital role in applications such as mining technology and geotechnical engineering. Although there are standards for measuring the quasistatic shear strength of brittle materials, the shear behaviour of rocks under dynamic loads is not well understood.

Huang *et al.* designed a ring device for an SHPB system for determining the dynamic shear strength of rocks (see Figure 66) [232]. They used thin disc specimens to minimize bending stresses. Fine-grained isotropic sandstone was used to demonstrate the measurement principle. They found that the shear strength of rocks increased with increasing loading rate. This device is applicable to fine-grained rocks with intermediate hardness.

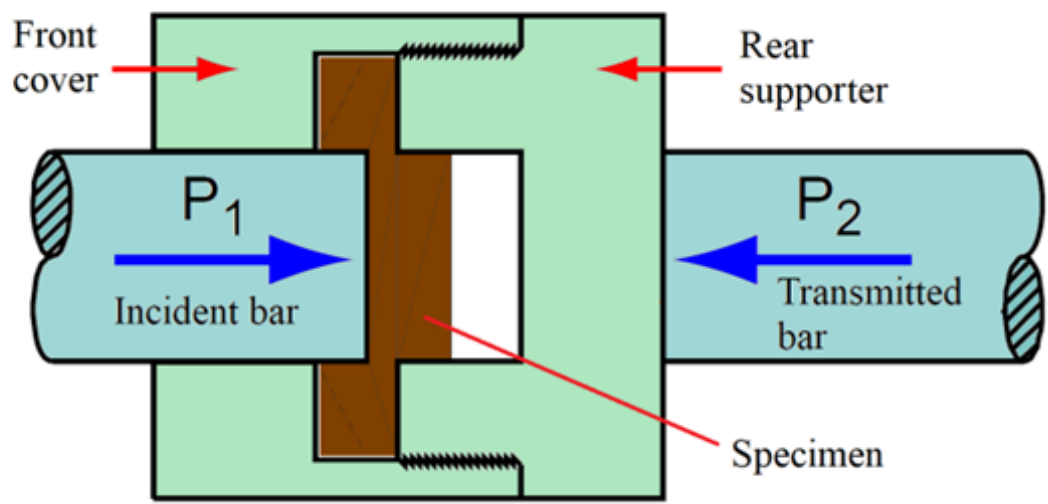

*Figure 66. Schematic diagram of ring shear test apparatus for used in an SHPB.*

## 2.7 For dynamic fracture toughness

A number of standard methods have been proposed to determine the fracture toughness of type I rock specimens. These include the short-rod method (SR), the chevron-notched specimen bend method (CB), and the cracked chevron-notch Brazilian disc (CCNBD) (see designations in Table 2). Depending on the angle between the notch and loading planes, type I or II failure is realized.

To study the type I fracture toughness of rocks, we use modifications of the Brazilian disc method with different notches loaded with compression pulses (Figure 67), as well as a semicircular specimen with different notches (Figure 68).

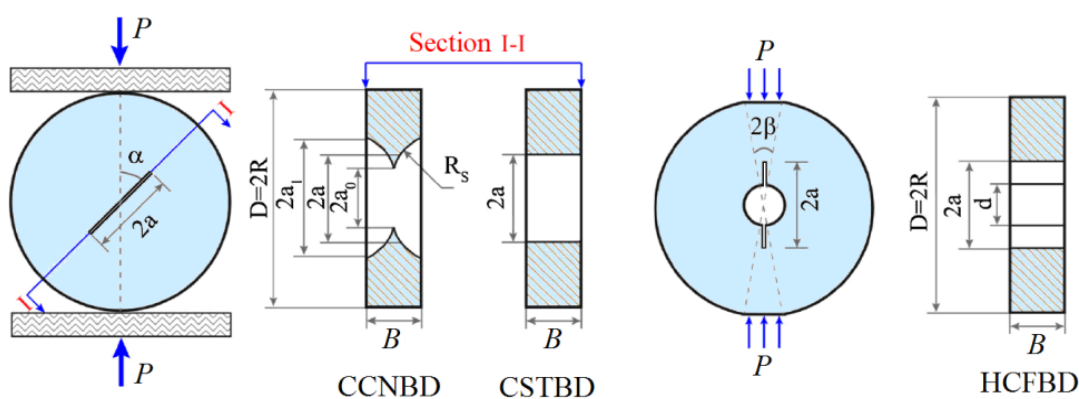

*Figure 67. Various compression test schemes for Brazilian disc specimens to determine fracture toughness.*

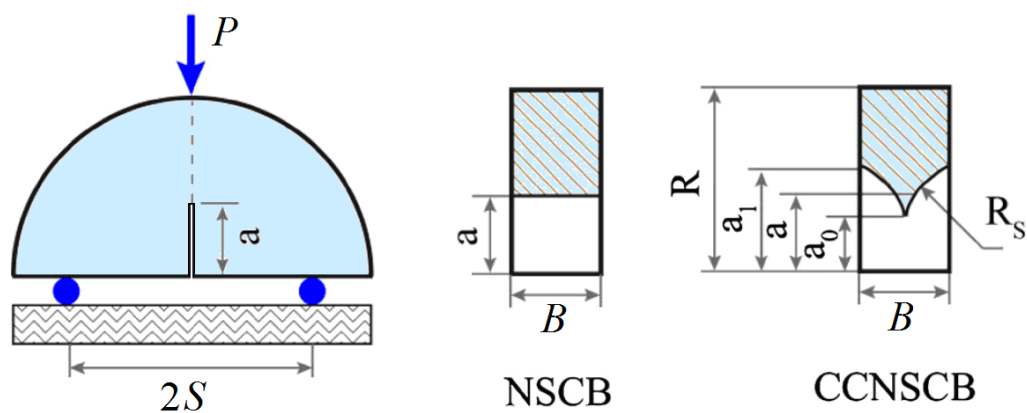

*Figure 68. Test schemes for bending semicircular specimens to determine the fracture toughness.*

Once the size of a brittle specimen is determined, the time $t_f$ at which fracture begins is the only important factor affecting the measurement. The rate of loading is usually expressed in terms of the fracture toughness:

$$\dot{K}_I^{dyn} = K_{Id} / t_f \ .$$  (99)

The application of the digital image correlation (DIC) method for measuring the fracture process of a specimen with a notched semi-circular bend (NSCB) was presented by Gao *et al.* [317]. The NSCB method was proposed by the International Society for Rock Mechanics (ISRM) to measure the impact strength of a rock fracture with a given external load and geometric parameters. Using the DIC method in combination with ultra-high-speed photography, it is possible to measure a large number of parameters associated with the fracture of an NSCB specimen loaded using an SHPB system (Figure 69).

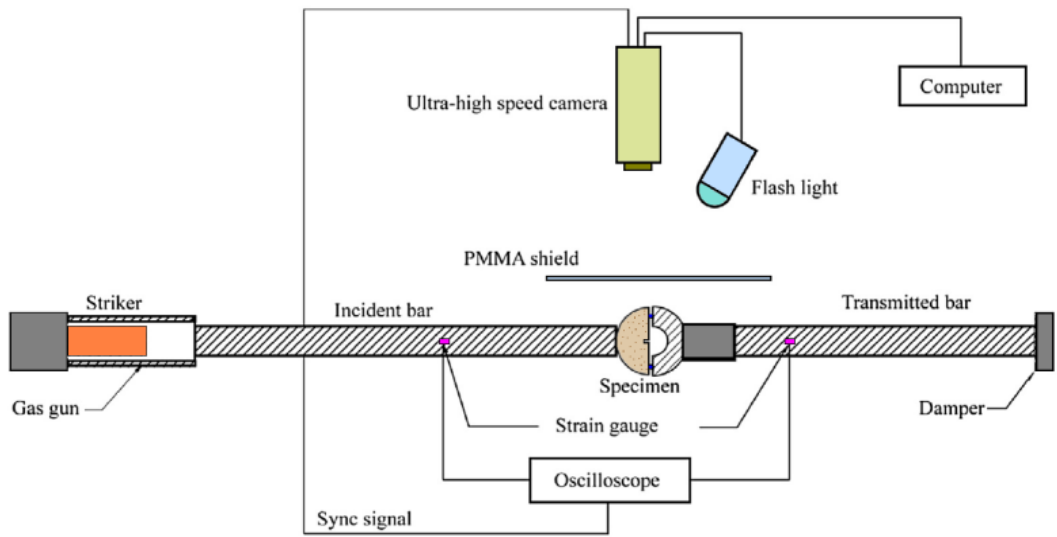

*Figure 69. Scheme for dynamically testing rock specimens in fracture.*

Using the DIC method, we determined the displacement and strain fields during the specimen fracture process. Then the location of the crack tip, the fracture toughness, and the crack propagation velocity were determined from the strain fields. Compared to traditional NSCB tests, the DIC optical method provides much more information about the fracture process.

To provide a three-point bending load at the end of the output bar, an adapter with two contact supports was used [317] (Figure 70).

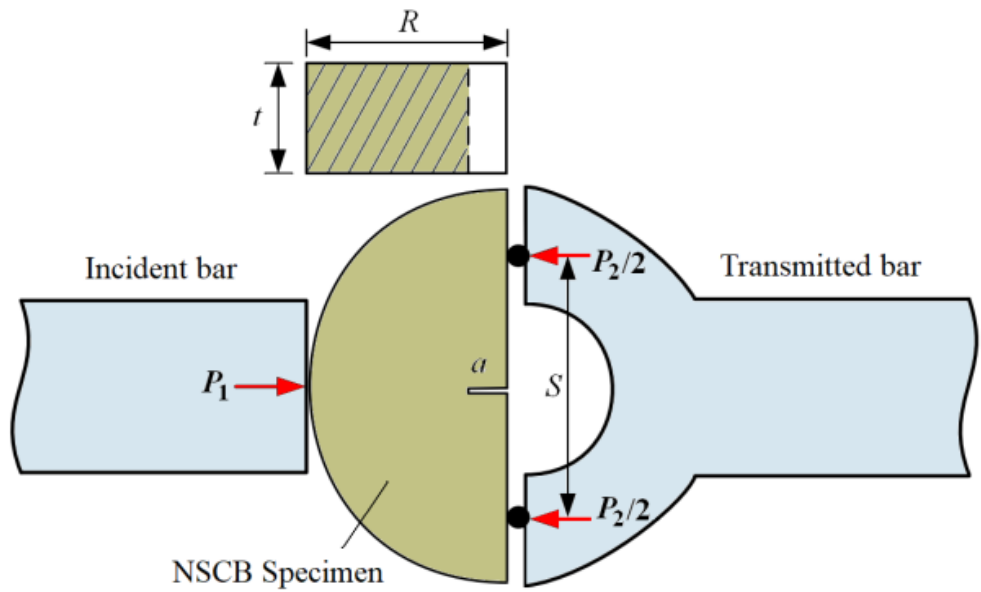

*Figure 70. Loading scheme for a semicircular specimen with a notch.*

The short rod method is also used to estimate the fracture toughness in mode I. The method was proposed by Barker [318, 319] and after the studies by Ouchterlony [239, 320], it became the method recommended by ISRM.

When making a short-rod specimen, a rectangular gripper groove with a width $T$ is cut at one end of the specimen, and two slits with a width $t$ are cut with opposite angles to form a triangular ligament (Figure 71).

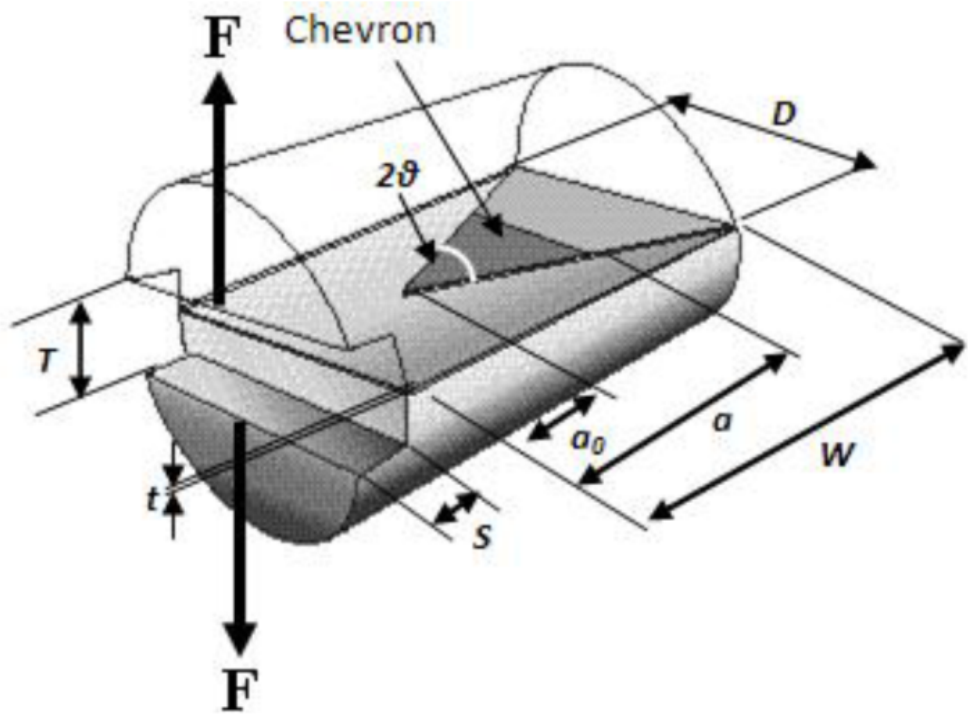

*Figure 71. Dimensions and loading scheme of a short-rod specimen.*

During testing, the specimen is positioned between the end of the output bar and the wedge fixed to the end of the input bar. The compression wave that is transmitted through the wedge causes the middle triangular part of the specimen to open and break (Figure 72). In specimens with this configuration, a pre-grown fatigue crack is not required.

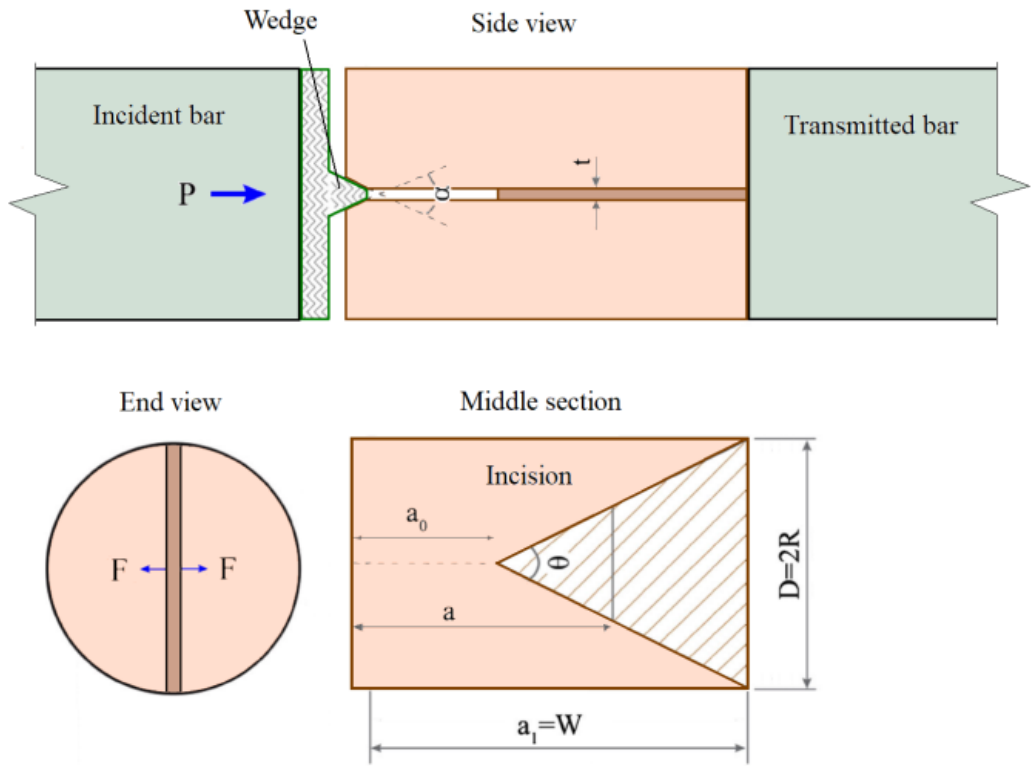

*Figure 72. Testing scheme for a modified short-rod specimen.*

## 2.8 In combined compression-shear loading

Rocks in the Earth's crust are in both a complex stress state and a complex environment with defects such as cracks, pores, etc. The mechanical behaviour of rocks depends on both the amount of load and the rate of deformation. Studies of determining the ratios and the strength properties of rock masses in static and dynamic complex stress states are of great relevance in the fields of mine construction, water management and hydropower, geotechnical engineering, and earthquake prediction [194, 321]. In particular, engineering structures for rock masses under complex loads require precise rock characteristics, taking into account the triaxial stress state and the influence of the strain rate. Special efforts have been made to improve experimental methods for studying the load-dependent properties of rocks under dynamic loading.

Xu *et al.* [274] proposed a new technology for loading a rock specimen with a combination of compression and shear stress based on a modification of the SHPB method in which the ends of the input and output bars have counter-inclined surfaces (Figure 73). Other modifications include the incorporation of a pulse shaper, an input bar with a gradually increasing cross-section, and a specially designed specimen configuration outlined within the red dotted curve in Figure 73 (shown in more detail in Figure 74*a*).

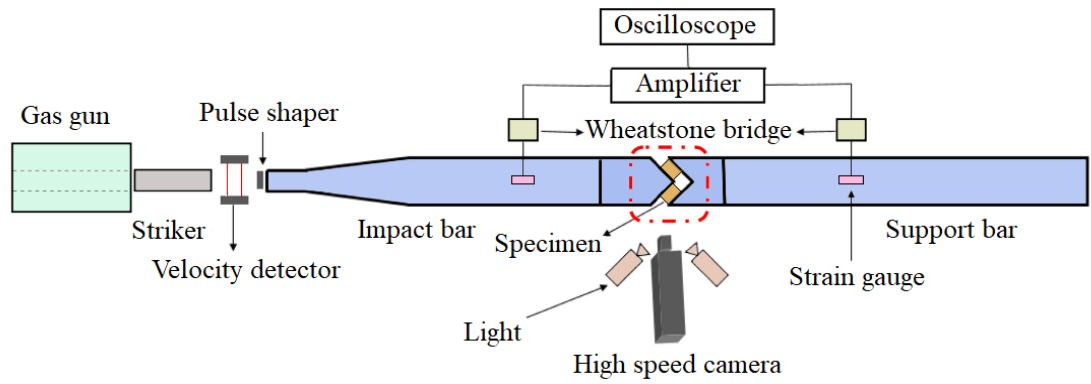

*Figure 73. Schematic diagram for combined compression-shear testing in an SHPB.*

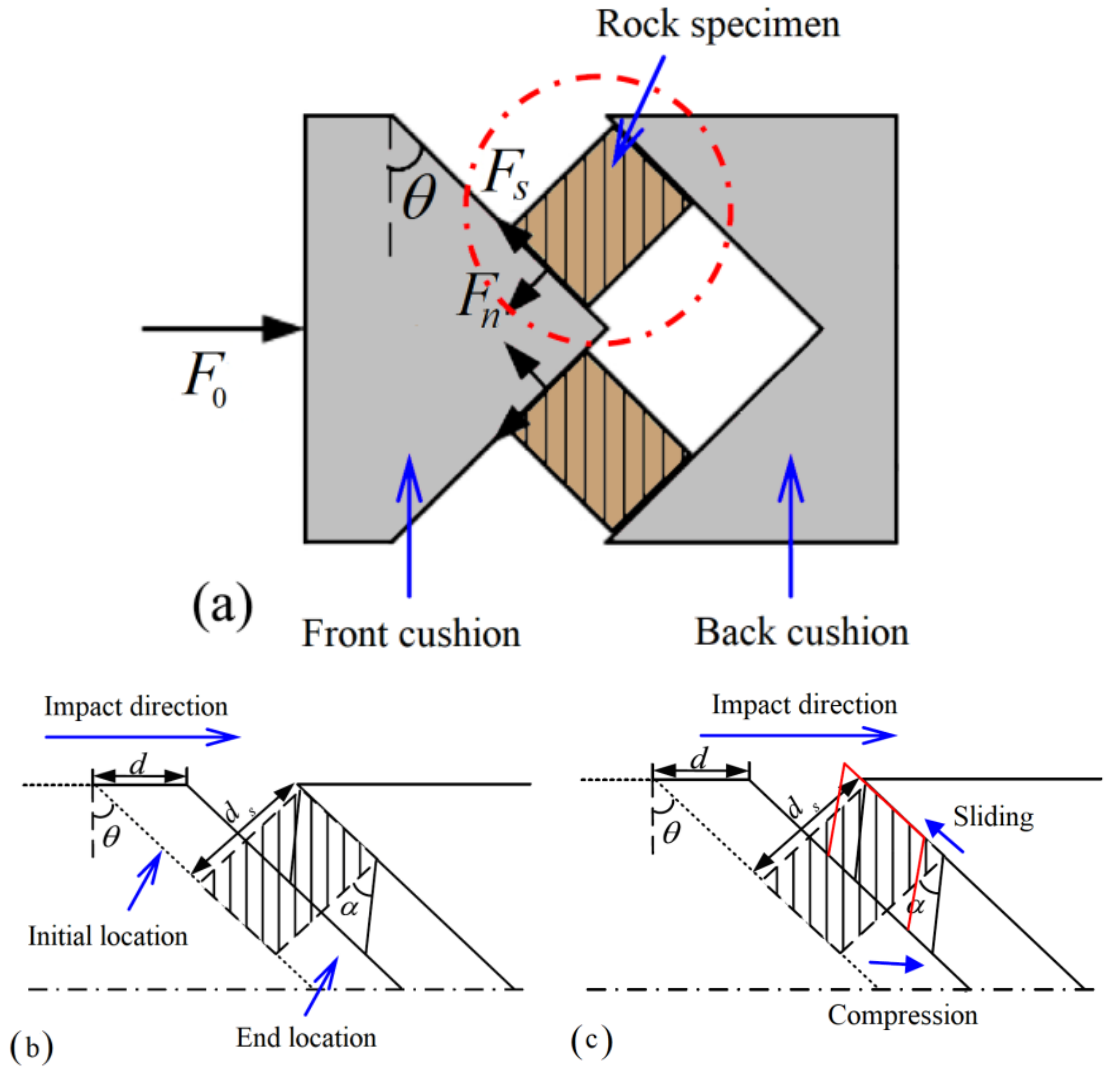

*Figure 74. Analysis of forces and strains of a specimen under combined compression-shear loading.*

In this configuration, two cylindrical or cubic rock specimens are located between the inclined end surfaces of the input and output bars and are subjected to compression loading. When a state of stress equilibrium is reached in the test specimens, reflection and transmission of stress waves at the internal inclined surfaces of the input and output bars can be ignored. Thus, although the region where the specimens are is under uniaxial compression, the rock specimens within are under a combined compression-shear load.

The analysis of forces and strains for two symmetrically arranged specimens is shown in Figure 74*a*. The analysis of specimen deformation without taking into account surface slip (Figure 74*b*) and with surface slip (Figure 74*c*) is presented below. The dotted line indicates the initial location of the inclined loading surface.

Using steel as the standard material, the method is calibrated by a series of quasistatic and dynamic experiments. A reliable method of data processing for the normal stress ($\sigma_n$) and shear stress ($\tau$) distribution on an inclined surface is proposed. The results show that the associated compression and shear responses of steel exhibit a clear strain rate effect.

When the angle of inclination of the loading surface ($\theta$) changes, which represents the angle between the direction of impact and the direction of the normal to the rock specimen, the compression and shear components imposed on the specimens have different values. The length of the specimens is determined by the required loading rate and the angle of inclination $\theta$. The experimental results should be checked for each loading rate and angle of inclination by comparing the loading, reflected and transmitted pulses.

Xu *et al.* performed a series of experiments to study the behaviour of granite with regard to compression and shear under quasistatic loading (strain rates ~$3 \cdot 10^{-5}$ s$^{-1}$) and dynamic loading (strain rates of 50 s$^{-1}$ and 100 s$^{-1}$) [274]. They performed tests with five tilt angles: 0˚, 15˚, 30˚, 45˚ and 60˚. The slopes of the separated normal and shear stress-strain curves show a clear dependence on the load and the strain rate. These two effects become more important as the tilt angle increases. The criterion for the fracture of a granite specimen at different strain rates can be described by the Drucker-Prager model. The strength of the granite also showed an obvious dependence on the rate at which the load changed. This method is useful for studying the dynamic properties of rocks in complex stress states.

**2.9 In multiaxial loading**
In many mining applications, such as underground blasting, it is extremely important to know the dynamic strength of rocks when they are stretched in confined conditions.

A multiaxial load that restricts the radial breaking of a specimen can be classified as an axial constraint, a lateral constraint, or a three-axis constraint. These are very important for underground mining engineering tasks. The mass of rock around an opening underground can be divided into three zones depending on the distance to the hole. The states of constrained stress vary from a predominantly hydrostatic state in the far zone to a triaxial state in the intermediate zone and to a tensile state in the near zone. In order to effectively account for the dynamic responses of rocks underground, it is essential that rock specimens are tested under dynamic loading in all three of these stress states.

Two approaches to creating multi-axis loading of a specimen in the SHPB have been developed: true triaxial loading and multiaxial loading with restriction of the radial expansion of the specimen. For the first, Cadoni & Albertini developed a real three-axis loading device (Figure 75) [269].

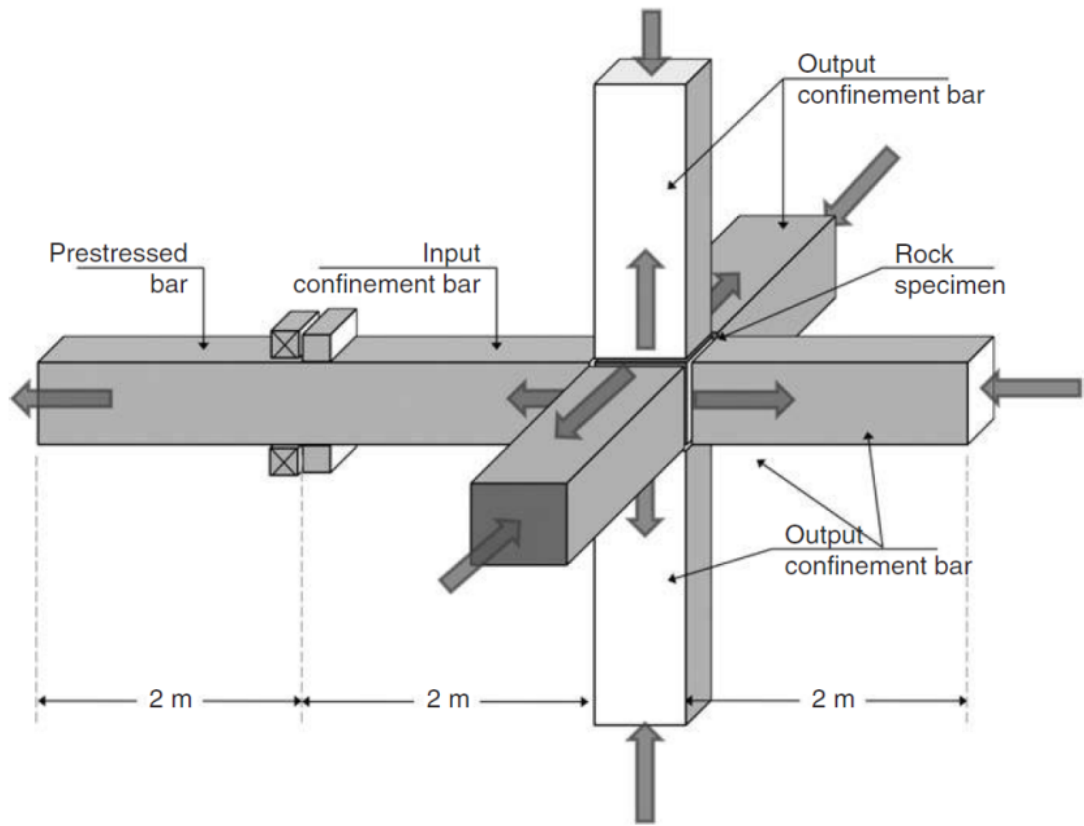

*Figure 75. Scheme for testing rock specimens in three-axis compression.*

However, it is difficult to synchronize very short duration dynamic loads applied in more than one direction. Therefore, Cadoni & Albertini apply static compression to two planes of their cubic specimens using hydraulic jacks and dynamic loading in the third axis by a compression wave excited by rapidly unloading a statically-stressed rod [269].

Various methods have been used to laterally constrain deformation in SHPB tests of brittle solids: a hydraulic pressure chamber [226, 322, 323] or a passive thick confining jacket [120, 324]. For fine-grained brittle solids, such as ceramics, other types of constraints are possible, such as a thermally shrunk-on metal sleeve [325-327] and planar confinement [328].

Lindholm *et al.* performed a pioneering study on the dynamic testing of rocks under hydrostatic compression (both radial and axial) and proposed a system for measuring the dynamic properties of rocks under three-axis confinement [329]. It consists of an SHPB system with two hydraulic cylinders, one of which encloses the specimen to limit its radial expansion. This cylinder creates confining stresses in the transverse direction, while the axial cylinder creates an axial confining stress on the rear end of the output bar (Figure 76).

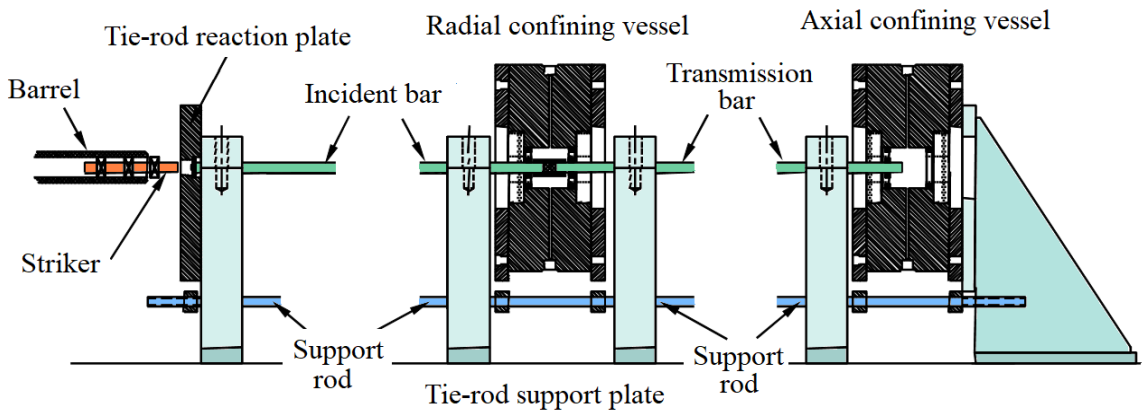

*Figure 76. Scheme for testing rock specimens in three-axis compression*

Using a very similar idea, Frew *et al.* developed a method that can apply a hydrostatic constraint with four tie rods connecting two cylinders (Figure 77) [330]. The way to achieve this constraint is to subject a cylindrical rock specimen to a compressive lateral and axial hydraulic pressure, and then maintain this fluid pressure in both cylinders. In their paper, Frew *et al.* presented dynamic diagrams for limestone under compression at hydrostatic pressures of up to 200 MPa and strain rates of 400 s⁻¹.

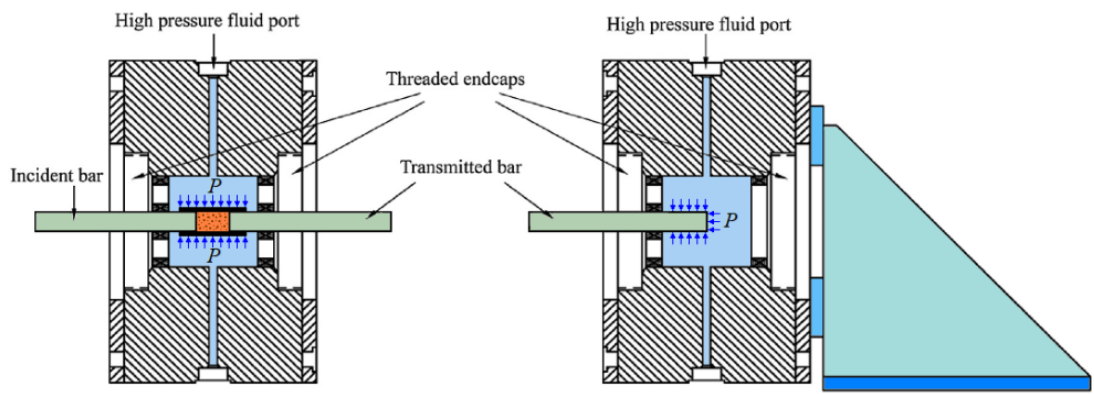

*Figure 77. Diagram of hydrostatic strain limitation.*

Various measurement techniques are also discussed by Frew *et al.* [330] for testing rocks using the SHPB method: X-ray micro-computed tomography (CT), laser gap gauge (LGG), digital image correlation (DIC), moiré, caustics, photoelastic coatings, and dynamic infrared thermography.

To simulate the stress-strain state of underground rocks, Wu *et al.* first applied hydrostatic pressure to Brazilian disc specimens and then applied a dynamic load using an SHPB (see Figure 78) [331]. In order to ensure the dynamic force balance in the SHPB experiment, a pulse shaper was used so as to generate a slowly increasing stress pulse.

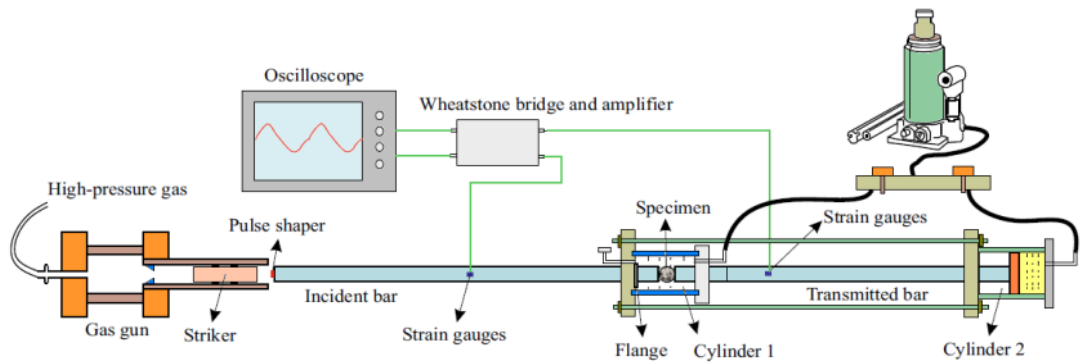

*Figure 78. Apparatus for applying hydrostatic compression to Brazilian disc rock specimens under dynamic compression.*

Wu *et al.*'s hydraulic system consists of: cylinder 1 that creates a lateral confining pressure on the rock specimen and cylinder 2 that applies an axial preload to the specimen through the output bar. Their hydraulic system is similar to Frew *et al.*'s previously described hydraulic system for performing dynamic compression tests under hydrostatic compression [330]. However, there is a major difference between the two designs namely the the compressed threaded connecting rods in Wu *et al.*'s design are much shorter and, therefore, they are less susceptible to damage from longitudinal bending.

Wu *et al.* dynamically tested five groups of Canadian granite specimens (having a quasistatic tensile strength of 12.8 MPa measured using the Brazilian test) in hydrostatic compression of 0, 5, 10, 15 and 20 MPa at different loading rates using an SHPB [331]. Their results show that the dynamic tensile strength increases with hydrostatic pressure. They also observed that at a given pressure, the dynamic tensile strength increases with increasing loading rate. In addition, the increment in dynamic tensile strength was found to decrease with pressure, similar to the known quasistatic tensile behaviour of rock subject to hydrostatic pressure. The tested specimens were analyzed using X-ray microcomputer tomography. They noted that the observed pattern of cracks is consistent with the experimental results.

### 2.10 A summary of the results of dynamic strength testing of brittle materials using the Kolsky bar method

Analysis of the available results of dynamic tests in compression, tension and shear of various brittle materials performed using the Kolsky method and its various modifications have allowed us to establish the following.

Dynamic tests of fine-grained concrete during compression and tension have shown that the strength of this material increases with strain rate (or a decrease in the loading time). The main measure of the behaviour of concrete under dynamic loading is the Dynamic Increase Factor (DIF), i.e. the ratio of the strength of concrete under dynamic loading to its quasistatic strength. The criterion for evaluating the duration of resistance to concrete overload (compared to its quasistatic strength) is the time of failure delay.

Figure 79 was compiled by Bischoff & Perry [195] and summarises the results of compression tests of various concretes carried out by a number of researchers [260-262, 332-359]. It presents in graphic form the dependence of the DIF on the strain rate. A description of the test rigs, geometric dimensions, shape, maximum size of aggregate, water/cement ratio, their age, quasistatic strength, conditions of hardening, DIF, and strain rate during tests and other more detailed information about the experimental results may also be found in their review paper [195]. Figure 79 also shows plots of the analytical dependences of the DIF on the strain rate proposed in [360] by the European International Committee on Concrete (Comité Euro-International du Béton, or CEB) for describing the behaviour under dynamic compression of concretes that have quasistatic strengths of 20 MPa and 50 MPa.

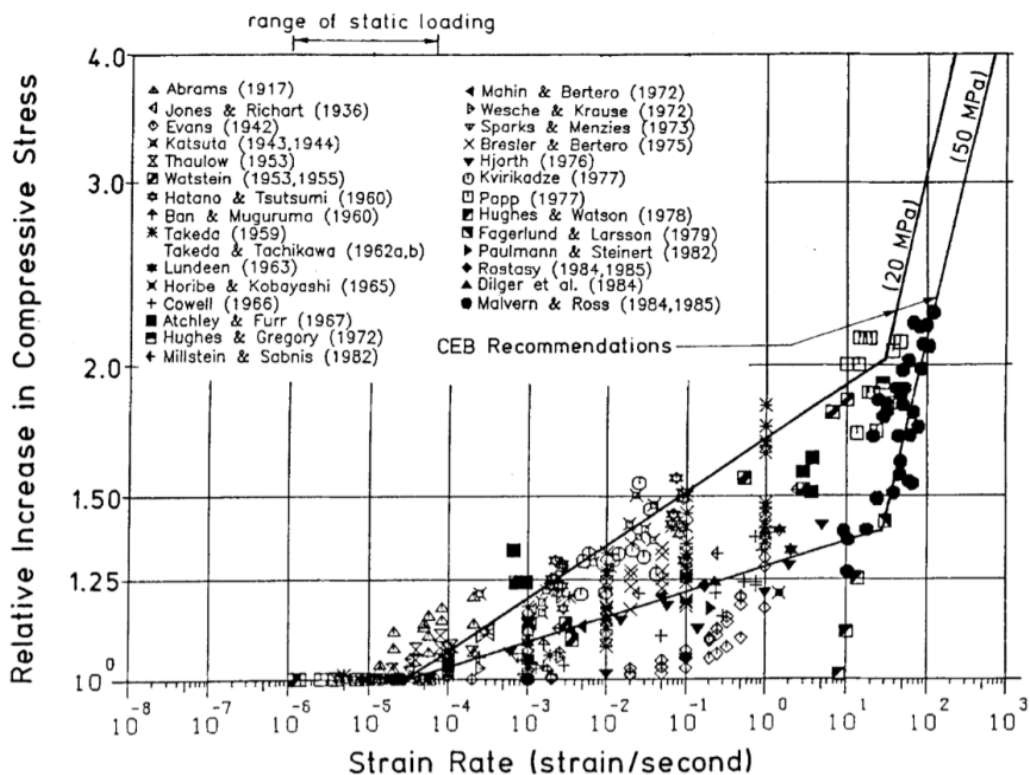

*Figure 79. Plots of the dependence on the strain rate of DIF values for concrete obtained by some researchers when testing various concretes in compression. From [195].*

Schuler & Hansson reported an investigation of the fracture of high-strength concrete using a an SHPB modified for performing spall experiments at strain rates of up to 100 s$^{-1}$ [361]. They measured the tensile strength and fracture energy. The fracture energy is of particular interest because it is a key parameter for the laws of damage to brittle materials.

In Figure 80, a comparison is made between the results obtained by Schuler *et al.* and other researchers for the normalized tensile strength of fine-grained concrete MB35 (35 MPa) [297]. The difference in the effect of the strain rate between 35 MPa concrete and 130 MPa concrete (HPC) studied in the work is also shown.

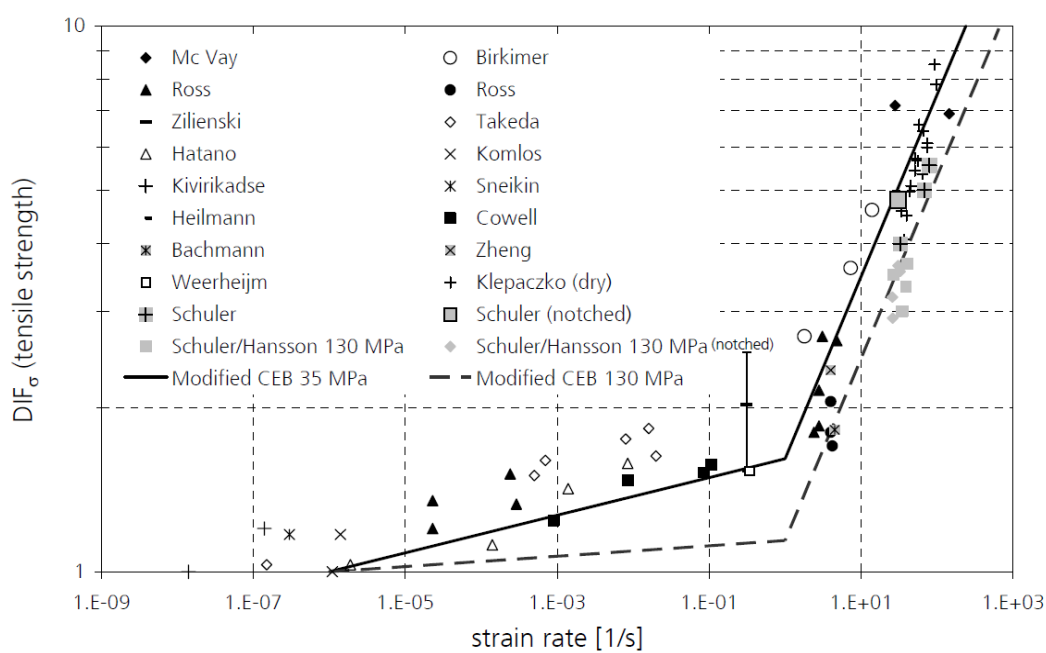

*Figure 80. Plot showing the results obtained by a number of researchers of the effect of strain rate on the DIF values for various concretes tested in tension. From [361].*

The results for the fracture energy are shown in Figure 81. There is only a small amount of data in the literature. Weerheijm measured the fracture energy in direct tensile tests using an SHPB [362]. In spall tests, Schuler & Hansson found that there was no significant difference between 35 MPa and 130 MPa (HPC) concretes using notched specimens [361]. Notched specimens are suitable for measuring fracture energy, since the crack surface is equal to the cross-section in the notch.

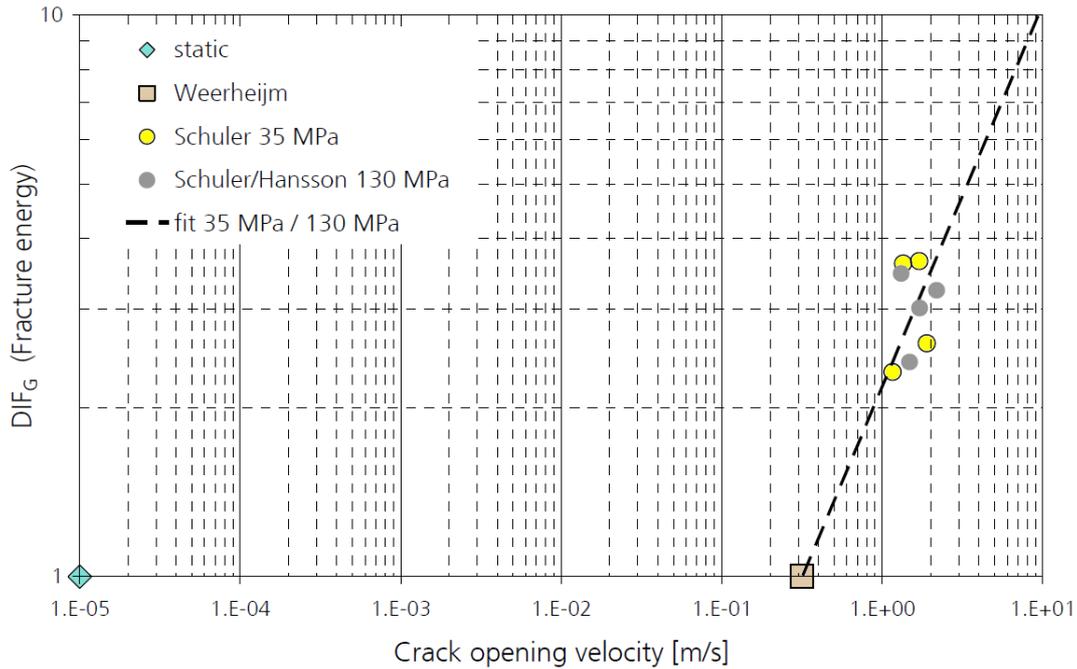

*Figure 81. Plot of the data obtained by some researchers of the dependence of DIF values on the strain rate when testing the tensile strength of various concretes. From [361].*

Zhang & Zhao plotted the normalized dynamic tensile strength obtained by various methods as a function of both the strain rate (Figure 82) and the stress growth rate (Figure 83) [194]. It can be seen that the strain rates of direct tension (DT) results are higher than the rates obtained by indirect tension (In-DT) methods, since the specimen sizes in the first case are usually smaller.

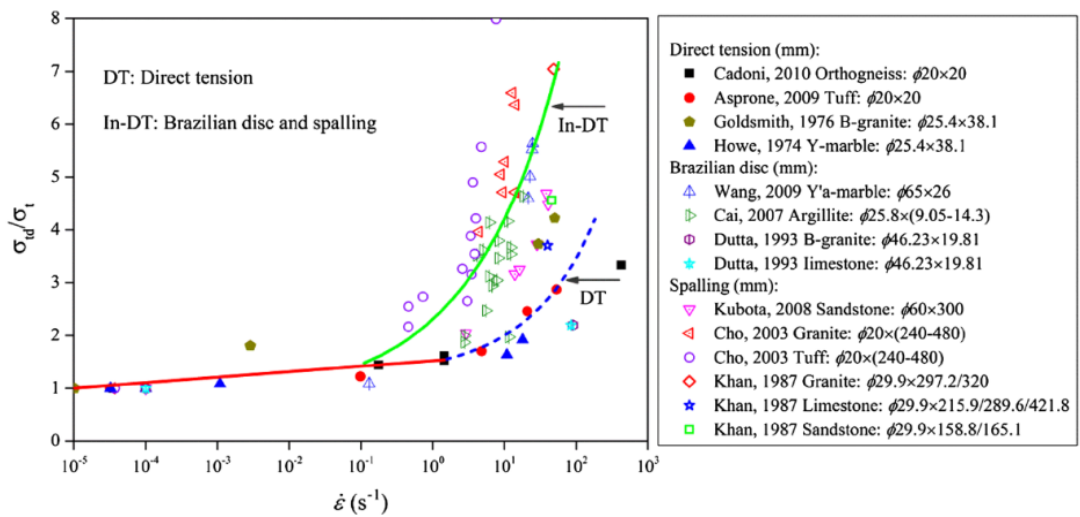

*Figure 82. Plot of the dependence of normalized dynamic tensile strength of various rocks on the strain rate. From [194].*

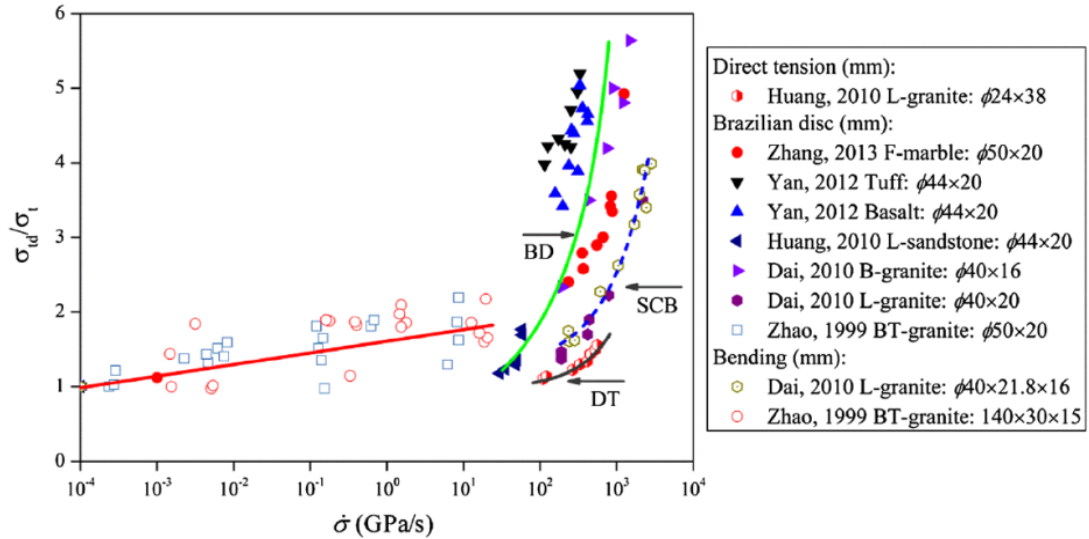

*Figure 83. Normalized dynamic tensile strength of various rocks as a function of stress rate. From [194].*

Figures 82 & 83 shows that the dynamic tensile strength increases sharply above a critical strain rate ($100$ s$^{-1}$) and stress growth rate ($10^2$ GPa/s). This critical strain rate is consistent with the values in semi-empirical equations previously published for concrete [266, 363].

The dynamic crack resistance $K_{Ic}$ (fracture toughness) is determined by the time $t_f$ of the beginning of the fracture and is given by $K_{Id}\left(\dot{K}_I^{dyn}\right) = K_I^{dyn}\left(t_f\right)$. Bazant *et al.* performed bending tests on limestone using the SENB method (see Table 2) using three different specimen sizes [364]. In their experiments, they changed the loading rate by four orders of magnitude at quasistatic loads so that the $t_f$ ranged from 2 to 82,500 s.

They found that the fracture toughness increased slightly with increasing loading rate. Due to the complexity of data processing, only limited data was obtained for intermediate loads.

Figure 84 shows the normalized impact strength for dynamic crack growth as a function of the normalized loading rate for a number of materials [194]. The results were obtained using various methods: SR, NSCB, CCNSCB, and WLCT (see Table 2).

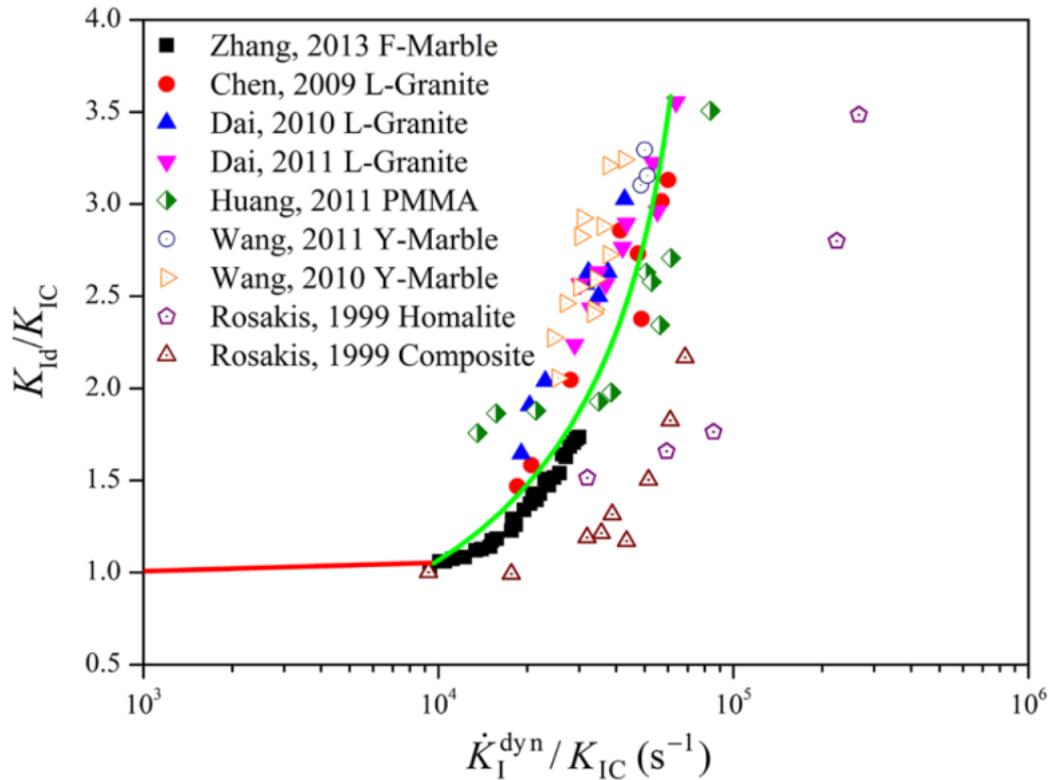

*Figure 84. Normalized dynamic fracture toughness as a function of the normalized loading rate. From [194].*

SENB tests were performed to determine the fracture toughness of shale using a Charpy impact machine with a loading rate of about $10^4$ MPa.m$^{0.5}$s$^{-1}$ [240], granite using a pneumatic hydraulic machine in the range of $10^{-1}$-$10^3$ MPa.m$^{0.5}$s$^{-1}$ [198], and rock using a machine with a reduced weight at a loading rate of $10^3$ MPa.m$^{0.5}$s$^{-1}$ [365].

At higher strain rates, the SEND method has been modified into a dynamic test using a single Hopkinson rod load configuration with a loading rate of about $10^5$ MPa.m$^{0.5}$s$^{-1}$ [250].

**CONCLUSIONS**

As follows from the presented analysis, a complete understanding of the dynamic mechanical behaviour of brittle materials depends on reliable experimental methods that ensure the reliability of the results of test procedures, and effective numerical modelling.

For testing brittle materials, pneumatic-hydraulic, fully gas-powered or drop-weight type machines are usually used as loading devices. At high strain rates, the main tool for studying the dynamic behaviour of materials of different physical nature under different types of stress and strain is the Kolsky bar method which uses two Hopkinson bars in numerous modifications.

When testing brittle materials (concretes, ceramics, rocks), special attention should be paid to ensuring that the strain rate and stress balance on the specimen ends remain constant for most of the test. It is necessary to minimize the effects of heterogeneity of the stress-strain state in the specimen during testing. The strain rate during testing should be well-controlled. For testing brittle materials (rocks), which have an almost linear dependence of the stress-strain curve up to fracture, we need a non-dispersive gradually-increasing loading pulse.

Mechanical loads on engineering structures are usually not uniaxial, and, in addition, the development and calibration of equations of state of materials for numerical modelling require experimental data for various types of stress-strain states, including triaxial and combined, over a wide range of strain rates.

Analysis of the results obtained during testing of brittle materials has shown that almost all the materials considered show a positive sensitivity to the strain rate to a greater or lesser extent, both in uniaxial tests (compression, tension) and in the study of crack resistance parameters.